\documentclass[12pt]{article}
\usepackage[latin9]{inputenc}
\usepackage{geometry}
\usepackage{array}
\usepackage{booktabs}
\usepackage{longtable}
\usepackage{calc}
\usepackage{amsmath}
\usepackage{amsthm}
\usepackage{amssymb}
\usepackage{stmaryrd}
\usepackage{setspace}
\usepackage[authoryear]{natbib}
\usepackage[unicode=true,
 bookmarks=false,
 breaklinks=false,pdfborder={0 0 1},colorlinks=false]
 {hyperref}
\hypersetup{
 colorlinks,citecolor=blue,pdftex}

\makeatletter

\providecommand{\tabularnewline}{\\}

\usepackage{amsfonts}
\usepackage{amsthm}
\usepackage{lscape}
\usepackage{appendix}
\usepackage{dcolumn}
\usepackage{rotating}
\usepackage{comment}
\usepackage{color}

\usepackage{diagbox}
\usepackage[flushleft]{threeparttable}
\usepackage{subfigure}
\usepackage{pgf}
\usepackage{tikz}\usepackage{caption}
\usepackage{tkz-tab}
\usepackage{tikz-3dplot}
\usetikzlibrary{arrows.meta, positioning}
\usetikzlibrary{calc}
\usetikzlibrary{arrows, automata}

\usepackage{changepage}
\usepackage{booktabs}

\usepackage{courier}

\newcolumntype{d}[1]{D{.}{.}{#1}}
\newcolumntype{t}[1]{D{,}{,}{#1}}
\newcolumntype{i}[1]{D{.}{}{#1}}

\theoremstyle{plain}

\newtheorem{innercustomas}{Assumption}\newenvironment{myas}[1]{\renewcommand\theinnercustomas{#1}\innercustomas}{\endinnercustomas}

\newtheorem{innercustomprompt}{Prompt}\newenvironment{myprompt}[1]{\renewcommand\theinnercustomprompt{#1}\innercustomprompt}{\endinnercustomprompt}

\numberwithin{equation}{section}

\makeatother

\begin{document}
\title{Mining Causality:\\
AI-Assisted Search for Instrumental Variables\thanks{I am grateful to Guido Imbens for his encouragement to turn my early ideas and findings into this paper.
I also thank Susan Athey, Orazio Attanasio, Kevin Chen, Davide Coluccia,
Phil Haile, Stefan Hubner, Paul Hufe, Hide Ichimura, Pat Kline, Michal
Kosinski, Lihua Lei, Michael Leung, Benjamin Manning, Adam McCloskey,
Iacopo Morchio, Mark Rosenzweig, Jesse Shapiro, Hans Sievertsen, Jann
Spiess, Pietro Spini, Kevin Tran and Yanos Zylberberg and participants
at seminars at Stanford, UCL, Bristol, and at the IAAE 2025, the Causal Data Science Meeting 2024, the International Workshop on Computational Economics and Econometrics 2025, and the Liverpool Econometrics Workshop 2026 for useful
feedback.}}
\author{Sukjin Han\thanks{School of Economics, University of Bristol. \UrlFont\protect\href{mailto:sukjin.han@gmail.com}{sukjin.han@gmail.com}}}
\date{First Version: September 21, 2024 \\ This Version: \today}

\maketitle
\vspace{-0.2cm}

\begin{abstract}
The instrumental variables (IVs) method is a leading empirical strategy
for causal inference. Finding IVs is a heuristic and creative process,
and justifying their validity---especially exclusion restrictions---is
largely rhetorical. We propose using large language models (LLMs)
to search for new IVs through narratives and counterfactual reasoning,
similar to how a human researcher would. The stark difference, however,
is that LLMs can dramatically accelerate this process and explore
an extremely large search space. We introduce a \emph{discovery pipeline} that searches for potentially novel and valid IVs using two-step and role-playing prompting strategies. We contend that these strategies simulate the endogenous decision-making of economic agents and social actors and ground language models in real-world scenarios, thereby masking the IV discovery task itself. We apply our method to three canonical areas in economics: returns to schooling, demand and supply, and peer effects. We then introduce an \emph{evaluation pipeline} in which we conduct expert surveys and train an LLM judge on the survey data. This evaluation reveals that the IV candidates discovered by our method appear both novel and likely valid, whereas a more direct search tends to return instruments already established in the literature.
\vspace{0.1in}


\end{abstract}

\section{Introduction\label{sec:Introduction}}

Endogeneity is the major obstacle in conducting causal inference
in observational settings. Since the credibility revolution \citep{angrist2010credibility}, researchers
in social science have developed
various identification strategies to overcome endogeneity by restoring
versions of quasi-experiments. A leading strategy is the instrumental
variables (IVs) method. Over decades, researchers with their ingenuity
have discovered IVs in various settings and justified their satisfaction
of \emph{exclusion restrictions} (e.g., IVs are conditionally exogenous
of latent variables). With its various applicability, the IVs method
has prevailed across all subfields of social science and beyond \citep[e.g.,][]{imbens1994identification,Blundell:2003wi,heckman2005structural,hernan2006instruments}.\footnote{See \cite{mogstad2024instrumental} for a more recent survey.}

Exclusion restrictions are fundamentally untestable assumptions.\footnote{An exception is a favorable situation where one enjoys over-identifying
restrictions. We discuss this point in our context below. Unlike
the exclusion restriction, the IV relevance is testable from data
\citep{olea2013robust,stock_yogo_2005}.} Often, in justifying them, researchers resort to \emph{rhetorical}
arguments specific to each setting. This non-statistical process follows
the discovery of potential candidate IVs, which itself requires researchers'
\emph{counterfactual reasoning} and creativity---and sometimes luck.
These elements all contribute to the heuristic processes employed
by human researchers.

We demonstrate that large language models (LLMs) can facilitate the
discovery of new IVs. Considering that narratives are the primary
method of supporting IV exclusion, we believe that LLMs, with sophisticated
language processing abilities, are well-suited to assist in the search
for new valid IVs and justify them rhetorically, just as human researchers
have done for decades. The stark difference, however, is that LLMs
can accelerate this process at an exponentially faster rate and explore
an extremely large search space, to an extent that human researchers
cannot match. It is now recognized that LLMs are capable of conducting counterfactual
reasoning---or, perhaps more precisely, exploring alternative scenarios---which
makes them a promising tool for causal inference \citep{jin2023cladder,jin2023can, kiciman2023causal}.

There are at least four benefits to pursuing this machine-assisted approach
to discovering IVs. First, researchers can conduct a systematic search
at a speedy rate, while adapting to the particularities of their
settings. Second, interacting with these tools can inspire ideas for
possible domains for novel IVs. Third, the systematic search could
increase the possibility of obtaining multiple IVs, which would then
enable formal (i.e., statistical) testing of their validity via over-identifying
restrictions. Fourth, having a list of candidate IVs would increase
the chances of finding actual data that contain IVs or guide the construction
of such data, including the design of (natural) experiments to generate IVs.

We introduce a \emph{discovery pipeline} that searches for potentially novel and valid IVs. We show how to construct prompts that guides LLMs to search
for candidates for valid IVs. The text representation of exclusion
restrictions (among others) is the main component of the prompts.
We propose a two-step approach in prompting that divides a discovery
task into multiple subtasks, and thus separates counterfactual statements
of different complexities. At the same time, we propose using role-playing
prompts, arguing that they align with the very source of endogeneity,
namely, agents' decisions.\footnote{Decisions of economic agents or social actors have been at the root of challenges for
causal analyses \citep{heckman1979sample,manski1993identification}.} By doing so, we equip LLMs with the perspective of agents, enabling
them to mimic agents' endogenous decision-making processes and gather
contextual information in realistic scenarios. Another benefit of two-step,
role-playing prompts is that they help navigate language models through
the realm of real-world scenarios, rather than anchor them within
the narrow realm of academic discourses on IVs. Each stage's prompt focuses only on a portion of the IV assumptions, framed as an
agent's real-world problem, thereby masking the IV-discovery purpose and maximizing the chance of finding novel IVs.\footnote{Even if LLMs circumvent this safeguard and exhibit \emph{memorization} from academic texts, we still
find value in the procedure as long as the list of discovered IV candidates
includes those that are recognized as new by researchers.}

To prove the proposed concept and illustrate the actual performance
of an LLM, we conduct discovery exercises using OpenAI's ChatGPT-4
(GPT4), one of the leading LLMs, to find IVs in six examples in three well-known areas of empirical economics: returns to schooling, demand and supply, and
peer effects. In all six examples, GPT4 produced a list of candidate
IVs, some of which appear to be new in the literature and provided
rationale for their validity. The list also contains IVs that are
popularly used in the literature. In the peer
effect example, we also demonstrate that the proposed method can be
effective in exploring relatively new topics for empirical research,
which may in turn increase the possibility of finding novel IVs.

We then introduce an \emph{evaluation pipeline} based on expert surveys and an LLM judge. In the survey, we elicit validity and novelty scores from empirical economists; these reveal that some of the discovered IV candidates in each domain are novel and at least as valid as the most established instruments in the relevant literature. To assess the discovery strategies systematically, we train and calibrate an LLM judge on the survey data and compare role-playing against direct prompting (i.e., providing the definition of valid IVs and revealing the discovery task), as well as multi-step against single-step approaches. Role-playing yields IVs that appear both novel and likely valid, whereas direct prompting tends to return instruments already established in the literature. Multi-step prompting further improves the search, though the gain over single-step is more modest.

From a broader perspective, the proposal is to systematically ``search for exogeneity.'' In this spirit, the procedure can be extended to other causal inference methods, such as (i) searching for proper control variables in regression and difference-in-differences analyses and (ii) searching for new running variables in regression discontinuity designs. We construct relevant prompts, apply them to well-known examples in the literature, and successfully identify new variables.

A list of candidate IVs or control variables produced as a result
of the proposed method is far from absolute. Rather, it serves
as a valuable benchmark that inspires empirical researchers about
which types of variables to consider and which domains to explore.
The dialogue carried out with LLMs in the process can also help researchers
solidify arguments or counterarguments for the validity of variables.
After all, artificial intelligence (AI)---like any machines---cannot be the ultimate authority
(at least not yet). We believe a human researcher assisted by AI can
choose research designs and conduct causal inference more effectively.

We contribute to a growing economics literature
on using machine learning and AI to assist research processes. This agenda views these methods as not only data-processing
and prediction tools but also as tools that can improve conventional research practices
themselves \citep{mullainathan2017machine,athey2019machine,gentzkow2019text,dell2025deep,ludwig2025large}. Such practices include the creative and heuristic aspects of research, such as hypothesis generation and empirical design. 
The use of AI tools (e.g., self-supervised learning, deep learning, generative models) in these stages of scientific discovery has become increasingly common in the natural sciences \citep[see the comprehensive survey in \emph{Nature} by][]{wang2023scientific}, but remains less common in the social sciences. 
In interesting work, \cite{ludwig2024machine} and \cite{mullainathan2024predictive} use
generative models to systematically produce hypotheses that are comprehensible
by humans in otherwise daunting settings. While we do not explicitly seek to generate
hypotheses, the variables we discover implicitly maintain
a range of hypotheses about their validity. Regarding the use of foundation models such as LLMs in economic research, \cite{du2024labor} use fine-tuned LLMs to predict job transitions and understand career trajectories
in labor economics, and \cite{manning2024automated} propose to use LLMs to automate the
entire process of social scientific research, from data generation
to testing causal hypotheses. We employ LLMs in statistical causal
inference by incorporating specific structure from econometric assumptions
and allowing for human intervention in discovery processes. A recent survey by \cite{korinek2023generative} demonstrates how economists can integrate generative AI as a collaborator throughout their research workflows, including the ideation stage.

This paper also relates to the approach of using LLMs in causal discovery \citep[see][for a recent survey and references therein]{wan2024bridging}. However, the fundamental difference of our approach to this line of work is that we use LLMs to systematically discover variables with particular causal structure rather than using LLMs to find causal links among a \emph{pre-determined} set of variables.

The paper is mainly organized in two parts: discovery and evaluation. Part I proposes the discovery pipeline. Section \ref{sec:Notation-and-IV}
states the IV assumptions, and Section \ref{sec:Prompt-Construction}
proposes the main idea of IV discovery along with the prompting strategies.
Section \ref{sec:Discovered-IVs} provides the examples of discovered
IV candidates in various domains. Section \ref{sec:Adversarial-Large-Language} discusses the use of an adversarial LLM to
refine the discovery process. Part II provides the evaluation pipeline and evidence that support the discovery method of Part I. Section \ref{sec:Expert-Evaluation-of} contains the expert evaluation of the IV candidates discovered in Section \ref{sec:Discovered-IVs}. Section \ref{sec:LLM_judge} then builds an LLM judge and evaluates different discovery strategies, which demonstrates that alternative prompting strategies that disclose the IV-search purpose lead to less novel candidates. It also checks whether datasets are available for the novel candidate IVs discovered. This ends Part II. Section \ref{sec:Variables-Search-in} extends the approach to other causal inference settings. Section \ref{sec:Conclusions} concludes. The Appendix includes alternative prompting strategies for IV search, additional prompts and findings in the IV discovery examples, and details of LLM judge construction.

\section{IV Assumptions\label{sec:Notation-and-IV}}

We first formally state our discovery goal. Let $Y$ be the outcome
of interest, $D$ be the potentially endogenous treatment, $\mathcal{Z}_{K}\equiv\{Z_{1},...,Z_{K}\}$
be the list of IVs $Z_{k}$'s with $K$ being the desired number of
IVs to discover, and $X$ be the covariates. Let $Y(d,z_{k})$ be
the counterfactual outcome given $(d,z_{k})$. We say $Z_{k}$ is a valid IV if
it satisfies the following assumptions:

\begin{myas}{REL}[Relevance]\label{as:REL}Conditional on $X$, the
distribution of $D$ given $Z_{k}=z_{k}$ is a nontrivial function
of $z_{k}$.\end{myas}

\begin{myas}{EX}[Exclusion]\label{as:EX}For any $(d,z_{k})$, $Y(d,z_{k})=Y(d)$.\end{myas}

\begin{myas}{IND}[Independence]\label{as:IND}For any $d$, $Y(d)\perp Z_{k}$
conditional on $X$.\end{myas}

The goal of our exercise is to search for IVs that satisfy Assumptions
\ref{as:REL}, \ref{as:EX} and \ref{as:IND}. Suppressing $X$, Figure \ref{fig:DAG} depicts the causal directed
acyclic graph (DAG) that implies \ref{as:REL}, \ref{as:EX} and \ref{as:IND}
and with $Y(d)$ being a transformation of latent confounders $U$.
This diagram is useful in describing our procedures.
\begin{figure}
\begin{centering}
\begin{tikzpicture}   
\tikzstyle{node} = [circle, draw, minimum size=20pt, inner sep=0pt]   
\tikzstyle{dashed_node} = [circle, draw, dashed, minimum size=20pt, inner sep=0pt]   
\tikzstyle{arrow} = [->, thin]   
\tikzstyle{dashed_arrow} = [->, thin, dashed]
\node[node] (Z) {$Z_k$};   
\node[node, right=of Z] (D) {$D$};   
\node[node, right=of D] (Y) {$Y$};   
\node[dashed_node, right=0.13cm of D, yshift=0.9cm] (U) {$U$};
\draw[arrow] (Z) -- (D);   
\draw[arrow] (D) -- (Y);   
\draw[dashed_arrow] (U) -- (D);   
\draw[dashed_arrow] (U) -- (Y);    
\end{tikzpicture}
\vspace{3pt}
\par\end{centering}
\caption{Causal DAG for a Valid IV ($X$ suppressed)}
\label{fig:DAG}

\end{figure}

\section{Prompt Construction\label{sec:Prompt-Construction}}

We propose a two-step approach for IV discovery. In Step 1, we prompt
an LLM to search for IVs that satisfy a verbal description of \ref{as:REL}
and \ref{as:EX} (i.e., \raisebox{-0.3\height}{
\begin{tikzpicture}[scale=0.5]   
\tikzstyle{node} = [circle, draw, minimum size=20pt, inner sep=0pt]
\tikzstyle{arrow} = [->, thin]
\node[node] (Z) {$Z_k$};   
\node[node, right=0.5cm of Z] (D) {$D$};   
\node[node, right=0.5cm of D] (Y) {$Y$}; 
\draw[arrow] (Z) -- (D);   
\draw[arrow] (D) -- (Y);
\end{tikzpicture}
}). In Step 2, we prompt the LLM to refine the search by selecting---among
the IVs found in Step 1---those that satisfy a verbal description
of \ref{as:IND} (i.e., \raisebox{-0.3\height}{
\begin{tikzpicture}[scale=0.5]   
\tikzstyle{node} = [circle, draw, minimum size=20pt, inner sep=0pt]   
\tikzstyle{dashed_node} = [circle, draw, dashed, minimum size=20pt, inner sep=0pt]   
\tikzstyle{arrow} = [->, thin]
\node[node] (Z) {$Z_k$};   
\node[dashed_node, right=0.5cm of Z] (U) {$U$};   
\end{tikzpicture}
}).\footnote{In the Appendix,
we present a three-step prompting that focuses on each of \ref{as:REL}, \ref{as:EX} and \ref{as:IND} in each step.} In both steps, the prompts will involve counterfactual statements.
In each step, we ask the LLM to provide rationale for its responses.
This feature is useful for the user to understand the LLM's reasoning.

We propose a two-step approach for several reasons: First, LLMs
are known to yield better performance when handling subtasks step-by-step,
focusing on important details in interpreting the prompts and avoiding
errors \citep{wu2022ai}. Second, this approach creates more room
for the user to inspect intermediate outputs. In particular, Step 2 involves more complex counterfactual
statements than Step 1, allowing the user (and the LLM) to apply varying
degrees of attention when fine-tuning is needed. Third, intermediate
outputs themselves can provide information and offer insights. Finally,
this approach---especially when each step is executed in a separate session---significantly reduces the likelihood that LLMs recognize
the task as IV discovery and generate text from relevant academic
sources.

Alongside the two-step method, we propose a role-playing approach.
When prompts are structured as role-plays, LLMs tend to gather better
contextual information, exhibit improved reasoning performances, and generate more tailored and unique responses \citep{kong2023better}.
In fact, in most scenarios, the explanatory variable $D$ represents
an economic agent's decision, which naturally facilitate role-playing.
Additionally, role-playing prompts are more effective in guiding LLMs
to respond as the relevant economic agent rather than as a researcher
searching for IVs.

\subsection{Step 1: Prompts to Search for IVs\label{subsec:Step-1:-Prompts}}

For Step 1, Prompt \ref{P1} is a role-playing prompt that queries
the search for $K_{0}$ IVs (obtaining $\mathcal{Z}_{K_{0}}$) that
satisfy verbal versions of \ref{as:REL} and \ref{as:EX} (with no
$X$). In all prompts below, each bracketed term represents a user
input:\texttt{ {[}treatment{]}} is the treatment $D$, \texttt{{[}agent{]}}
is the economic agent whose decision is $D$, \texttt{{[}scenario{]}}
is the specific setting of interest, \texttt{{[}outcome{]}} is the
outcome $Y$, and \texttt{{[}K\_0{]}} is the desired number of variables
$K_{0}$. When prompting, we ask the LLM to play the role of \texttt{{[}agent{]}}
to make a \texttt{{[}treatment{]}} decision in a hypothetical \texttt{{[}scenario{]}}.
Examples of these inputs are given in Section \ref{sec:Discovered-IVs}.

\begin{myprompt}{1}\label{P1}\sloppy\texttt{you are {[}agent{]} who needs to make a {[}treatment{]} decision
in {[}scenario{]}. what are factors that can determine your decision but do not directly affect your {[}outcome{]}, except through {[}treatment{]} (that is, factors that affect your {[}outcome{]} only through {[}treatment{]})? list
{[}K\_0{]} factors that are quantifiable. explain the answers.}\end{myprompt}

There are at least two variants of Prompt \ref{P1} that may be useful
in certain scenarios. First, instead of ``\texttt{list {[}K\_0{]}
factors that are quantifiable}'' one may simply write ``\texttt{list
{[}K\_0{]} factors}'' or even ``\texttt{list {[}K\_0{]} factors
that are hard to quantify}.'' This would return candidates of IVs
that are harder to measure but can inspire creative data collection
(e.g., text, images, or other unstructured data). Second, one can
expand Prompt \ref{P1} to be more specific about categorizing factors
for relevant parties in a given setting. For example, in the schooling
scenario (Section \ref{subsec:Returns-to-Education}), we request
separate lists for student and school factors. This approach
can facilitate the evaluation of the results.

\subsection{Step 2: Prompts to Refine the Search for IVs\label{subsec:Step-2:-Prompts}}

Take the set of IVs, $\mathcal{Z}_{K_{0}}\equiv\{Z_{1},...,Z_{K_{0}}\}$,
obtained by running Prompt \ref{P1} in Step 1. Next, for Step 2, Prompt \ref{P2x} is a role-playing prompt that queries
the search for $K$ IVs (obtaining $\mathcal{Z}_{K}$, $K\le K_{0}$)
within $\mathcal{Z}_{K_{0}}$ that satisfy a verbal version of \ref{as:IND}.
Below, \texttt{{[}confounders{]}} is the user input
for unobserved confounders of concern and \texttt{{[}K{]}} is the
user choice of $K$. In addition, we introduces the notion of conditioning variables in the prompt to reflect covariates $X$ in \ref{as:IND};\footnote{The Appendix presents an alternative approach to introducing covariates, namely, jointly searching for IVs and covariates
that satisfy \ref{as:REL}--\ref{as:IND}.} role-playing prompts are suitable for this purpose. Although \ref{as:REL} also involves conditioning
on $X$, we find that results are not sensitive to a relevant modification
of Prompt \ref{P1}. Prompt \ref{P2x} qualifies \emph{both} \texttt{{[}agent{]}}
and \texttt{{[}scenario{]}} by \texttt{{[}covariates{]}}, the pre-determined
user choice of covariates.

\begin{myprompt}{2$_{x}$}\label{P2x}\sloppy
\texttt{suppose you are {[}agent{]} in {[}scenario{]} with {[}covariates{]}. among
the {[}K\_0{]} factors listed above, choose {[}K{]} factors that are
most likely to be unassociated with {[}confounders{]}, which determine
your {[}outcome{]}. the chosen factors can still influence your {[}treatment{]}. for
each chosen factor, explain your reasoning.}\end{myprompt}

Unlike Prompt \ref{P1}, this prompt contains a statement about variables
typically unobserved to researchers, which may pose challenges. We
believe that incorporating the researcher's prior knowledge on latent
confounders helps simplify the overall search process and yield more
desirable results. For instance, in the schooling scenario, one can specify ``\texttt{innate
ability and personality and school quality}.''\footnote{An alternative option is to systematically
search for possible unobserved confounders.} In Prompt \ref{P2x}, we use the
term ``\texttt{unassociated}.'' If the LLM ever captures the nuance
of this word, it reflects the mean independence version of \ref{as:IND},
making the search easier.

{\sloppy The recommended approach for incorporating \texttt{{[}covariates{]}}
is to assign specific values for the covariates. For instance, in
the schooling scenario, one can write ``\texttt{suppose you are an
asian female high school student from california who considers attending
a private college}.''\footnote{One can run multiple queries across different covariate values
for robustness, although this does not appear to be necessary in most
cases unless extreme values are assigned in the initial run.} Alternatively, one can simply use terms like ``\texttt{specific}''
or ``\texttt{particular}'' along with the name of chosen covariates
(e.g., ``\texttt{suppose you are a high school student with specific
gender, race, and regional origin who considers attending a college
of specific type}'').\par}

There are useful variants of Prompt \ref{P2x}. First, one can omit
\texttt{{[}K{]}} and instruct the LLM to ``\texttt{choose all factors}''
from $\mathcal{Z}_{K_{0}}$ that are likely to satisfy \ref{as:IND},
allowing the LLM to determine $K$ independently; we apply this strategy
in all examples later. Second, as a sanity check, one can direct the
LLM to select elements in $\mathcal{Z}_{K_{0}}$ that \emph{violate}
\ref{as:IND} in addition to those that satisfy it. This can be achieved
by adding ``\texttt{also choose factors that are, in contrast, associated
with {[}confounders{]}}.'' To gain further insights, the user can
request explanations for factors that she identifies as valid IVs
in the initial set $\mathcal{Z}_{K_{0}}$ from Step 1, but which are somehow
not included in the final set $\mathcal{Z}_{K}$ by the LLM. We apply
the last approach to the application in Section \ref{subsec:Demand-Estimation}.

\section{Discovered IV Candidates\label{sec:Discovered-IVs}}

Using Prompts \ref{P1} and \ref{P2x},
we aim to identify candidates for IVs in six domains of three canonical areas
in economics: returns to schooling, demand and supply, and peer effects.
These examples are chosen for their significance in the literature and commonly employ the IVs method as an empirical strategy. This exercise is to evaluate LLM performance and demonstrate the proposed method's practicality.

In all the examples, LLMs appear to discover
new candidates for IVs and candidates that are related to well-known
IVs in the literature. Also, many candidates demonstrate high
levels of specificity to their context.
We report the full results for the first three examples. For the last three, we only provide summaries and include the comprehensive results in the Appendix. For each example,
we present results from the \emph{initial} \emph{single} run of the
prompts without any curation or further refinement. Results across sessions are largely consistent, although
they can vary when different values of $K_{0}$ and $K$ are chosen. In all the examples, we choose $K_{0}=40$ and let
$K$ be determined in the discovery process. We use GPT4 as our LLM.\footnote{We focus on GPT4 as the other versions of ChatGPT (e.g., GPT4o) produce similar findings. The comparison of prompting strategies in Section \ref{sec:LLM_judge} uses GPT4o.}

\subsection{Returns to Education\label{subsec:Returns-to-Education}}

Estimating the causal effects of educational
attainment (e.g., college attendance, years of schooling) on earnings has been a significant objective in economics.
The main latent confounders in this setting is unobserved individual
and school characteristics (e.g., student ability and personality,
school quality) that affect both the schooling decisions and future
earnings. To address this endogeneity and recover meaningful causal
effects (e.g., local average treatment effects \citep{imbens1994identification}),
IVs such as distance to schools, tuition fees, and compulsory schooling
laws have been widely used in the literature \citep{card1999causal}. 

\subsubsection{Returns to College Attendance\label{subsec:Returns-of-College}}

As the first example, we focus on the returns to college attendance.
The following is the prompts we use. We explicitly request separate lists for individual and school factors.

\begin{myprompt}{1.1}\label{P1-1}\sloppy\texttt{you are a high school graduate. you need to make a college
attendance decision. what would be factors (factors of schools and
factors of yourself) that can determine your decision but that do
not directly affect your future earnings, except through college attendance
(that is, that affect your earnings only through college attendance)? list
forty factors that are quantifiable, twenty for school factors and
twenty for factors of yourself. explain the answers.}\end{myprompt}

\begin{myprompt}{2$_{x}$.1}\label{P2x-1}\sloppy\texttt{suppose you are a student with family income \$10K per year,
who is asian female from california, whose parents have college education,
who is catholic. among the forty factors listed above, choose all
factors that are not associated with your innate ability and personality
and school quality, which determine earnings. create separate lists
for school factors and factors of yourself. for each factor chosen,
explain your reasoning.}\end{myprompt}

From running Prompts \ref{P1-1} and \ref{P2x-1}, the LLM produces candidates for valid IVs and its rationale for the suggestions (Table \ref{tab:schooling}). The list includes IVs that are already established in the literature, such as \emph{distance to college} \citep{card1995using,card2001estimating} and \emph{family background} \citep{card1995using,ashenfelter1997estimates}, as well as IVs that seem to be new (to our best
knowledge): student-related factors as \emph{personal savings for education} and various school characteristics such as \emph{campus crime rates}, \emph{diversity of student body}, \emph{availability of online courses}, \emph{campus housing capacity} and \emph{proximity to urban centers}. The newly found variables have potential
to be valid, especially after being conditioned on covariates. Producing all these results
took less than one minute in total.

\begin{table}[h!]
\begin{centering}
\begin{tabular}{|c|>{\centering}p{6cm}|>{\raggedright}p{8cm}|}
\hline 
\# & IVs Suggested & \centering{}Rationale Provided\tabularnewline
\hline 
\hline 
1 & {\footnotesize{}Distance from Home to College\\(\cite{card1995using}, 1955; \cite{card2001estimating}, 3134)} & {\footnotesize{}This is purely geographical and independent of personal
attributes. }\tabularnewline
\hline 
2 & {\footnotesize{}Personal Savings for Education } & {\footnotesize{}While this might be influenced by a student's ability
to save, it's largely determined by external economic conditions and
family support. }\tabularnewline
\hline 
3 & {\footnotesize{}Number of Siblings Attending College\\(\cite{griliches1977estimating}, 2393)} & {\footnotesize{}This depends on family decisions and the educational
pursuits of siblings, unrelated to the student's personal capabilities
or characteristics. }\tabularnewline
\hline 
4 & {\footnotesize{}Parents' Educational Background\\(\cite{card1995using}, 1955; \cite{ashenfelter1997estimates}, 335)} & {\footnotesize{}A demographic factor that precedes the student's own
choices and abilities. }\tabularnewline
\hline 
5 & {\footnotesize{}Religious Affiliations } & {\footnotesize{}A personal and family cultural factor, not necessarily
a direct influence on personal academic or career potential. }\tabularnewline
\hline 
\hline 
6 & {\footnotesize{}Campus Crime Rates } & {\footnotesize{}Safety measures and local crime statistics are external
to the student's attributes. }\tabularnewline
\hline 
7 & {\footnotesize{}Diversity of Student Body } & {\footnotesize{}Reflects the school's admissions policies and demographic
outreach, independent of any one student's traits. }\tabularnewline
\hline 
8 & {\footnotesize{}Library Size (Volumes Held) } & {\footnotesize{}A resource factor that's determined by the institution's
funding and priorities. }\tabularnewline
\hline 
9 & {\footnotesize{}Availability of Online Courses } & {\footnotesize{}Indicates the institution's commitment to flexible
learning options, independent of student characteristics. }\tabularnewline
\hline 
10 & {\footnotesize{}Campus Facilities Quality (Gyms, Labs, etc.) } & {\footnotesize{}Dependent on the school's investment in infrastructure. }\tabularnewline
\hline 
11 & {\footnotesize{}Campus Housing Capacity } & {\footnotesize{}Dictated by the university's accommodations and facilities,
not by student traits. }\tabularnewline
\hline 
12 & {\footnotesize{}Proximity to Urban Centers } & {\footnotesize{}Geographic location of the school relative to cities. }\tabularnewline
\hline 
13 & {\footnotesize{}Environmental Sustainability Rating } & {\footnotesize{}Reflects the institution's environmental policies
and infrastructure. }\tabularnewline
\hline 
14 & {\footnotesize{}Technology Integration in Classrooms } & {\footnotesize{}Indicates the level of modern technological adoption
by the school. }\tabularnewline
\hline 
\end{tabular}
\par\end{centering}
\vspace{0.2cm}
\caption{Returns to College: Suggested IVs and Rationale for Validity}
\label{tab:schooling}

\noindent\begin{minipage}[t]{1\columnwidth}%
\medskip{}

\emph{\scriptsize{}Notes}{\scriptsize{}: IV candidates for college attendance
are presented. All IV candidates are discovered and explained by GPT4 from a
single run of Prompts \ref{P1-1} and \ref{P2x-1} with $K_{0}=40$
and $K$ left unspecified. The first 5 rows are categorized by GPT4
to be student-related factors, and the next 9 rows to be school-related
factors. When the responses
match the IVs described in the literature exactly, we include the corresponding references and their
citation counts in parenthesis. The rationale given by LLM can
be elaborated further by requesting it in the same session. The total running time was less than 1 minute.}{\scriptsize\par}%
\end{minipage}
\end{table}

\subsubsection{Returns to Years of Education\label{subsec:Returns-to-Years}}

{\sloppy As a second example, we consider the returns to years of schooling.
We omit the prompts as they are similar as before, except that we
impose the following role: ``\texttt{you are a student beginning
high school in the united states. you need to make a decision on how
many more years you will stay in school}.'' \par}

Table \ref{tab:schooling-1} presents the list of candidate IVs (together with the LLM's reasoning). Among them are instruments well documented in the literature, such as \emph{state education laws} \citep{angrist1991does,staiger1997instrumental} and \emph{distance to college} \citep{card1995using,card2001estimating}, alongside several that appear novel and interesting: regulatory and contextual factors such as \emph{credits required for graduation}, \emph{transportation availability}, \emph{local education funding per student}, \emph{local college enrollment rates}, and \emph{local unemployment rates}.

\begin{table}
\begin{centering}
\begin{tabular}{|c|>{\centering}p{6cm}|>{\raggedright}p{8cm}|}
\hline 
\# & IVs Suggested & \centering{}Rationale Provided\tabularnewline
\hline 
\hline 
1 & {\footnotesize{}Credits Required for Graduation} & {\footnotesize{}This is a regulatory requirement and does not depend
on your personal qualities or the inherent quality of your school
but can determine the length of time you stay in school. }\tabularnewline
\hline 
2 & {\footnotesize{}State Education Laws\\(\cite{angrist1991does}, 3951; \cite{staiger1997instrumental}, 10656)} & {\footnotesize{}Laws mandating education until a certain age or credits
affect all students uniformly, irrespective of individual abilities
or school quality. }\tabularnewline
\hline 
3 & {\footnotesize{}Distance to Nearest College/University\\(\cite{card1995using}, 1955; \cite{card2001estimating}, 3134)} & {\footnotesize{}The geographical proximity to higher education institutions
is an external factor that doesn't depend on your ability but can
influence the decision to pursue further education based on accessibility. }\tabularnewline
\hline 
4 & {\footnotesize{}Transportation Availability} & {\footnotesize{}The ease of getting to and from educational institutions
is a logistical concern that affects your ability to attend and continue
education, independent of personal academic qualities or school resources. }\tabularnewline
\hline 
5 & {\footnotesize{}Local Education Funding per Student} & {\footnotesize{}This reflects the financial investment in education
at a community or state level, impacting the resources available to
you, which can influence educational continuation decisions, yet is
not a reflection of personal ability. }\tabularnewline
\hline 
6 & {\footnotesize{}Local College Enrollment Rates} & {\footnotesize{}Community trends in college enrollment can create
a social environment that encourages further education, which is more
about local culture and less about individual capacity. }\tabularnewline
\hline 
7 & {\footnotesize{}Local Unemployment Rates} & {\footnotesize{}Higher local unemployment might incentivize staying
in school longer as a way to defer entering a tough job market, a
decision influenced by economic conditions rather than personal academic
prowess. }\tabularnewline
\hline 
\end{tabular}
\par\end{centering}
\vspace{0.2cm}
\caption{Returns to Years of Schooling: Suggested IVs and Rationale for Validity}
\label{tab:schooling-1}

\noindent\begin{minipage}[t]{1\columnwidth}%
\medskip{}

\emph{\scriptsize{}Notes}{\scriptsize{}: IV candidates for years of schooling
are presented. All IV candidates are discovered and explained by GPT4 from a
single run of Prompts \ref{P1-1} and \ref{P2x-1} (modified for this
example) with $K_{0}=40$ and $K$ left unspecified. When the responses
match the IVs described in the literature exactly, we include the corresponding references and their
citation counts in parenthesis. The rationale given by LLM can
be elaborated further by requesting it in the same session. The total running
time was less than 1 minute.}{\scriptsize\par}%
\end{minipage}
\end{table}

\subsection{Demand and Supply}

\subsubsection{Demand Estimation\label{subsec:Demand-Estimation}}

Consider estimating demand for consumers in a given market. In estimating
the effect of price on demand, the main concern is that price is endogenous
as an equilibrium outcome. Researchers have used supply-side
IVs that are excluded from the demand equation in a simultaneous system
for supply and demand \citep{angrist2000interpretation} or motivated
by structural models \citep{berry1995automobile}.

Below are the prompts we use, focusing on the setting of \citet{angrist2000interpretation}.\footnote{Note that
in Prompt \ref{P2-2-2}, covariate information is omitted.}

\begin{myprompt}{1.2}\label{P1-2-2}\sloppy\texttt{you are a dealer at a fish market. you need to set the prices
of fish. what would be factors that can determine your decision but
that do not directly affect the customers' demand for fish, except
through the price you set (that is, that affect the demand only through
fish prices). list forty factors that are quantifiable. explain your
answer.}\end{myprompt}

\begin{myprompt}{2.2}\label{P2-2-2}\sloppy\texttt{suppose you are a dealer at the fish market who is selling
fish and setting its prices on a day of the week. among the factors
listed above, choose all factors are not influenced by fish market
conditions or customers' characteristics that determine demand for
fish. for each factor chosen, explain your reasoning.}\end{myprompt}

Table \ref{tab:demand} contains the candidate IVs from running Prompts \ref{P1-2-2} and \ref{P2-2-2}. There are many \emph{supply-side factors} (e.g.,
\emph{fuel costs}), which are reasonable candidates for IVs.
The \emph{weather conditions} variable used as IVs in \cite{angrist2000interpretation}
appears in the list. Interestingly, another supply-side factor, \emph{labor costs}, produced by Prompt \ref{P1-2-2} is not included
in the final list produced by Prompt \ref{P2-2-2}. When asked ``\texttt{explain
why you didn't include ``labor costs'' in the final list},'' the LLM
responded that ``labor costs are somewhat flexible and can be adjusted
in response to changes in market conditions and customer demand, making
them more dynamic than some of the other factors listed.''

\begin{table}
\begin{centering}
\begin{tabular}{|c|>{\centering}p{4cm}|>{\raggedright}p{10cm}|}
\hline 
\# & IVs Suggested & \centering{}Rationale Provided\tabularnewline
\hline 
\hline 
1 & {\footnotesize{}Fuel Costs} & {\footnotesize{}Costs for fuel typically depend on broader economic
factors such as oil prices and are not directly affected by market
demand for fish or specific customer characteristics. }\tabularnewline
\hline 
2 & {\footnotesize{}Fishing Equipment Depreciation} & {\footnotesize{}The depreciation of fishing gear like nets and boats
is largely a function of use and time, rather than market conditions
or customer demand. }\tabularnewline
\hline 
3 & {\footnotesize{}Weather Conditions\\(\cite{angrist2000interpretation}, 454)} & {\footnotesize{}Weather affects fishing operations directly but is
not influenced by market conditions or the specific characteristics
of customers. }\tabularnewline
\hline 
4 & {\footnotesize{}Regulatory Costs} & {\footnotesize{}Costs associated with licensing and compliance with
fishing regulations are dictated by regulatory bodies and do not fluctuate
based on customer demand or market conditions.}\tabularnewline
\hline 
5 & {\footnotesize{}Interest Rates} & {\footnotesize{}The rates charged on loans for purchasing boats or
equipment are determined by financial markets and policy, rather than
direct market conditions of the fish market or customer demands. }\tabularnewline
\hline 
6 & {\footnotesize{}Insurance Costs} & {\footnotesize{}Premiums for insuring fishing equipment and operations
are typically based on risk assessments and sector-wide data, rather
than being directly influenced by day-to-day market conditions or
characteristics of customers. }\tabularnewline
\hline 
7 & {\footnotesize{}Utility Costs at Sales Points} & {\footnotesize{}The cost of utilities like electricity and water at
sales points tends to be fixed or based on usage rates that are independent
of market demand specifics. }\tabularnewline
\hline 
8 & {\footnotesize{}Economic Conditions} & {\footnotesize{}Broader economic factors that affect overall spending
and investment patterns influence operational costs but are not dictated
by the fish market conditions or customer preferences. }\tabularnewline
\hline 
9 & {\footnotesize{}Technological Advances} & {\footnotesize{}Investments in technology to improve fishing or sales
operations are usually planned based on long-term business strategy
and efficiency gains rather than immediate market conditions or specific
customer demographics. }\tabularnewline
\hline 
10 & {\footnotesize{}Government Subsidies} & {\footnotesize{}Subsidies to the fishing industry are determined by
government policies, which are independent of daily market conditions
or customer demands in the fish market. }\tabularnewline
\hline 
11 & {\footnotesize{}Tariffs on Imports} & {\footnotesize{}Tariffs imposed on imported fish are a matter of international
trade policy and do not change based on daily fluctuations in market
demand or customers characteristics. }\tabularnewline
\hline 
12 & {\footnotesize{}Employee Training Costs} & {\footnotesize{}Costs of training employees in handling and selling
fish relate to operational efficiency, which are not directly influenced
by the day's market conditions or customer demands. }\tabularnewline
\hline 
\end{tabular}
\par\end{centering}
\vspace{0.2cm}
\caption{Demand Estimation: Suggested Candidates for IVs}
\label{tab:demand}

\noindent\begin{minipage}[t]{1\columnwidth}%
\medskip{}

\emph{\scriptsize{}Notes}{\scriptsize{}: IV candidates for price are presented.
All IV candidates are discovered and explained by GPT4 from a single run of
Prompts \ref{P1-2-2} and \ref{P2-2-2} with $K_{0}=40$ and $K$
left unspecified. When the responses
match the IVs described in the literature exactly, we include the corresponding references and their
citation counts in parenthesis. The rationale given by LLM can
be elaborated further by requesting it in the same session. The total running time was less than 1 minute.}{\scriptsize\par}%
\end{minipage}
\end{table}

\subsubsection{Production Function Estimation\label{subsec:Production-Functions}}

Consider estimating a production function that captures the causal
relationship between inputs and outputs. The key identification challenge
is that input decisions can be correlated with unobserved productivity
shocks, which directly influence outputs. To address this, IVs such
as input prices have been proposed in the literature \citep{griliches1998production},
which have been subsequently criticized \citep{olley1996dynamics,levinsohn2003estimating,ackerberg2015identification}.

In the prompts, we explicitly request separate lists for market
factors and firm and manager factors. For this example and all examples below, the actual prompts are contained in the Appendix.

The LLM finds sixteen IV candidates (Table \ref{tab:production} in the Appendix). Notably, IVs commonly suggested in the literature (i.e., \emph{input prices}) do not appear among the LLM's final selections, echoing similar concerns in the literature \citep{olley1996dynamics,levinsohn2003estimating,ackerberg2015identification}. Instead, the model suggests factors that may influence input prices, such as \emph{interest rates}, \emph{exchange rates}, \emph{tariffs and trade policies}, \emph{transportation costs}, \emph{real estate prices}, and the firm's \emph{financial health}, several of which could plausibly be exogenous. There are also other instruments that fall into regulatory factors (e.g., \emph{environmental regulations}) and managerial factors (e.g., \emph{risk tolerance}, \emph{strategic objectives}). The model justifies these by arguing that such variables shape firms' input choices and long-term positioning but do not directly influence short-run productivity shocks. However, some of these long-term decisions may in fact affect long-term outputs, which may or may not be relevant to the short-run outcomes of concern.

\subsection{Peer Effects\label{subsec:Peer-Effects}}

Suppose we are interested in the causal effects of peers on an individual's
outcomes within a social network. We consider two well-known examples:
(i) the effects of peer farmers on the adoption of new farming technologies
\citep{foster1995learning,conley2010learning}; (ii) the effects of
peers on teenage smoking \citep{gaviria2001school}. In both examples,
the main source of endogeneity is latent factors that determine the
formation of network (e.g., latent homophily). To address this, the
literature on peer effects sometimes uses friends of friends as IVs
\citep{bramoulle2009identification,angrist2014perils}.

It is worth noting that, in these examples, we prompt the LLM to role-play from the peer's perspective, which is more convenient than prompting from that of the individual whose outcome is of concern.

\subsubsection{Effects of Peer Farmers on New Technology Adoption\label{subsec:Effects-of-Peer-1}}

Interestingly, some IVs that
are suggested in the literature (e.g., \emph{friends of friends}) are not
chosen by the LLM (Table \ref{tab:peer1} in the Appendix). It can be because the model either views them as invalid
or is incapable of identifying them. On the other hand, the variable \emph{success stories
documented and shared} seems
to relate to the IV used in \cite{conley2010learning}, which exploits
\emph{variation in the presence of experienced farmers}. Additionally, there
are other IVs that seem to be new, notably \emph{access to credit} and \emph{engagement in social
media or digital
platforms}. Finally, the LLM fails
to identify the IV used in \cite{foster1995learning}, namely, \emph{endowed
land size}. The validity of this variable is justified in their paper
by the specific historical backgrounds of the Indian villages studied.
This implies that, for some IVs, providing LLMs with institutional
details can be necessary; this can be implemented via LLM's system messages.

\subsubsection{Effects of Peer Teenagers on Smoking Behavior\label{subsec:Effects-of-Peer}}

We choose a teenager in urban Indonesia
for its relevance, given that the teenage smoking rate in Indonesia
has been recently reported as one of the world's highest \citep{fithria2021indonesian}.
We consider a social media network in the scenario to illustrate the
effectiveness of our approach in exploring relatively recent topics
in the literature, thereby highlighting the potential to discover
novel IVs.

It is important to note that,
given that the prompts are written from the perspective of peers,
the variables in the table should be understood as factors influencing
\emph{peers} of the focal individual. Given that the setup incorporates
modern elements such as social media, we identify many potentially
new and interesting IVs, particularly from the social media category such as \emph{engagement rate}
and \emph{frequency of posts} (Table \ref{tab:peer2} in the Appendix). Interestingly, \emph{access to smartphones
among peers} can be viewed as a ``friends
of friends'' IV.\footnote{There were other ``friends of friends'' IVs that are produced from
Step 1 but did not survive Step 2.}

\section{Adversarial Large Language Models\label{sec:Adversarial-Large-Language}}

One way to refine the responses of an LLM is to request another LLM to play the role of an adversary
and review the responses produced by the first LLM, namely the defender
LLM. In the adversarial stage, we fully disclose the IV discovery
task and ask the adversarial LLM to provide counter-arguments, but
without using econometric jargon. These counter-arguments are then
given to the defender LLM, which is asked to refine the previous answers.
Overall, this process produces more sophisticated responses. We find
this to be a useful exercise, which mimics the mental process of a
human researcher.

The following is an example prompt given to the adversarial LLM. In
the prompt, \texttt{{[}the list of variables by Defender{]}} is the
list provided the defender LLM, generated from running, for example,
Prompts \ref{P1}--\ref{P2x}.

\begin{myprompt}{AD1}\label{P_AD1}\sloppy\texttt{you are a researcher who wants to find instrumental variables
to estimate the effect of {[}treatment{]} on {[}outcome{]}. below
is a list of candidate instrumental variables. for each variable in
the list, provide arguments as to why it may not be a valid instrument:}~\\
\texttt{{[}the list of variables by Defender{]}}\end{myprompt}

Then, the counter-arguments generated from Prompt \ref{P_AD1} are
presented to the defender LLM using the following prompt, which follows
Prompts \ref{P1}--\ref{P2x}.

\begin{myprompt}{3}\label{P3}\sloppy\texttt{below are counter-arguments for each of your previous answers. based
on these arguments, revise your selection and provide a list:}~\\
\texttt{{[}the arguments by Adversary{]}}\end{myprompt}

When running Prompt \ref{P3}, it is important to encourage the defender
to create a list, otherwise, it might become pessimistic and reject
all the previous selections.

We apply this adversarial process to our examples in Section \ref{sec:Discovered-IVs}.
Overall, the defender provides more sophisticated answers in the end.
In the ``Returns to College'' example (Section \ref{subsec:Returns-of-College}),
the variables related to \emph{geographical proximity} or \emph{transportation}
have been retained. The other variables related to \emph{school facilities}
(e.g., \emph{library size}, \emph{technology integration in classrooms}) have been
modified to factors that may better satisfy exogeneity (e.g., \emph{the
percentage of renewable energy used on campus}, \emph{the number of green
spaces on campus}, \emph{campus medical facilities}). Interestingly, many variables now
appear to be weak IVs, which is sensible. In the ``Returns to Years
of Education'' examples (Section \ref{subsec:Returns-to-Years}),
the \emph{state laws} variable survived, but was renamed to the more
descriptive \emph{mandatory minimum years of schooling required by state}.
Other variables were also further refined. \emph{GED program availability}
emerged. In the ``Demand Estimation'' example (Section \ref{subsec:Demand-Estimation}),
the \emph{weather condition} variable was refined to \emph{global climate
patterns (e.g., El Ni\~{n}o)}, because the counter-argument was
that ``bad weather can also discourage customers from coming to the
market,'' which is a compelling argument. Many other variables were refined to reflect \emph{global and
macroeconomic conditions}. Again, these variables suffer from being
weak IVs or having less individual variation.

\section{Expert Evaluation: Discovered IV Candidates\label{sec:Expert-Evaluation-of}}

Although evaluating the validity of IVs is intrinsically elusive,
we attempt to validate our approach through expert evaluation. To
that end, we conducted a survey of experts in various fields of empirical
economics---where IVs are heavily used---to elicit their evaluation
of the IV candidates discovered by LLMs. We received evaluations from nine experts, each of whom evaluated all 70 IV candidates.

For each of the six domains in Section \ref{sec:Discovered-IVs},
we collected two metrics through the survey: the IV validity score
(1 = least valid, 5 = most valid) and the non-novelty score (1 = least
likely to exist in the previous literature, 5 = most likely to exist).\footnote{A similar design is used by \citet{si2025novel}, who ask expert researchers to score the novelty and feasibility of research ideas generated by LLMs.}
Lower non-novelty scores would correspond to \textit{higher} IV novelty.
The validity scores should be interpreted with care. First, the difficulty
of finding valid IVs varies across domains, and thus the scores should
be interpreted relative to a benchmark. We define the benchmark as
the average validity of the most established IVs (non-novelty score
= 5) in each domain. Second, although the average score may be lower,
individual IVs found by LLMs may be viewed as novel and valid. Having
these points in mind, for each domain, we report (i) the average validity
score given non-novelty score $E[\text{validity}|\text{non-novelty}=k]$
for each $k=1,\ldots,5$ and (ii) the names of all IVs such that $E[\text{validity}|\text{non-novelty}=k,\text{ IV name}]\ge\text{benchmark}$ for $k=1$ or for $k=2$,
where $\text{benchmark}=E[\text{validity}|\text{non-novelty}=5]$.\footnote{For each IV and each $k$, validity is averaged over the experts who gave the IV the non-novelty score $k$. IVs with references in Tables \ref{tab:schooling}--\ref{tab:demand} (i.e., those verified to be established) are excluded from (ii).}
IVs meeting this criterion can be viewed as relatively novel yet at
least as valid as existing ones.

Tables \ref{tab:eval-1-2}--\ref{tab:eval-1} report the results.
Overall, we observe the following: (i) No IV is perfect; even the
most established IVs sometimes averaged a validity score below 3 on
our scale.\footnote{The literature recognizes that mildly
invalid (or only plausibly exogenous) IVs are ubiquitous \citep{conley2012plausibly,deaner2025trade}.} (ii) Finding valid IVs is challenging for both LLMs and human researchers
(the latter reflected in the established IVs). (iii) The LLM has the
capability of identifying IVs that are viewed as both novel and more
valid than the most established benchmarks. (iv) The LLM discovered
the greatest number of valid IV candidates in the most novel domain (i.e., the
example of peer effects on teenage smoking in social-media networks),
which was also the most challenging domain with the lowest benchmark
score.
\begin{table}
\begin{centering}
\setlength{\tabcolsep}{3.5pt}
\begin{tabular}{ccccccc}
\toprule 
Domain$^{\dagger}$ & $k=1$ & $k=2$ & $k=3$ & $k=4$ & $k=5$ (benchmark) & \# of Novel IVs $\geq$ Benchmark$^{\dagger\dagger}$\tabularnewline
\midrule 
1 & 1.87 & 2.23 & 2.72 & 2.17 & 2.63 & 3\tabularnewline
2 & 3.00 & 2.10 & 2.39 & 2.97 & 3.19 & 2\tabularnewline
3 & 1.92 & 2.39 & 2.63 & 3.76 & 3.00 & 4\tabularnewline
4 & 1.92 & 2.22 & 2.70 & 3.19 & 3.72 & 1\tabularnewline
5 & 2.36 & 2.26 & 2.96 & 2.39 & 3.67 & 1\tabularnewline
6 & 2.10 & 2.03 & 2.53 & 3.00 & 2.00 & 10\tabularnewline
\bottomrule
\end{tabular}
\par\end{centering}
\caption{Average Validity Across Novelty ($k=1$: most novel; $k=5$: least
novel)}

\label{tab:eval-1-2}

\noindent\begin{minipage}[t]{1\columnwidth}%
\vspace{-0.2cm}

$^{\dagger}${\scriptsize{}Domain ID: 1. Returns to College Attendance; 2.
Returns to Years of Education; 3. Demand
Estimation; 4. Production Functions; 5. Peer Effects (technology adoption); 6. Peer Effects
(teenage smoking). }$^{\dagger}$$^{\dagger}${\scriptsize{}Number of IVs with $k=1$ or $2$ whose validity score is at least the benchmark; these IVs are listed in Table \ref{tab:eval-1}.}{\scriptsize\par}%
\end{minipage}
\end{table}

\medskip{}
\begin{table}
\noindent \begin{centering}
\begin{tabular}{>{\centering}p{4.2cm}>{\centering\arraybackslash}p{9.2cm}}
\toprule 
Domain$^{\dagger}$ (Benchmark) & Novel IVs Suggested\tabularnewline
\midrule 
1 (2.63)$^{\dagger}$$^{\dagger}$ & {\footnotesize{}Campus Housing Capacity (+0.70)$^{\dagger}$$^{\dagger}$$^{\dagger}$}\tabularnewline
 & {\footnotesize{}Technology Integration in Classrooms (+0.37)}\tabularnewline
 & {\footnotesize{}Campus Crime Rates (+0.03)}\tabularnewline
\midrule
2 (3.19) & {\footnotesize{}Credits Required for Graduation (+0.31)}\tabularnewline
 & {\footnotesize{}Transportation Availability (+0.31)}\tabularnewline
\midrule
3 (3.00) & {\footnotesize{}Fuel Costs (+1.00)}\tabularnewline
 & {\footnotesize{}Tariffs on Imports (+1.00)}\tabularnewline
 & {\footnotesize{}Technological Advances (+0.00)}\tabularnewline
 & {\footnotesize{}Government Subsidies (+0.00)}\tabularnewline
\midrule 
4 (3.72) & {\footnotesize{}Exchange Rates (+0.28)}\tabularnewline
\midrule 
5 (3.67) & {\footnotesize{}Access to Credit (+0.33)}\tabularnewline
\midrule
6 (2.00) & {\footnotesize{}Access to Smartphones Among Peers (+1.50)}\tabularnewline
 & {\footnotesize{}Content Reach (+1.33)}\tabularnewline
 & {\footnotesize{}Engagement Rate (+1.00)}\tabularnewline
 & {\footnotesize{}Local Regulations on Tobacco Advertising (+1.00)}\tabularnewline
 & {\footnotesize{}Number of Platforms Used (+0.40)}\tabularnewline
 & {\footnotesize{}Frequency of Posts (+0.33)}\tabularnewline
 & {\footnotesize{}Visibility at School Events (+0.25)}\tabularnewline
 & {\footnotesize{}School Policies on Smoking (+0.00)}\tabularnewline
 & {\footnotesize{}History of Disciplinary Actions at School (+0.00)}\tabularnewline
 & {\footnotesize{}Academic Performance (+0.00)}\tabularnewline
\bottomrule
\end{tabular}
\par\end{centering}
\caption{IVs with $k=1$ or $2$ Outperforming Benchmark ($k=1$: most novel;
$k=5$: least novel)}
\label{tab:eval-1}

\noindent\begin{minipage}[t]{1\columnwidth}%
\vspace{-0.2cm}

$^{\dagger}${\scriptsize{}Domain ID: 1. Returns to College Attendance; 2.
Returns to Years of Education; 3. Demand
Estimation; 4. Production Functions; 5. Peer Effects (technology adoption); 6. Peer Effects
(teenage smoking). }$^{\dagger}$$^{\dagger}${\scriptsize{}The validity
score of benchmark is in parenthesis. }$^{\dagger}$$^{\dagger}$$^{\dagger}${\scriptsize{}The
improvement of validity score over benchmark is in parenthesis.}{\scriptsize\par}%
\end{minipage}
\end{table}

\section{LLM Judge Evaluation: Comparison to Direct Approaches\label{sec:LLM_judge}}

\subsection{Possible Alternative Approaches\label{subsec:alternatives}}

One alternative to the proposed prompting strategy (i.e., two-step,
role-playing prompts) would be to explicitly mention the term ``instrumental
variables'' and inform LLMs that the ultimate objective is to ``search
for instrumental variables.'' Additionally, one can verbally train
LLMs to understand the IV assumptions before conducting the search.
Unfortunately, we find that this direct approach tends to suffer from
\emph{memorization}, where LLMs primarily focus on sources from existing
academic articles and textbooks relevant to IVs. One can easily verify
that, when applying prompts from the direct approach to the examples
in Section \ref{sec:Discovered-IVs}, GPT4 generates answers that
are reminiscent of established IVs in the literature. Moreover, when
additionally prompted with ``\texttt{did you have sources for your
previous responses?},'' GPT4 responds by citing well-known references
in each application that propose IVs. In contrast, when the same question
is asked after the proposed procedure, GPT4 responds that it ``didn't
have specific sources.'' When pressed to provide sources from the
web, in the returns to education application, for example, it cites
academic references from behavioral and social sciences on students'
decision-making, including qualitative and case studies.

Motivated by this, we compare the performance of the following five strategies to understand the contribution of each aspect of the proposed approach:
\begin{enumerate}
    \item Role-playing and two-step (the proposed approach);
    \item Role-playing and one-step;
    \item Direct and two-step;
    \item Direct and one-step (the direct approach discussed above);
    \item Direct and one-step with an emphasis on novelty.
\end{enumerate}
The one-step role-playing strategy merges Prompts \ref{P1} and \ref{P2x} into a single prompt. The direct strategies first provide the LLM with the definition of a valid IV and then ask for IVs in the given setting, either in one step or in two steps that mirror Prompts \ref{P1} and \ref{P2x}; the last strategy additionally asks for IVs that have not been used in the existing literature. The prompts are given in Appendix \ref{app:judge-prompts}. The comparison of strategies 1 vs. 3 and 2 vs. 4 would reveal the effect of the role-playing prompt on performance, while the comparison of 1 vs. 2 and 3 vs. 4 would show the effect of the two-step prompt.

For each of the six domains in Section \ref{sec:Discovered-IVs}, we run each strategy ten times to account for the stochasticity of LLM responses.\footnote{We use GPT4o with temperature 0.7 and fix $K=20$ in all strategies so that they are compared on an equal footing; the candidates therefore differ from those evaluated in Section \ref{sec:Expert-Evaluation-of}.} Each run generates twenty IV candidates. We then score each discovered IV candidate in terms of validity and non-novelty. There are in total $6000$ ($=$ 5 (strategies) $\times$ 6 (domains) $\times$ 10 (runs) $\times$ 20 (IVs)) candidates to score, and so conducting an additional expert survey for them would be exceedingly expensive. Therefore, in what follows, we introduce an evaluation pipeline that draws a more systematic comparison of the alternative strategies at significantly lower cost.

\subsection{Construction of LLM Judge}

We develop a procedure to build an \emph{LLM judge} trained and calibrated to the small-scale human expert survey we conducted in Section \ref{sec:Expert-Evaluation-of}.\footnote{In addition to the nine experts in Section \ref{sec:Expert-Evaluation-of}, we use the scores of two experts whose validity scores were incomplete and who were thus dropped in Section \ref{sec:Expert-Evaluation-of}.} Specifically, the LLM judge is an LLM that plays the role of an econometrician. It is instructed that it is an academic economist evaluating candidate IVs, and it scores each candidate on the same two scales as the experts, validity and non-novelty. Its scores are then calibrated to predict the experts' scores. The judge is only a proxy for expert judgment, but it is substantially cheaper and more scalable than a survey. The construction consists of three steps; the prompts and further details on the construction and evaluation of the LLM judge are given in Appendix \ref{app:judge-prompt}.

\noindent\emph{Step 1}. To account for the small sample size of the expert survey, we denoise the expert scores by hierarchical Bayesian shrinkage \citep{li2011bayesian}.\footnote{One can use empirical Bayes instead, although the hyperparameters can be poorly estimated given the small sample size.} Each expert was asked to score every IV in the survey, but the experts differ in leniency and the domains differ in difficulty. For each of the validity and non-novelty scores, we therefore estimate an ordered-logit model in which the score of IV $i$ by expert $j$ depends on the latent quality of the IV, $\theta_{i}=\alpha_{d(i)}+u_{i}$ (where $\alpha_{d(i)}$ is a domain effect and $u_i$ is an IV-specific deviation) and on the leniency of the expert, $b_{j}$. Specifically, the probability that expert $j$ gives IV $i$ the score $s\in\{1,\ldots,5\}$ is $\Lambda(\kappa_{s}-\theta_{i}-b_{j})-\Lambda(\kappa_{s-1}-\theta_{i}-b_{j})$, where $\Lambda$ is the logistic distribution function and $-\infty=\kappa_{0}<\kappa_{1}<\cdots<\kappa_{4}<\kappa_{5}=\infty$ are cutoffs. The model is estimated by Markov chain Monte Carlo with Gaussian priors on $\alpha_{d}$, $u_{i}$, and $b_{j}$ whose variances are estimated from the data.\footnote{The prior on $u_{i}$ shrinks the latent quality of an IV with few or noisy scores toward the mean of its own domain (rather than the overall mean).} The judge is then calibrated to predict $\hat{m}_{i}\in[1,5]$, the posterior mean of the expected score of IV $i$ by a typical expert (i.e., with $b_{j}=0$).

\noindent\emph{Step 2}. We prompt the judge to evaluate candidate IVs. The system message instructs the LLM that it is an academic economist evaluating candidate IVs, explicitly states the definition of a valid IV (i.e., verbal versions of \ref{as:REL}--\ref{as:IND}), and gives the definitions of the two scores, validity and non-novelty, closely following the survey questionnaire.\footnote{To be consistent with the expert survey, we ask the judge to score the IV itself rather than the quality of its rationale.} We give the LLM the domain information (i.e., the setting, the treatment decision, and the source of endogeneity) and implement \emph{few-shot learning} \citep{brown2020language} by providing, as few-shot examples, all but one of the expert-scored IVs of the same domain with their denoised scores (i.e., $N_{d}-1$ examples, where $N_{d}$ is the number of expert-scored IVs in domain $d$). It is then asked to score the remaining expert-scored IV in that domain.\footnote{Restricting the examples to the candidate's own domain keeps the scores relative to the domain, as in the survey, and fixes the set of examples, so that every candidate of a domain is scored with an identical prompt.} We use GPT5 as the judge, with one independent query per candidate.

\noindent\emph{Step 3}. We calibrate the LLM judge by mapping the judge's raw score $k\in\{1,\ldots,5\}$ to the expert scale by a monotone calibration function $\hat{g}(k)$. Among affine, isotonic, and ordered-logit functions, leave-one-out cross-validation selects the affine mapping for validity and the ordered-logit mapping for non-novelty.

\begin{figure}
\begin{centering}
\includegraphics[width=0.9\textwidth]{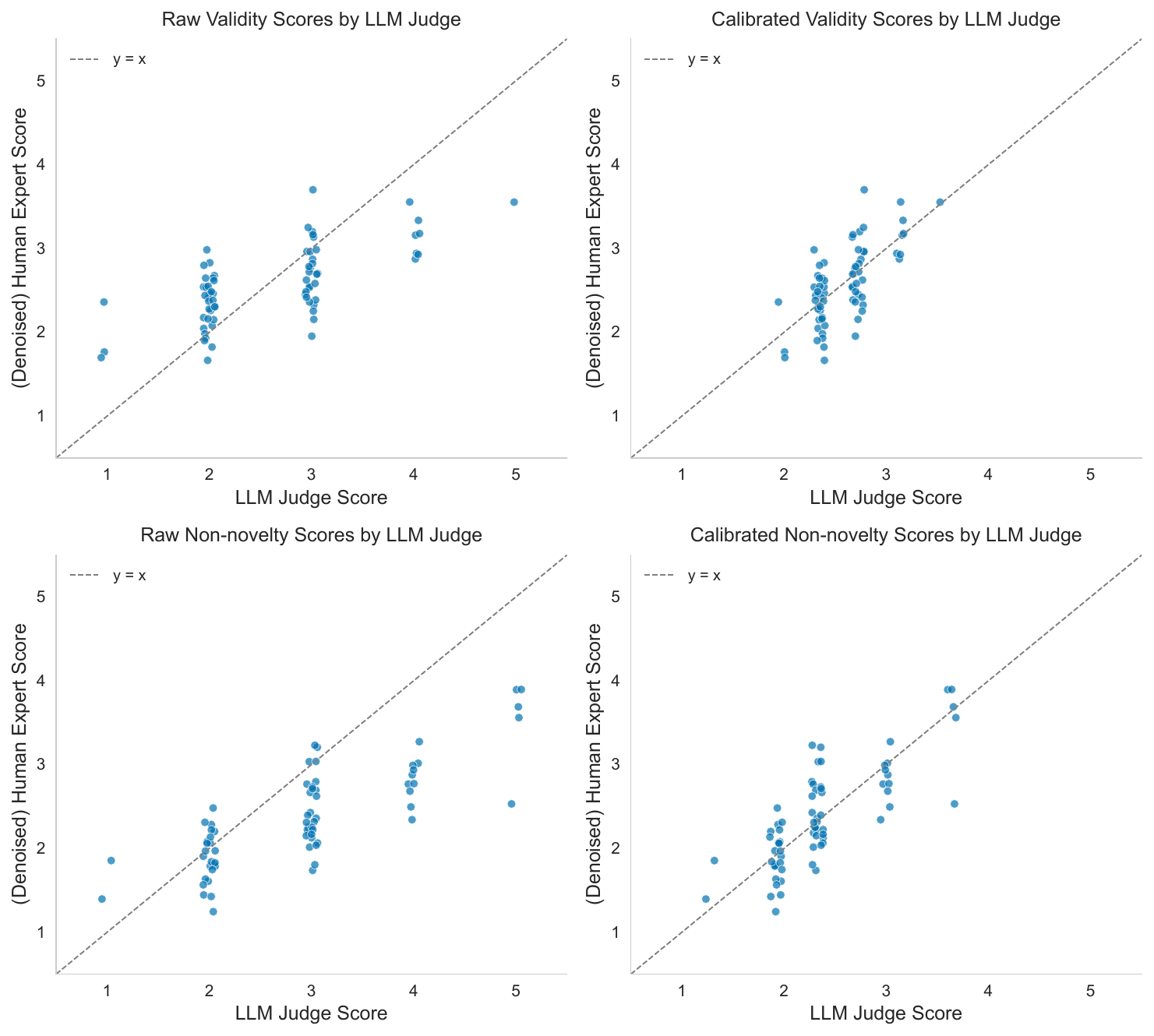}
\par\end{centering}
\caption{LLM Judge versus Human Expert Scores of the Survey IVs}
\label{fig:judge-scatter}

\noindent\begin{minipage}[t]{1\columnwidth}%
\vspace{-0.2cm}

\emph{\scriptsize{}Notes}{\scriptsize{}: Each point is one of the 70 IVs in the expert survey of Section \ref{sec:Expert-Evaluation-of}. The horizontal axis is the score of the LLM judge, raw (left panels) or calibrated by $\hat{g}$ (right panels); the vertical axis is the denoised expert score $\hat{m}_{i}$. Top panels: validity; bottom panels: non-novelty. The dashed line is the 45-degree line. Points are jittered horizontally for visibility.}{\scriptsize\par}%
\end{minipage}
\end{figure}

We evaluate the LLM judge using the hold-out IVs in the expert survey that are removed from the few-shot examples. Figure \ref{fig:judge-scatter} plots the judge's scores against the denoised expert scores. Unlike its raw scores (left panels), the judge's calibrated scores (right panels) track the experts reasonably well.  As one would expect, the judge is better at telling whether an IV is established in the literature than at scoring its validity, which is the more elusive task for human experts as well.

Once calibrated and evaluated, the judge is fixed and applied to all candidate IVs produced with the five strategies, using the full set of expert-scored examples of each domain (i.e., all $N_{d}$ expert-scored IVs of the domain). Because $\hat{g}$ is strictly increasing, the calibrated scores take five values per dimension: $\hat{g}(1),\ldots,\hat{g}(5)$ are 1.95, 2.34, 2.73, 3.11, and 3.50 for validity and 1.29, 1.92, 2.32, 2.99, and 3.64 for non-novelty, so that a calibrated score at or above $\hat{g}(k)$ means that the raw score is at least $k$. In what follows, we call a candidate \emph{novel} if its non-novelty score is $k=1$ or $2$, and, as in Section \ref{sec:Expert-Evaluation-of}, we compare validity with a benchmark. Since the judge rarely assigns $k=5$, we define the benchmark of a strategy in a domain as the average calibrated validity of that strategy's candidates with $k=4$ or $5$ in the domain.

\subsection{Performance Comparison}

Using the LLM judge, we produce validity and non-novelty scores for all candidate IVs and compare the strategies. Table \ref{tab:judge-summary} reports summary statistics of the scores by strategy. Overall, the direct strategies produce candidates that are on average more valid but much less novel than those of the role-playing strategies. Their average validity is higher by 0.18--0.26 on the 1--5 scale (less than one step of the raw scale), but only 7--18\% of their candidates are novel, compared with a third of the candidates of the role-playing strategies; conversely, 41--51\% of their candidates are scored as established IVs, compared with 21--23\% (Table \ref{tab:judge-summary}). This is consistent with the memorization discussed in Section \ref{subsec:alternatives}.

\begin{table}
\begin{centering}
{\footnotesize
\begin{tabular}{lccccccc}
\toprule 
 & & \multicolumn{2}{c}{Validity} & \multicolumn{2}{c}{Non-novelty} & \multicolumn{2}{c}{Share of candidates (\%)$^{\dagger\dagger}$}\tabularnewline
\cmidrule(lr){3-4}\cmidrule(lr){5-6}\cmidrule(lr){7-8}
Strategy$^{\dagger}$ & $N$ & Mean & SD & Mean & SD & Novel & Established\tabularnewline
\midrule 
Two-step role-play & 1200 & 2.54 & 0.28 & 2.35 & 0.44 & 34 & 23\tabularnewline
One-step role-play & 1200 & 2.56 & 0.29 & 2.35 & 0.42 & 31 & 21\tabularnewline
Two-step direct & 1200 & 2.74 & 0.29 & 2.59 & 0.49 & 18 & 44\tabularnewline
One-step direct & 1200 & 2.77 & 0.30 & 2.68 & 0.45 & 7 & 51\tabularnewline
One-step direct (novelty-encouraged) & 1200 & 2.80 & 0.30 & 2.58 & 0.42 & 9 & 41\tabularnewline
\bottomrule
\end{tabular}
}
\par\end{centering}
\caption{Summary Statistics of Calibrated Validity and Non-novelty Scores by Strategy}
\label{tab:judge-summary}

\noindent\begin{minipage}[t]{1\columnwidth}%
\vspace{-0.2cm}

$^{\dagger}${\scriptsize{}Strategies correspond to the numbered list in Section \ref{sec:LLM_judge}: 1. two-step role-play; 2. one-step role-play; 3. two-step direct; 4. one-step direct; 5. one-step direct with an emphasis on novelty.} {\scriptsize{}$N$ is the number of scored candidates (6 domains $\times$ 10 runs $\times$ 20 candidates; one-step role-play runs whose answer was parsed into more than 20 rows are capped at their first 20 candidates after removing rows that merely repeat a candidate's explanation). Scores are the calibrated scores of the LLM judge (Section \ref{sec:LLM_judge}). }$^{\dagger}$$^{\dagger}${\scriptsize{}Shares of candidates with non-novelty score $k=1$ or $2$ (novel) and $k=4$ or $5$ (established).}{\scriptsize\par}%
\end{minipage}
\end{table}

Table \ref{tab:judge-eval-1-2} reports, for each strategy and domain, the average calibrated validity across non-novelty scores in the format of Table \ref{tab:eval-1-2}, together with the number of novel candidates that are at least as valid as the benchmark. These candidates are listed in Tables \ref{tab:judge-eval-1-two_step_roleplay}--\ref{tab:judge-eval-1-direct_novel} in Appendix \ref{app:judge-tables}, the analogs of Table \ref{tab:eval-1}. Because a strategy often proposes the same IV under several names across runs (e.g., \emph{campus safety} and \emph{campus safety statistics}), we group the candidate IVs of each domain into \emph{instrument families}, namely clusters of candidates whose meanings are close (measured by text embeddings), and we count and list families rather than names.\footnote{Similarly, \citet{zhang2025noveltybench} evaluate the diversity of LLM outputs by partitioning repeated generations into classes of functionally equivalent responses and counting the distinct classes.}

\begin{table}
\begin{centering}
{\small\setlength{\tabcolsep}{4pt}
\begin{tabular}{cccccccc}
\toprule 
Domain$^{\dagger}$ & $k=1$ & $k=2$ & $k=3$ & $k=4$ & $k=5$ & Benchmark$^{\dagger\dagger}$ & \# of Novel IVs $\geq$ Benchmark$^{\dagger\dagger\dagger}$\tabularnewline
\midrule 
\multicolumn{8}{l}{\emph{Two-step role-play}}\tabularnewline
1 & 2.34 & 2.38 & 2.33 & 2.47 & 2.84 & 2.59 & 1\tabularnewline
2 & -- & 2.33 & 2.39 & 2.30 & 2.15 & 2.28 & 23\tabularnewline
3 & -- & 2.68 & 2.76 & 2.97 & -- & 2.97 & 0\tabularnewline
4 & -- & 2.42 & 2.66 & 2.89 & -- & 2.89 & 0\tabularnewline
5 & -- & 2.49 & 2.50 & 2.53 & -- & 2.53 & 9\tabularnewline
6 & -- & 2.55 & 2.41 & 2.32 & -- & 2.32 & 27\tabularnewline
\midrule
\multicolumn{8}{l}{\emph{One-step role-play}}\tabularnewline
1 & 2.34 & 2.39 & 2.38 & 2.56 & 2.90 & 2.71 & 3\tabularnewline
2 & -- & 2.30 & 2.35 & 2.43 & 2.47 & 2.43 & 0\tabularnewline
3 & -- & 2.67 & 2.74 & 2.94 & -- & 2.94 & 0\tabularnewline
4 & -- & 2.34 & 2.65 & 2.89 & -- & 2.89 & 0\tabularnewline
5 & -- & 2.49 & 2.56 & 2.61 & -- & 2.61 & 8\tabularnewline
6 & -- & 2.51 & 2.41 & 2.38 & -- & 2.38 & 15\tabularnewline
\midrule
\multicolumn{8}{l}{\emph{Two-step direct}}\tabularnewline
1 & -- & 2.73 & 2.72 & 2.88 & 3.12 & 2.97 & 0\tabularnewline
2 & -- & 2.46 & 2.55 & 2.74 & 3.12 & 2.88 & 0\tabularnewline
3 & -- & 2.63 & 2.84 & 3.00 & -- & 3.00 & 0\tabularnewline
4 & -- & -- & 2.81 & 2.90 & -- & 2.90 & 0\tabularnewline
5 & -- & 2.53 & 2.57 & 2.67 & -- & 2.67 & 2\tabularnewline
6 & -- & 2.52 & 2.42 & -- & -- & -- & --\tabularnewline
\midrule
\multicolumn{8}{l}{\emph{One-step direct}}\tabularnewline
1 & -- & 2.73 & 2.66 & 2.88 & 3.14 & 2.98 & 0\tabularnewline
2 & -- & 2.44 & 2.54 & 2.65 & 3.14 & 2.79 & 0\tabularnewline
3 & -- & 2.47 & 2.80 & 2.99 & 3.11 & 2.99 & 0\tabularnewline
4 & -- & -- & 2.85 & 2.93 & -- & 2.93 & 0\tabularnewline
5 & -- & 2.34 & 2.55 & 2.65 & -- & 2.65 & 0\tabularnewline
6 & -- & 2.48 & 2.73 & 2.98 & -- & 2.98 & 0\tabularnewline
\midrule
\multicolumn{8}{l}{\emph{One-step direct (novelty-encouraged)}}\tabularnewline
1 & -- & 2.58 & 2.72 & 2.91 & 3.16 & 2.97 & 0\tabularnewline
2 & -- & 2.44 & 2.53 & 2.82 & 3.06 & 2.88 & 0\tabularnewline
3 & -- & 2.49 & 2.87 & 3.13 & 3.31 & 3.13 & 0\tabularnewline
4 & -- & 2.73 & 2.89 & 2.97 & -- & 2.97 & 0\tabularnewline
5 & -- & -- & 2.71 & 2.78 & 3.11 & 2.79 & 0\tabularnewline
6 & -- & 2.46 & 2.78 & 2.73 & -- & 2.73 & 4\tabularnewline
\bottomrule
\end{tabular}
}
\par\end{centering}
\caption{LLM Judge: Average Calibrated Validity Across Novelty by Strategy ($k=1$: most novel; $k=5$: least novel)}
\label{tab:judge-eval-1-2}

\noindent\begin{minipage}[t]{1\columnwidth}%
\vspace{-0.2cm}

$^{\dagger}${\scriptsize{}Domain ID: 1. Returns to College Attendance; 2. Returns to Years of Education; 3. Demand Estimation; 4. Production Functions; 5. Peer Effects (technology adoption); 6. Peer Effects (teenage smoking).} $^{\dagger}$$^{\dagger}${\scriptsize{}Benchmark $=E[\text{validity}\mid\text{non-novelty}\ge4]$: because the LLM judge rarely assigns $k=5$, the benchmark pools $k=4$ and $5$ (cf. $k=5$ in Table \ref{tab:eval-1-2}). Cells are averages of calibrated validity over all candidates in the bin, pooled across the 10 runs; ``--'' marks an empty bin. Non-novelty bins are the judge's raw scores. }$^{\dagger}$$^{\dagger}$$^{\dagger}${\scriptsize{}Number of instrument families with $k=1$ or $2$ whose average calibrated validity is at least the benchmark, as listed in Tables \ref{tab:judge-eval-1-two_step_roleplay}--\ref{tab:judge-eval-1-direct_novel} in Appendix \ref{app:judge-tables}; ``--'' where the benchmark is undefined. }{\scriptsize{}Strategies correspond to the numbered list in Section \ref{sec:LLM_judge}: 1. two-step role-play; 2. one-step role-play; 3. two-step direct; 4. one-step direct; 5. one-step direct with an emphasis on novelty.}{\scriptsize\par}%
\end{minipage}
\end{table}

As in the expert evaluation, validity tends to increase with non-novelty within each strategy (Table \ref{tab:judge-eval-1-2}). Among novel candidates, however, the average validity is similar across strategies (about 2.5), so the higher overall validity of the direct strategies reflects their larger share of established candidates. There are also exceptions where the novel candidates are on average more valid than the established ones: the domain of peer effects on teenage smoking for both role-playing strategies and, for the two-step role-playing strategy, the domain of returns to years of education.

\begin{figure}
\begin{centering}
\subfigure[All candidates]{\includegraphics[width=0.8\textwidth]{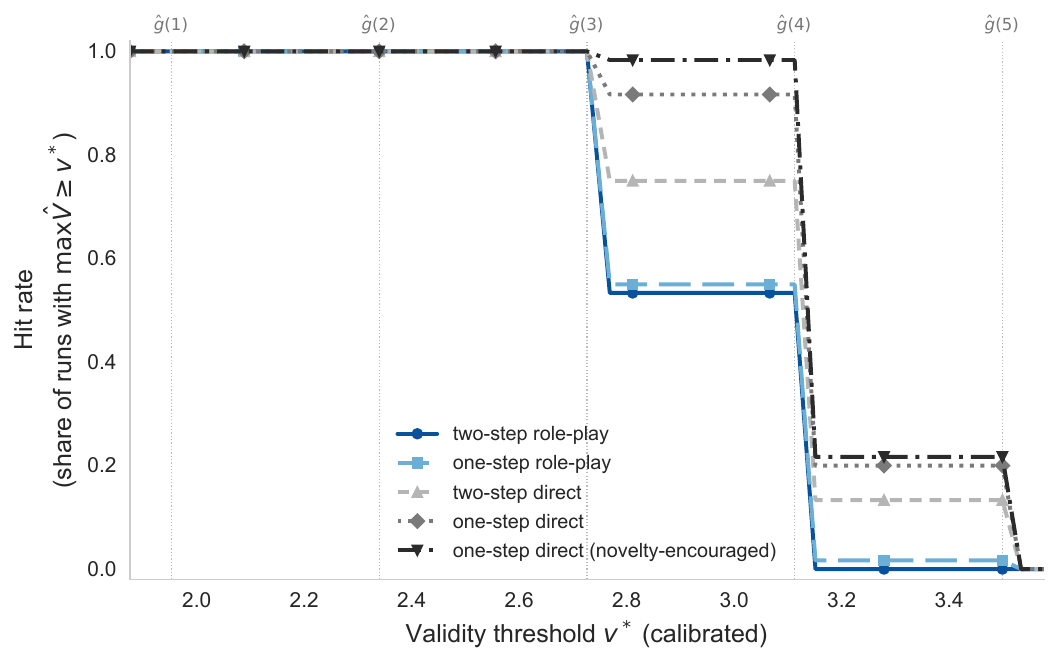}}

\subfigure[Novel candidates only ($k=1$ or $2$)]{\includegraphics[width=0.8\textwidth]{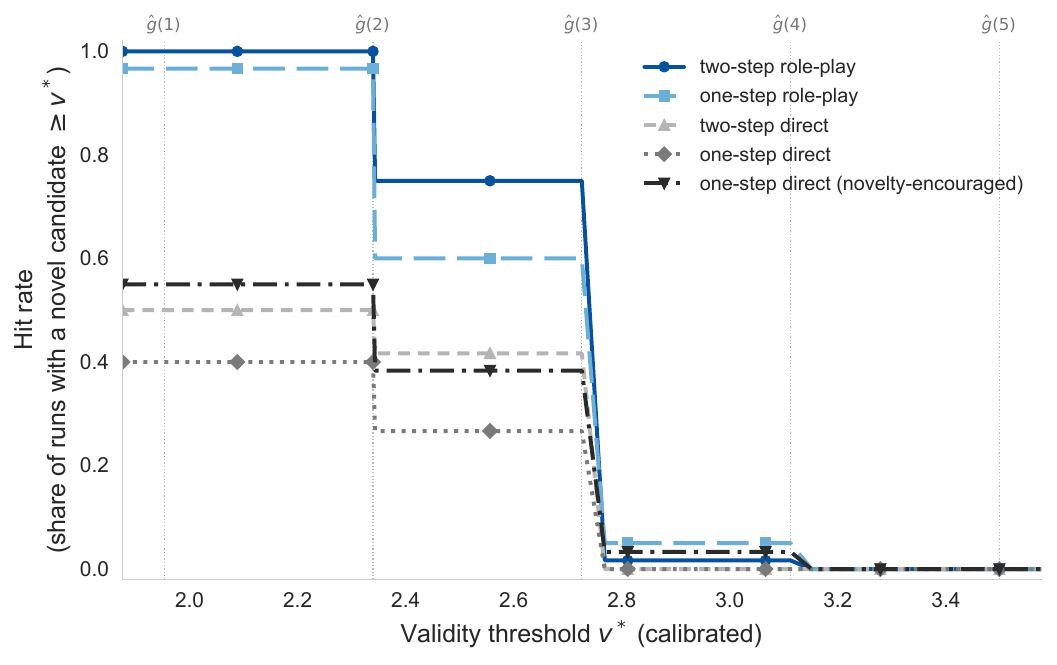}}
\par\end{centering}
\caption{Hit-Rate Curves by Strategy}
\label{fig:hit-rate}

\noindent\begin{minipage}[t]{1\columnwidth}%
\vspace{-0.2cm}

\emph{\scriptsize{}Notes}{\scriptsize{}: The hit rate at threshold $v^{*}$ is the share of runs (60 per strategy: 6 domains $\times$ 10 runs) containing at least one candidate whose calibrated validity is at least $v^{*}$; panel (b) counts only candidates with non-novelty score $k=1$ or $2$. Vertical lines mark $\hat{g}(k)$, the calibrated value of the judge's raw score $k$, so the hit rate at $\hat{g}(k)$ equals the share of runs with at least one candidate whose raw validity score is at least $k$. }{\scriptsize{}Strategies correspond to the numbered list in Section \ref{sec:LLM_judge}: 1. two-step role-play; 2. one-step role-play; 3. two-step direct; 4. one-step direct; 5. one-step direct with an emphasis on novelty.}{\scriptsize\par}%
\end{minipage}
\end{figure}

Importantly, the proposed strategy discovers far more IV candidates that are novel yet at least as valid as the benchmark than the other strategies: 60 families, compared with 26 for the one-step role-playing strategy and at most 4 for the direct strategies (last column of Table \ref{tab:judge-eval-1-2}). As in Section \ref{sec:Expert-Evaluation-of}, these IVs concentrate in the domains with the lowest benchmarks, most of all peer effects on teenage smoking, which is viewed as the most novel domain.\footnote{Note that the benchmark is that of each strategy, and it is low for the role-playing strategies in these domains (about 2.3, i.e., a raw validity score of about 2), so that many of these families are novel candidates of modest validity; a comparison with a common threshold is given by the hit rates below.}

Finally, Figure \ref{fig:hit-rate} reports, for each strategy, the \emph{hit rate}, namely the share of runs that produce at least one candidate whose calibrated validity is at least a threshold $v^{*}$, for all candidates (panel (a)) and for novel candidates only (panel (b)).\footnote{The hit rate is analogous to the \emph{pass@k} metric that \citet{chen2021evaluating} use to evaluate LLM code generation, namely the share of problems for which at least one of the generated code samples passes the unit tests.} The hit rate is a natural metric for a discovery tool, because a researcher needs only one valid IV from a run to proceed. The hit rates lead to the same conclusion from the researcher's perspective. Every strategy produces a candidate with a raw validity score of at least 3 in every run, and at higher thresholds the direct strategies have higher hit rates because they reproduce established IVs (Figure \ref{fig:hit-rate}(a)). However, once the candidate is required to be novel, the ranking is reversed: the proposed strategy produces a novel candidate with a raw validity score of at least 3 in 75\% of the runs, the one-step role-playing strategy in 60\%, and the direct strategies in 27--42\% (Figure \ref{fig:hit-rate}(b)).

To understand the separate effects of role-playing and two-step, we compare strategies 1 vs. 3 and 2 vs. 4 for the former and strategies 1 vs. 2 and 3 vs. 4 for the latter. Both role-playing and two-step raise the share of novel candidates (Table \ref{tab:judge-summary}), the number of novel candidates above the benchmark (Table \ref{tab:judge-eval-1-2}), and the hit rate for novel candidates (Figure \ref{fig:hit-rate}(b)). However, role-playing dominates the improvement: for example, the hit rate for novel candidates with a raw validity score of at least 3 increases by 33 percentage points with role-playing and by 15 percentage points with two-step. Because role-playing is introduced to mask the IV discovery task, its effect includes the effect of masking. Notably, explicitly asking for novel IVs in the direct prompt has only a modest effect overall: the share of novel candidates rises only slightly (Table \ref{tab:judge-summary}), the novelty-encouraged strategy finds few novel candidates above the benchmark (Table \ref{tab:judge-eval-1-2}), and the hit-rate improvement is not substantial (Figure \ref{fig:hit-rate}).

We believe that this additional evidence shows the effectiveness of the proposed pipeline for IV discovery. Importantly, we view this evaluation pipeline as part of the overall pipeline that empirical researchers can use in IV discovery tasks. The above exercises provide useful implications for the researchers to refine the search outputs. First, running multiple sessions and collating IV candidates into families can be helpful. This is especially the case given the correlation of LLM responses. Second, given this aspect, one would want to identify candidates that are scored as both novel and valid by the LLM judge across families (as in Table \ref{tab:judge-eval-1-2}) or runs (as in Figure \ref{fig:hit-rate}).

\subsection{Data Availability for Candidate IVs\label{subsec:data-availability}}

One may wonder whether this discovery exercise remains only rhetorical without being grounded in real data. We believe that the availability of data is not the most important criterion in the search process, because the discovery exercise can inspire researchers to design experiments and collect novel datasets. Nonetheless, it would be useful to check whether the IV candidates proposed by LLMs have corresponding datasets readily available online.

To that end, we take all candidate IV names from the two-step role-play strategy that were scored by the LLM judge as both novel ($k=1$ or $2$) and at least moderately valid (raw validity score of at least 3) at least once across the ten runs. This gives 87 distinct names across the six domains, which the family clustering groups into 47 instrument families.

For each candidate, we ask an LLM with web-search access (GPT5) to locate an actual dataset that contains a measure of the candidate together with the treatment, outcome, and covariates of the domain, either in a single source or in two sources that can be linked. The LLM reports the web page of the dataset and how to obtain the data, or else reports that no such dataset was found.\footnote{The LLM is instructed never to invent a dataset or a web address, and we check automatically whether each reported address responds.} Tables \ref{tab:data-1}--\ref{tab:data-6} in Appendix \ref{app:judge-tables-list} list the results by domain.

A dataset was found for 30 of the 47 IV families: 4 of 4 families for returns to college, 1 of 1 for returns to years of education, 8 of 12 for demand estimation, 5 of 7 for production functions, 5 of 9 for peer effects on technology adoption, and 7 of 14 for peer effects on teenage smoking.\footnote{The datasets are mostly public-use surveys and replication packages (e.g., the National Longitudinal Surveys of Youth and the longitudinal studies of the National Center for Education Statistics merged with institutional data in the schooling domains, the Fulton fish market data, the replication package of a social-network field experiment for technology adoption, and the Add Health and Notre Dame NetHealth network studies for teenage smoking), and in more than half of the cases the candidate must be merged from a second source. See Appendix \ref{app:judge-tables-list} for details.} Data are scarcest in the most novel domain, peer effects on teenage smoking in social-media networks, where many candidates describe features of online behavior that existing surveys do not measure. Of the 42 datasets reported for individual candidate names, an independent check by a second LLM, which visited the reported web pages, confirmed 32, judged 8 plausible, and refuted 2; in most cases the candidate is measured by a proxy rather than directly. We view these results as encouraging: for a majority of the novel and valid candidates, an empirical analysis can start from existing data. If finding datasets is the ultimate goal, then one may want to increase the number of IV candidates the discovery pipeline produces (which is 20 in the current exercise) to maximize the chance. One can also consider including data availability as an additional criterion for the adversarial LLM discussed in Section \ref{sec:Adversarial-Large-Language}.

Note that we do not impose the data availability criterion directly in the search prompt for two reasons. First, it is difficult to impose such a criterion when the LLM plays the role of an economic agent rather than a researcher. Second, imposing it would discourage the LLM from finding truly novel, unexplored IV candidates. Nonetheless, integrating the post-search data-availability check into the proposed pipeline of discovery and evaluation can still make the pipeline relevant to real data.

\section{Variables Search in Other Causal Inference Methods\label{sec:Variables-Search-in}}

In this section, we demonstrate how prompting strategies similar to
those for the IV discovery can be used to find (i) control variables
under which treatments are conditionally independent (i.e., exogenous);
(ii) control variables under which parallel trends are likely to hold
in difference-in-differences; and (iii) running variables in regression
discontinuity designs.

\subsection{Conditional Independence\label{subsec:Conditional-Independence}}

Using the same notation as in Section \ref{sec:Notation-and-IV},
consider a conditional independence (CI) assumption that assigns a
more crucial role to the vector of control variables $X\equiv(X_{1},...,X_{L})$:

\begin{myas}{CI}\label{as:CI} For any $d$, $D\perp Y(d)|X$.\end{myas}

Assumption \ref{as:CI} is commonly introduced in causal inference
settings, especially when combined with machine learning to estimate
nuisance functions; e.g., debiased/double machine learning methods
\citep{chernozhukov2024applied}. More traditionally, this assumption
is closely related to matching and propensity score matching techniques
\citep{heckman1998matching}. The mean independence version of \ref{as:CI}
(i.e., $E[Y(d)|D,X]=E[Y(d)|X]$) is relevant to regression methods.

We propose using LLMs to systematically search for $X$ that satisfies
a verbal version of \ref{as:CI}. The prompt writing is slightly simpler
than that for IVs. In particular, we construct prompts that solicit
the relationship between $X$ and $D$ (Step 1) and $X$ and $Y(d)$
(Step 2). Therefore, \emph{only} the second-step prompt involves a
counterfactual statement. Let $L_{0}$ be the number of controls
to be found in Step 1 ($L_{0}\ge L$). One may want to choose the
value of $L_{0}$ to be larger than one would normally use for $K_{0}$
and leave $L$ unspecified.\smallskip{}

\begin{myprompt}{C1}[Search for Control Variables]\label{PC1}\sloppy\texttt{you are {[}agent{]} who needs to make a {[}treatment{]} decision
in {[}scenario{]}. what factors determine your decision? list {[}L\_0{]}
factors that are quantifiable. explain the answers.}\end{myprompt}

\begin{myprompt}{C2}[Refine Control Variables]\label{PC2}\sloppy\texttt{among the {[}L\_0{]} factors listed above, choose all factors
that directly determine your {[}outcome{]}, not only indirectly through
{[}treatment{]}. the chosen factors can still influence your {[}treatment{]}. for
each factor chosen, explain your reasoning.}\end{myprompt}

The prompts are constructed to search for confounders and need to
be controlled for. Researchers sometimes mistakenly control for ``colliders''
and/or ``mediators'' \citep{pearl2000causality}, which are intended
to be excluded from the search. Note that Prompts \ref{PC1}--\ref{PC2}
can also be adapted to jointly search for covariates and latent confounders
in the IV search. In this case, one can distinguish $X$ from latent
confounders by referring to the former as ``\texttt{quantifiable}.''
Also, one may want to use the phrase ``\texttt{demographic factors}''
to refer to $X$, as they are common control variables in many empirical
applications.

\subsection{Difference in Differences\label{subsec:Difference-in-Differences}}

The difference-in-differences (DiD) method is popular in empirical
research, partly due to the simplicity and intuitiveness of its main
assumption, namely, the parallel trend assumption (stated below).
However, this assumption is not directly testable and typically hard
to justify \citep{ghanem2022selection,rambachan2023more}. It is believed
that conditioning on the right control variables can make this assumption
more justifiable, which can motivate the search for such controls.

\begin{myas}{PT}\label{as:PT}$E[\Delta Y(0)|D,X]=E[\Delta Y(0)|X]$
where $\Delta Y(0)\equiv Y_{after}(0)-Y_{before}(0)$.\end{myas}

{\sloppy Assumption \ref{as:PT} can be viewed as a mean independence version
of \ref{as:CI}, where the counterfactual outcome is replaced with
the temporal difference of counterfactual (untreated) outcomes before
and after the event. Therefore, Prompts \ref{PC1}--\ref{PC2} can
be directly used to search for $X$ that satisfy a verbal version
of \ref{as:PT}. This can be done by inputting ``\texttt{average
temporal changes in {[}outcome\_t{]} during the time of no {[}treatment{]}}''
for \texttt{{[}outcome{]}} in Prompt \ref{PC2}, where \texttt{{[}outcome\_t{]}}
refers to $Y_{t}$ for $t\in\{before,after\}$. The example of such
prompts is constructed to revisit the classical empirical example,
namely, the effects of minimum wage on the fast food industry's labor
markets \citep{card1994minimum}.\par}

The following prompts are used to produce the results in Table \ref{tab:DID}.

\begin{myprompt}{C1.1}[Example: Minimum Wage]\label{PR1-1}\sloppy\texttt{you are the policymaker in the department of labor, deciding
whether to increase the minimum wage or not and to which state to
introduce this minimum wage law. what factors determine your decision? list
forty factors that are quantifiable. explain the answers.}\end{myprompt}

\begin{myprompt}{C2.1}[Example: Minimum Wage]\label{PR2-1}\sloppy\texttt{among the forty factors listed above, choose all factors that
directly determine the temporal changes in average wages at fast food
restaurants, not only indirectly through the minimum wage law. the
chosen factors can still determine your decision of introducing minimum
wage law. for each factor chosen, explain your reasoning.}\end{myprompt}

{\sloppy Table \ref{tab:DID} contains the control
variables suggested by the LLM, conditional on which the parallel trend
is likely to hold, and the LLM's rationale. On the list, \#3, 4, 7, 10,
11 are particularly interesting and \#11 seems particularly novel.
In the table, the first four rows (\#1, 2, 3, 4) are chosen by the LLM
from an additional prompt that emphasizes the requirement with respect
to $\Delta Y(0)$: ``\texttt{be sure to choose all factors that do
not determine the average wage level but only determine the temporal
changes in average wages}.'' Nonetheless, controls that satisfy the
mean version of \ref{as:CI} with the level, $Y_{t}(0)$ for $t\in\{before,after\}$,
are also valid controls for \ref{as:PT}.\par}

\begin{table}
\begin{centering}
\begin{tabular}{|c|>{\centering}p{3.5cm}|>{\raggedright}p{10cm}|}
\hline 
\# & Control Variables Suggested & \centering{}Rationale Provided\tabularnewline
\hline 
\hline 
1 & {\footnotesize{}Inflation Rates} & {\footnotesize{}Higher inflation can decrease the real income of workers
if wages don't adjust accordingly, making a case for introducing a
minimum wage to maintain purchasing power.}\tabularnewline
\hline 
2 & {\footnotesize{}Consumer Price Index (CPI)} & {\footnotesize{}A rising CPI without wage increases can reduce workers'
real earnings, suggesting a minimum wage could help offset cost of
living increases.}\tabularnewline
\hline 
3 & {\footnotesize{}Job Vacancy Rates} & {\footnotesize{}Low vacancies could keep wages down due to ample labor
supply; high rates may drive wages up, affecting the urgency and approach
to minimum wage legislation.}\tabularnewline
\hline 
4 & {\footnotesize{}Labor Productivity Growth} & {\footnotesize{}This affects temporary wage increases linked to efficiency
improvements. Productivity gains can lead to bonus payments or higher
wages temporarily but don't define starting wage rates.}\tabularnewline
\hline 
\hline 
5 & {\footnotesize{}Employment Growth Rates} & {\footnotesize{}Robust job growth in the sector might show the industry
can support higher wages, while slow growth suggests caution in wage
hikes.}\tabularnewline
\hline 
6 & {\footnotesize{}Labor Force Participation Rate} & {\footnotesize{}High participation may suppress wages due to increased
labor supply, supporting the need for a minimum wage to ensure fair
earnings.}\tabularnewline
\hline 
7 & {\footnotesize{}Union Membership Rates} & {\footnotesize{}Low unionization in fast food often means weaker wage
bargaining, possibly justifying minimum wage laws to improve wage
standards.}\tabularnewline
\hline 
8 & {\footnotesize{}Turnover Rates} & {\footnotesize{}High turnover might indicate wage dissatisfaction,
pointing to the need for a minimum wage to stabilize the workforce.}\tabularnewline
\hline 
9 & {\footnotesize{}Corporate Profit Trends} & {\footnotesize{}If profits are high but wages are stagnant, it could
suggest that the industry can afford to pay more, bolstering the argument
for a minimum wage.}\tabularnewline
\hline 
10 & {\footnotesize{}Economic Diversity Score} & {\footnotesize{}Dependency on low-wage sectors like fast food in less
economically diverse areas might necessitate a minimum wage to ensure
living standards.}\tabularnewline
\hline 
11 & {\footnotesize{}Percentage of Workforce in Gig Economy} & {\footnotesize{}Increased gig work could pressure fast food employers
to offer competitive wages, influencing when and how to implement
minimum wage laws.}\tabularnewline
\hline 
\end{tabular}
\par\end{centering}
\vspace{0.2cm}
\caption{Difference-in-Differences for Minimum Wage: Suggested Control Variables}
\label{tab:DID}

\noindent\begin{minipage}[t]{1\columnwidth}%
\medskip{}

\emph{\scriptsize{}Notes}{\scriptsize{}: All controls are discovered
and explained by GPT4 from a single run of Prompts \ref{PC1}--\ref{PC2},
adapted to Assumption \ref{as:PT} with $L_{0}=40$ and $L$ left
unspecified. Among them, the first four row are factors that are chosen
from the additional emphatic prompt: ``}\texttt{\scriptsize{}be sure
to choose all factors that do not determine the average wage level
but only determine the temporal changes in average wages}{\scriptsize{}.''
The total running time was less than 1 minute.}{\scriptsize\par}%
\end{minipage}
\end{table}

\clearpage
\subsection{Regression Discontinuity\label{subsec:Regression-Discontinuity}}

Regression discontinuity designs (RDDs) are another well-known method
for causal inference that closely relates to the IVs method \citep{lee2010regression}.\footnote{For example, the fuzzy RDD estimand can be viewed as the two-stage
least squares estimand.} The key for this method to work is to find a running variable (i.e.,
assignment variable) that satisfies the following:

\begin{myas}{RD}\label{as:CI-1} There exists a variable $R_{j}$
and a cutoff $r_{0}$ such that $D=1$ if $R_{j}\ge r_{0}$ and $D=0$
if $R_{j}<r_{0}$.\end{myas}

One can use LLMs to systematically search for running variables $\{R_{1},...,R_{J}\}$
for a given $D$ and $Y$ of interest. We provide the example of prompts
here. It is worth noting that, unlike in all the previous cases, none
of the prompts below involve counterfactual statements. Therefore,
if LLMs outperform a traditional search for running variables, it
would be due to their automated and comprehensive search behavior. Similarly as above, we only specify initial $J_{0}$ and leave $J$
unspecified.\smallskip{}

\begin{myprompt}{R1}[Search for Running Variables]\label{PR1}\sloppy\texttt{you are {[}agent{]} who needs to make a {[}treatment{]} decision
in {[}scenario{]}. what would be the possible criteria based on which
your eligibility for {[}treatment{]} is determined? provide {[}J\_0{]}
of the most relevant criteria that are (1) quantifiable and (2) have
specific cutoffs determining eligibility. explain the answers.}\end{myprompt}

\begin{myprompt}{R2}[Refine Running Variables]\label{PR2}\sloppy\texttt{among the {[}J\_0{]} criteria listed above, choose all criteria
that involve continuous or ordered measures and have precise cutoffs
determining eligibility. also report the cutoff value for each criterion
from verifiable sources only (ensuring no fabricated or hypothetical
numbers are used). explain the answers.}\end{myprompt}

\begin{table}
\renewcommand{\arraystretch}{0.9}
\begin{centering}
\begin{tabular}{|c|>{\centering}p{2.5cm}|>{\centering}p{1.9cm}|>{\centering}p{2.3cm}|>{\raggedright}p{6.6cm}|}
\hline 
\# & {\footnotesize{}Outcome(s) (Country)} & {\footnotesize{}Treatment(s)} & {\footnotesize{}Suggested Running Variable (Same as the Literature)} & \begin{centering}
{\footnotesize{}Newly Suggested Running Variables}{\footnotesize\par}
\par\end{centering}
\centering{}{\footnotesize{}(Cutoffs for Eligibility)}\tabularnewline
\hline 
\hline 
1 & {\footnotesize{}Spending on schools, test scores (US)} & {\footnotesize{}State education aid} & {\footnotesize{}Relative average property values \citep{guryan2001does}} & {\footnotesize{}- Percentage of low-income students (e.g., Equity
Multiplier 2023-2024, above 70\%)}{\footnotesize\par}

{\footnotesize{}- Mobility rate (e.g., Equity Multiplier, above 25\%)}{\footnotesize\par}

{\footnotesize{}- Age (e.g., Transitional Kindergarten (TK) expansion
2023-24, 5th b-day by April 2)}{\footnotesize\par}

{\footnotesize{}- Local Control Funding Formula (LCFF) (California)$^{*}$}\tabularnewline
\hline 
2 & {\footnotesize{}College enrollment (US)} & {\footnotesize{}Financial aid offer} & {\footnotesize{}SAT scores, GPA \citep{vanderklaauw2002estimating}} & {\footnotesize{}- Expected family contribution (EFC) (e.g., the Pell
Grant 2023-2024: up to \$6,656)}\tabularnewline
\hline 
3 & {\footnotesize{}Overall insurance coverage (US)} & {\footnotesize{}Medicaid eligibility} & {\footnotesize{}Age \citep{card2004using}} & {\footnotesize{}- Federal Poverty Level (FPL) (e.g., Washington D.C. in 2023:
up to 215\% for childless adults and 221\% for parents/caretaker relatives; equiv. annual incomes up to
\$31,347 (household of 1) and \$54,940 (household of 3))}{\footnotesize\par}

{\footnotesize{}- Household Size (e.g., Modified Adjusted Gross Income
(MAGI) rules: expressed as \% of FPL, adjusted by 5\% FPL disregard)}\tabularnewline
\hline 
4 & {\footnotesize{}Employment rates (Italy)} & {\footnotesize{}Job training program} & {\footnotesize{}Attitudinal test score \citep{battistin2002testing}$^{\dagger}$} & {\footnotesize{}- Age (e.g., Bonus Giovani hiring incentive, hires Sept. 2024--Dec. 2025: below 35; source: Decree-Law 60/2024)}{\footnotesize\par}

{\footnotesize{}- Salary (e.g., EU Blue Card: at least the average
gross annual salary in Italy; source: Legislative Decree 152/2023)}\tabularnewline
\hline 
5 & {\footnotesize{}Re-employment probability (UK)} & {\footnotesize{}Job search assistance, training, education} & {\footnotesize{}Age at end of unemployment spell \citep{degiorgi2005long}} & {\footnotesize{}- Age (e.g., Jobseeker\textquoteright s Allowance
(JSA): 18 or over and under State Pension age; source: GOV.UK) }{\footnotesize\par}

{\footnotesize{}- Minimum Salary (e.g., Skilled Worker visa: at least
\pounds41,700 (since July 2025) or going rate for job type, whichever is higher; source:
GOV.UK)}{\footnotesize\par}

{\footnotesize{}- Residency Duration (e.g., income-based JSA, 2014--2026 (now abolished): at least 3 months
in the Common Travel Area prior to claim; source: SI 2013/3196)}\tabularnewline
\hline 
\end{tabular}
\par\end{centering}
\caption{Regression Discontinuity: Suggested Candidates for Running Variables}
\label{tab:RD}

\noindent\begin{minipage}[t]{1\columnwidth}%

\emph{\scriptsize{}Notes}{\scriptsize{}: All running variables are
discovered and explained by GPT4 from a single run of Prompts \ref{PR1}--\ref{PR2},
adapted to each context with $J_{0}=20$ and $J$ left unspecified.
All running variables used in the literature (Column 4) are also found
by GPT4, except \#4. The total running time for each row was less
than 1 minute (even with an automated web search for Prompt \ref{PR2}).
We verify the cutoffs against official sources and correct or update entries when necessary. The sources indicated are those given by GPT4, except for the corrected entries, for which we indicate the official sources we consulted. $*$: A formula,
not a running variable. $\dagger$: Not found by GPT4.}{\scriptsize\par}%
\end{minipage}
\end{table}

Note that when Prompt \ref{PR2} is run on GPT4, it will engage in
a series of automated web searches. The request for cutoff values
may lead the LLM to provide hypothetical numbers as possibilities.
When one wants to get the actual values from verifiable sources, it
is important to explicitly state that, as we do above. We apply Prompts
\ref{PR1}--\ref{PR2} to a range of famous examples in the literature
where RDDs are used as empirical strategies. Table \ref{tab:RD} presents
the results obtained by running the prompts, which are adapted to
each specific context and country of the empirical example. In most
cases, a handful of new possible running variables are suggested by
GPT4 with specific cutoffs obtained from web sources. Except for one
case (i.e., \#4), GPT4 also identifies the running variables used
in the literature.

\section{Conclusions\label{sec:Conclusions}}

This paper proposes the agenda to use LLMs to systematically search
for variables in designing causal inference. It merely serves as a
starting point, and there are potential next steps that can follow.
In constructing prompts for IVs, there are many possible ways for
sophistication: First, one can consider using previously known IVs
in the literature to guide LLMs to discover new ones. This can be
done by adding textual demonstration of how Assumptions \ref{as:REL}--\ref{as:IND}
are satisfied with known IVs \emph{before} starting the proposed prompts.
This approach would evoke few-shot learning in LLMs (as in the LLM judge construction),
which can enhance their performances. This approach would also
``orthogonalize'' the search \citep{ludwig2024machine} to focus
on novel IVs. One caveat of this approach is the difficulty of masking the discovery task. Second, the elaborated search can be directed toward
finding IVs that are more policy-relevant \citep{imbens1994identification,heckman2005structural}
by specifying targeted policies in the prompt. Third, the discovery results in Part I are from single sessions. Section \ref{sec:LLM_judge} shows one way of aggregating across sessions (i.e., grouping candidates into instrument families) to account for and potentially leverage the stochastic nature of LLMs' responses. Aggregating outputs across different LLMs (e.g., GPT and Claude) would also be beneficial.\footnote{Similar to running multiple sessions within an LLM, responses of multiple LLMs can be correlated and potentially biased \citep{kim2025correlated}. A suitable aggregation strategy akin to that discussed in Section \ref{sec:LLM_judge} can resolve the issue of correlation, and the expert survey and the LLM judge may be able to detect possible bias.}

One challenge of our approach---shared with other hypothesis-generation procedures---is that the performance metric is difficult to define due to the lack of ground truth for valid IVs. In fact, this is the very reason we propose using LLMs
in the first place: for any IVs found by human researchers or the
machine, there are only more compelling narratives or less compelling
ones. We partly address this challenge in the evaluation pipeline, through an expert survey in Section \ref{sec:Expert-Evaluation-of} and an LLM judge calibrated to the survey in Section \ref{sec:LLM_judge}. In later stages, when data eventually come into play, over-identification
tests can potentially be a fruitful framework for the evaluation of
LLMs. More broadly, evaluating hypothesis-generating systems remains an interesting open question. We still believe the paper's proposal serves as a thought-provoking starting point. It would be interesting to apply the proposed variable-search approach to other empirical settings and causal inference methods.

\newpage{}

\appendix

\section{Three-Step Prompts with System Messages\label{subsec:Three-Step-Prompts}}

We present three-step prompts, where each step corresponds to one of \ref{as:REL}, \ref{as:EX} and \ref{as:IND}. This version includes \emph{system messages} that establish
the scenario (including institutional details) and role of interest as context information and response formats before initiating
the main prompts.

\begin{myprompt}{A0}[System Messages]\label{P_3step_1-0}\sloppy\texttt{you are {[}agent{]} who needs to make a {[}treatment{]} decision
in {[}scenario{]}. you will be given questions on your decisions and
related outcomes. in answering the questions, always provide explanations
for your answers. each answer should be within five words and each
explanation should be within twenty words. do you understand your
role, the scenario you are in, and the instructions given to you?}\end{myprompt}

\begin{myprompt}{A1}[Search for Relevance]\label{P_3step_1-1}\sloppy\texttt{what are factors that can determine your decision? list {[}K\_0{]}
factors that are quantifiable. explain the answers.}\end{myprompt}

\begin{myprompt}{A2}[Search for Exclusion]\label{P_3step_1}\sloppy\texttt{among {[}K\_0{]} factors listed above, what are factors that
affect your {[}outcome{]} only through {[}treatment{]}? explain the
answers.}\end{myprompt}

\begin{myprompt}{A3}[Search for Independence]\label{P_3step_2}\sloppy\texttt{among the factors listed above, choose {[}K{]} factors that
are most likely to be unassociated with {[}confounders{]}, which determine
your {[}outcome{]}. the chosen factors can still influence your {[}treatment{]}. for
each factor chosen, explain your reasoning.}\end{myprompt}

\section{Alternative Prompts with Covariate Search\label{subsec:Alternative-Prompts-with}}

Instead of using user-specified covariates in Prompt \ref{P2x}, an
alternative way is to search for covariates in Step 1. Prompt \ref{P1xprime}
below is designed to jointly search for $(Z_{k},X)$ that satisfy
EX and \ref{as:REL2}: 

\begin{myas}{REL2}\label{as:REL2}(i) The distribution of $D$ given
$(Z_{k},X)=(z_{k},x)$ is a nontrivial function of $(z_{k},x)$ and
(ii) the distribution of $Y(d)$ given $X=x$ is a nontrivial function
of $x$.\end{myas}

\begin{myprompt}{1$_{x}$}[Search for IVs and Covariates]\label{P1xprime}\sloppy\texttt{you are {[}agent{]} who needs to make a {[}treatment{]} decision
in {[}setting{]}. what are factors that can determine your decision
but that do not directly affect your {[}outcome{]}, except through
{[}treatment{]} (that is, factors that affect your {[}outcome{]} only
through {[}treatment{]})? list {[}K\_0{]} factors. also, what are
your characteristics that directly influence {[}treatment{]} and directly
influence {[}outcome{]} (not just through {[}treatment{]})? list {[}L\_0{]}
characteristics. explain the answers.}\end{myprompt}

Instead of running Prompt \ref{P1} in Step 1, one can run Prompt
\ref{P1xprime} to find an initial set of covariates and select a subset
among them at her discretion to run Prompt \ref{P2x} in Step 2.

\section{Prompts Used in Applications (Section \ref{sec:Discovered-IVs})}\label{sec:Prompts-in-App2}

\subsection{Production Function Estimation\label{subsec:Production-Functions2}}

We choose $K_{0}=40$ and
let GPT4 choose $K$; we explicitly request separate lists for market
factors and firm and manager factors. Note that in Prompt \ref{P2x-2-1},
we use a loose description of covariates (unlike in Prompt \ref{P2x-1} in Section \ref{subsec:Returns-to-Education}
where we assign specific values).
\begin{myprompt}{1.2.1}[Example: Production Functions]\label{P1-2-1}\sloppy\texttt{you are a manager at a manufacturing firm. you need to make
a decision on how much labor and capital inputs to use to produce
outputs. what would be factors (factors of markets and economy and
factors of yourself) that can determine your decision but that do
not directly affect your output productions, except through the input
choices (that is, that affect your firm's outputs only through inputs)? list
forty factors that are quantifiable, twenty for market factors and
twenty for managerial factors. explain the answers.}\end{myprompt}

\begin{myprompt}{2$_{x}$.2.1}[Example: Production Functions]\label{P2x-2-1}\sloppy\texttt{suppose you are a manager at a firm with specific level of
capital intensity and specific scale of operations, which has a specific
market share in a specific industry. among the forty factors listed
above, choose all factors that are not influenced by productivity
shocks of your firm, which determine outputs. create separate lists
for market factors and managerial factors. for each factor chosen,
explain your reasoning.}\end{myprompt}

\subsection{Peer Effects\label{subsec:Peer-Effects2}}

\subsubsection{Effects of Peer Farmers on New Technology Adoption\label{subsec:Effects-of-Peer-12}}

It is worth noting that, in this example, the
role-playing is done from the peer's perspective, rather than from
the perspective of the individual whose outcome is of concern.

\begin{myprompt}{1.3.1}[Example: Peer Effects on Technology Adoption]\label{P1-3-1}\sloppy\texttt{you are a farmer in a village in rural india. you want to
influence your peer farmers in the same village to introduce a new
farming technologies that you introduced. what would be factors (factors
of farming and village, and factors of yourself) that can determine
your influence on peers but that do not directly affect your peers'
technology adoption decisions, except through your influence (that
is, that affect your peers' decisions only through your influence)? list
forty factors that are quantifiable. explain your answer.}\end{myprompt}

\begin{myprompt}{2$_{x}$.3.1}[Example: Peer Effects on Technology Adoption]\label{P2x-3-1}\sloppy\texttt{suppose you are a 40 year old male farmer of a specific crop
in the village in rural india. among the factors listed above, which
factors are not influenced by factors (e.g., similar background and
preferences) that brought you and your peers in the same neighborhood
and social network from the first place? for each factor chosen, explain
your reasoning.}\end{myprompt}

\subsubsection{Effects of Peer Teenagers on Smoking Behavior\label{subsec:Effects-of-Peer2}}

Again, in this example, the role-playing is
done from the peer's perspective.

\begin{myprompt}{1.3.2}[Example: Peer Effects on Smoking]\label{P1-3-2}\sloppy\texttt{you are a teenager in indonasia who smokes. you want to influence
your peers in your social media network to smoke. what would be factors
(factors of social media, your school and region, and factors of yourself)
that can determine your influence on peers but that do not directly
affect your peers' smoking decisions, except through your influence
(that is, that affect your peers' decisions only through your influence)? list
forty factors that are quantifiable. explain your answer.}\end{myprompt}

\begin{myprompt}{2$_{x}$.3.2}[Example: Peer Effects on Smoking]\label{P2x-3-2}\sloppy\texttt{suppose you are a teenage boy in urban indonesia who goes
to high school and is from middle-income family. among the factors
listed above, which factors are not influenced by factors (e.g., similar
background and preferences) that brought you and your peers in the
same social network from the first place? for each factor chosen,
explain your reasoning.}\end{myprompt}

\section{Additional Tables for Discovered IV Candidates\\ (Section \ref{sec:Discovered-IVs})}

\begin{table}
\begin{centering}
\begin{tabular}{|c|>{\centering}p{6cm}|>{\raggedright}p{8cm}|}
\hline 
\# & IVs Suggested & \centering{}Rationale Provided\tabularnewline
\hline 
\hline 
1 & {\footnotesize{}Interest Rates} & {\footnotesize{}Influence the cost of borrowing, affecting decisions
on how much capital to acquire or lease for production without altering
current productivity levels.}\tabularnewline
\hline 
2 & {\footnotesize{}Exchange Rates} & {\footnotesize{}Affect the price of imported inputs, which can change
input costs and investment decisions in foreign capital, again without
directly impacting the productivity of existing inputs.}\tabularnewline
\hline 
3 & {\footnotesize{}Tariffs and Trade Policies} & {\footnotesize{}Alter the cost structures for imported and exported
goods, impacting decisions on where to source inputs or sell outputs,
but do not affect the productivity of the inputs themselves.}\tabularnewline
\hline 
4 & {\footnotesize{}Regulatory Changes} & {\footnotesize{}Can necessitate changes in production processes or
input materials, which may affect costs and sourcing decisions without
directly impacting the productivity of the inputs once deployed.}\tabularnewline
\hline 
5 & {\footnotesize{}Transportation Costs} & {\footnotesize{}Affect the total cost of inputs and can influence
decisions regarding sourcing and logistics. These costs don't directly
alter how effectively inputs are converted into outputs.}\tabularnewline
\hline 
6 & {\footnotesize{}Real Estate Prices} & {\footnotesize{}Influence decisions about expanding or relocating
production facilities, affecting capital expenditure and operational
scale without altering current production efficiency.}\tabularnewline
\hline 
7 & {\footnotesize{}Environmental Regulations} & {\footnotesize{}Stricter environmental regulations may force a firm
to switch to greener, possibly more expensive inputs or production
technologies, affecting input choices and costs but not directly modifying
the productivity of any specific input set.}\tabularnewline
\hline 
8 & {\footnotesize{}Patents and Intellectual Property Rights} & {\footnotesize{}Governed by legal systems and not directly influenced
by a firm\textquoteright s internal productivity shocks.}\tabularnewline
\hline 
9 & {\footnotesize{}Political Stability} & {\footnotesize{}Affect market predictability and investment decisions,
impacting where and how much to invest in capital and labor, without
changing how these inputs produce once acquired.}\tabularnewline
\hline 
10 & {\footnotesize{}Global Economic Conditions} & {\footnotesize{}Influence market opportunities and risks associated
with foreign investments, guiding input acquisition strategies without
changing input productivity.}\tabularnewline
\hline 
\end{tabular}
\par\end{centering}
\vspace{0.2cm}
\renewcommand{\thetable}{A1}\caption{(a) Production Functions: Suggested Candidates for IVs (Market Factors)}
\label{tab:production}

\noindent\begin{minipage}[t]{1\columnwidth}%
\medskip{}

\emph{\scriptsize{}Notes}{\scriptsize{}: IV candidates for inputs are presented.
All IV candidates are discovered and explained by GPT4 from a single run of
Prompts \ref{P1-2-1} and \ref{P2x-2-1} with $K_{0}=40$ and $K$
left unspecified. Panel (a) is categorized by GPT4 to be market-related
factors, and Panel (b) to be managerial factors. The rationale given by LLM can
be elaborated further by requesting it in the same session. The total running
time was less than 1 minute.}{\scriptsize\par}%
\end{minipage}
\end{table}

\begin{table}
\begin{centering}
\begin{tabular}{|c|>{\centering}p{6cm}|>{\raggedright}p{8cm}|}
\hline 
\# & IVs Suggested & \centering{}Rationale Provided\tabularnewline
\hline 
\hline 
11 & {\footnotesize{}Capital Availability} & {\footnotesize{}Determines its ability to invest in both labor and
capital inputs, shaping the scale of operations and types of technology
employed, without directly changing how productive these inputs are}\tabularnewline
\hline 
12 & {\footnotesize{}Company\textquoteright s Risk Tolerance} & {\footnotesize{}Can decide the extent to which a firm is willing to
invest in new, potentially more efficient, but riskier technologies
or markets, affecting input decisions rather than the productivity
of current inputs}\tabularnewline
\hline 
13 & {\footnotesize{}Strategic Objectives} & {\footnotesize{}Long-term strategic objectives may dictate prioritizing
certain types of inputs or production scales, influencing the firm\textquoteright s
approach to markets and technology investments without affecting current
input productivity.}\tabularnewline
\hline 
14 & {\footnotesize{}Financial Health of the Company} & {\footnotesize{}The overall financial stability can limit or expand
the firm\textquoteright s ability to procure and utilize inputs optimally,
shaping how inputs are managed and financed rather than directly influencing
their productivity}\tabularnewline
\hline 
15 & {\footnotesize{}Compliance and Legal Considerations:} & {\footnotesize{}Driven by external legal requirements and internal
ethics, not by short-term productivity}\tabularnewline
\hline 
16 & {\footnotesize{}Corporate Social Responsibility (CSR) Initiatives} & {\footnotesize{}Strategic decisions about CSR are influenced by long-term
planning and brand image considerations.}\tabularnewline
\hline 
\end{tabular}
\par\end{centering}
\vspace{0.2cm}
\renewcommand{\thetable}{A1}\caption{(b) Production Functions: Suggested Candidates for IVs (Managerial
Factors)}
\label{tab:production-1}

\noindent\begin{minipage}[t]{1\columnwidth}%
\medskip{}

\emph{\scriptsize{}Notes}{\scriptsize{}: IV candidates for production inputs
are presented. All IV candidates are discovered and explained by GPT4 from a
single run of Prompts \ref{P1-2-1} and \ref{P2x-2-1} with $K_{0}=40$
and $K$ left unspecified. Panel (a) is categorized by GPT4 to be
market-related factors, and Panel (b) to be managerial. The rationale given by LLM can
be elaborated further by requesting it in the same session. The total
running time was less than 1 minute.}{\scriptsize\par}%
\end{minipage}
\end{table}

\begin{table}
\begin{centering}
\begin{tabular}{|c|>{\centering}p{4cm}|>{\raggedright}p{10cm}|}
\hline 
\# & IVs Suggested & \centering{}Rationale Provided\tabularnewline
\hline 
\hline 
1 & {\footnotesize{}Number of Training Sessions Attended} & {\footnotesize{}This reflects an individual\textquoteright s initiative
to seek additional education and may vary widely among farmers with
similar backgrounds due to personal choices, availability of time,
or perception of the value of formal education}\tabularnewline
\hline 
2 & {\footnotesize{}Number of Demonstrations Conducted} & {\footnotesize{}This is influenced by an individual\textquoteright s
willingness to share knowledge and the resources they are willing
or able to allocate for such activities, which might not be uniformly
distributed even among neighbors}\tabularnewline
\hline 
3 & {\footnotesize{}Increase in Yield After Technology Adoption} & {\footnotesize{}This factor depends significantly on how effectively
an individual implements and adapts new technologies to their specific
farming practices, which can differ greatly even among farmers growing
the same crops}\tabularnewline
\hline 
4 & {\footnotesize{}Access to Credit} & {\footnotesize{}Individual financial management skills, credit history,
and relationships with lending institutions can influence this factor,
which isn't necessarily shared among all peers in a community}\tabularnewline
\hline 
5 & {\footnotesize{}Participation in Local Decision Making Bodies} & {\footnotesize{}Involvement in local governance or committees often
depends on personal ambition, political skills, and community perception
of an individual's leadership qualities, which can vary independently
of shared initial factors}\tabularnewline
\hline 
6 & {\footnotesize{}Awards or Recognitions Received} & {\footnotesize{}Achieving recognition for farming practices or community
service is heavily dependent on individual effort and excellence and
is not uniformly influenced by community-shared characteristics}\tabularnewline
\hline 
7 & {\footnotesize{}Communications Skills Assessment} & {\footnotesize{}Communication effectiveness can be developed through
personal experience, education, and innate ability, which vary among
individuals regardless of their initial common background}\tabularnewline
\hline 
8 & {\footnotesize{}Number of Articles or Case Studies Published} & {\footnotesize{}This is influenced by an individual\textquoteright s
ability to write, research, and engage with broader audiences, skills
that are not common to all individuals in a network}\tabularnewline
\hline 
9 & {\footnotesize{}Engagement in Social Media or Technology Platforms} & {\footnotesize{}The level of engagement in modern communication and
technology platforms depends on personal interest in technology, availability
of resources, and individual attitudes toward innovation}\tabularnewline
\hline 
10 & {\footnotesize{}Success Stories Documented and Shared} & {\footnotesize{}The ability and willingness to document and share
success stories can be a personal choice influenced by an individual\textquoteright s
perspective on self-promotion and communication}\tabularnewline
\hline 
\end{tabular}
\par\end{centering}
\vspace{0.2cm}
\renewcommand{\thetable}{A2}\caption{Peer Effects on Technology Adoption: Suggested Candidates for IVs}
\label{tab:peer1}

\noindent\begin{minipage}[t]{1\columnwidth}%
\medskip{}

\emph{\scriptsize{}Notes}{\scriptsize{}: IV candidates for peers' decisions
are presented. Given the perspective in the prompts, the variables
should be understood as factors of }\emph{\scriptsize{}peers}{\scriptsize{}.
All IV candidates are discovered and explained by GPT4 from a single
run of Prompts \ref{P1-3-1} and \ref{P2x-3-1} with $K_{0}=40$ and
$K$ left unspecified. The rationale given by LLM can
be elaborated further by requesting it in the same session. The total running time was less than 1 minute.}{\scriptsize\par}%
\end{minipage}
\end{table}

\begin{table}
\begin{centering}
\begin{tabular}{|c|>{\centering}p{4cm}|>{\raggedright}p{10cm}|}
\hline 
\# & IVs Suggested & \centering{}Rationale Provided\tabularnewline
\hline 
\hline 
1 & {\footnotesize{}Engagement Rate} & {\footnotesize{}While the number of followers might be initially influenced
by shared backgrounds, the engagement rate depends more on the content
quality and how it resonates with the audience at any given time,
rather than the reasons why the audience initially formed}\tabularnewline
\hline 
2 & {\footnotesize{}Frequency of Posts} & {\footnotesize{}This is largely a personal choice and strategic decision,
independent of peer background. It reflects personal commitment to
the platform rather than shared interests or backgrounds}\tabularnewline
\hline 
3 & {\footnotesize{}Content Reach} & {\footnotesize{}This metric depends on the algorithm and how engaging
the content is, which can transcend the initial reasons for network
formation}\tabularnewline
\hline 
4 & {\footnotesize{}Number of Platforms Used} & {\footnotesize{}Choosing to diversify presence across various platforms
is a strategic decision that doesn\textquoteright t necessarily correlate
with shared backgrounds or interests}\tabularnewline
\hline 
\hline 
5 & {\footnotesize{}School Policies on Smoking} & {\footnotesize{}These are institutional decisions that apply to all
students, regardless of the factors that brought them together}\tabularnewline
\hline 
6 & {\footnotesize{}Local Regulations on Tobacco Advertising} & {\footnotesize{}Like school policies, these are determined by local
government and affect all residents uniformly, irrespective of social
network formations}\tabularnewline
\hline 
7 & {\footnotesize{}Access to Smartphones Among Peers} & {\footnotesize{}This might vary widely even within similar socio-economic
backgrounds due to individual family decisions or priorities}\tabularnewline
\hline 
\hline 
8 & {\footnotesize{}Visibility at School Events} & {\footnotesize{}Active participation and visibility in school events
are personal choices and efforts that go beyond shared backgrounds,
reflecting individual initiative}\tabularnewline
\hline 
9 & {\footnotesize{}History of Disciplinary Actions at School} & {\footnotesize{}This is generally a result of personal behavior and
choices rather than group influence}\tabularnewline
\hline 
10 & {\footnotesize{}Academic Performance} & {\footnotesize{}Although there could be a correlation with socio-economic
status, individual effort and capability play significant roles, making
this somewhat independent of why peers might group together initially}\tabularnewline
\hline 
11 & {\footnotesize{}Extracurricular Leadership Roles} & {\footnotesize{}Holding leadership positions is often based on personal
qualities, skills, and choices rather than the shared preferences
and backgrounds that might define a social network initially}\tabularnewline
\hline 
\end{tabular}
\par\end{centering}
\vspace{0.2cm}
\renewcommand{\thetable}{A3}\caption{Peer Effects on Smoking: Suggested Candidates for IVs}
\label{tab:peer2}

\noindent\begin{minipage}[t]{1\columnwidth}%
\medskip{}

\emph{\scriptsize{}Notes}{\scriptsize{}: IV candidates for peers' decisions
are presented. Given the perspective in the prompts, the variables
should be understood as factors of }\emph{\scriptsize{}peers}{\scriptsize{}.
All IV candidates are discovered and explained by GPT4 from a single run of
Prompts \ref{P1-3-2} and \ref{P2x-3-2} with $K_{0}=40$ and $K$
left unspecified. The first four rows concern social media factors;
the next three rows concern school and regional factors; and the last
four rows concern personal factors. The rationale given by LLM can
be elaborated further by requesting it in the same session. The total running time was less
than 1 minute.}{\scriptsize\par}%
\end{minipage}
\end{table}

\clearpage
\section{Details of LLM Judge Evaluation (Section \ref{sec:LLM_judge})\label{app:judge-tables}}

\subsection{Prompts for the Alternative Strategies\label{app:judge-prompts}}

We present the prompts of the five strategies compared in Section \ref{sec:LLM_judge} for the returns to college example (Section \ref{subsec:Returns-of-College}); the prompts for the other domains are constructed in the same way. In every strategy, the LLM is asked for twenty IV candidates in the final step.

The two-step role-playing strategy (1) uses Prompts \ref{P1-1} and \ref{P2x-1}, except that Prompt \ref{P2x-1} asks to ``\texttt{choose 20 factors}'' instead of ``\texttt{choose all factors}.'' The one-step role-playing strategy (2) merges the two prompts into Prompt \ref{P_E1}.

\begin{myprompt}{E1}[One-Step Role-Play]\label{P_E1}\sloppy\texttt{you are a high school graduate with family income \$10K per year, who is asian female from california, whose parents have college education, who is catholic. you need to make a college attendance decision. what would be factors (factors of schools and factors of yourself) that can determine your decision but that do not directly affect your future earnings, except through college attendance (that is, that affect your earnings only through college attendance), and that are not associated with your innate ability and personality and school quality, which determine earnings? choose 20 factors that are quantifiable. create separate lists for school factors and factors of yourself. for each factor chosen, explain your reasoning.}\end{myprompt}

The direct strategies (3)--(5) start with Prompt \ref{P_E2}, which gives the definition of a valid IV. The two-step direct strategy (3) then runs Prompts \ref{P_E3a} and \ref{P_E3b}, which mirror Prompts \ref{P1-1} and \ref{P2x-1}; the one-step direct strategy (4) runs Prompt \ref{P_E4} instead; and the novelty-encouraged strategy (5) runs Prompt \ref{P_E4} with the sentence in Prompt \ref{P_E5} appended. The covariates in these prompts are the characteristics of the student in Prompt \ref{P2x-1}.

\begin{myprompt}{E2}[Definition of IV]\label{P_E2}\sloppy\texttt{here is the definition of instrumental variables (IV): IV is used to estimate causal effect of a treatment on outcome when treatment is endogenous in the sense that there are unobserved confounders that simultaneously affect the treatment and outcome. IV is a variable that is correlated with the treatment and affects the outcome only through the treatment and IV is independent of the unobserved confounders. Do you understand the definition of IV?}\end{myprompt}

\begin{myprompt}{E3a}[Two-Step Direct, Step 1]\label{P_E3a}\sloppy\texttt{now, suppose the treatment is college attendance, the outcome is wages and the unobserved confounders are students' ability, personality and school quality. suggest me 40 candidate variables that are correlated with the treatment and that affect the outcome only through the treatment. the variables should be quantifiable. explain your answers.}\end{myprompt}

\begin{myprompt}{E3b}[Two-Step Direct, Step 2]\label{P_E3b}\sloppy\texttt{among the 40 variables listed above, choose 20 variables that are also independent of the unobserved confounders, so that they would be valid instrumental variables conditional on family income, race/ethnicity, gender, state of residence, parental education, religion. for each variable chosen, explain your reasoning.}\end{myprompt}

\begin{myprompt}{E4}[One-Step Direct]\label{P_E4}\sloppy\texttt{now, suppose the treatment is college attendance, the outcome is wages and the unobserved confounders are students' ability, personality and school quality. suggest me 20 instrumental variables that would be valid conditional on family income, race/ethnicity, gender, state of residence, parental education, religion}\end{myprompt}

\begin{myprompt}{E5}[Emphasis on Novelty]\label{P_E5}\sloppy\texttt{as much as possible, suggest instrumental variables that have not been used in the existing empirical literature on this question (that is, novel instruments), while making sure that they still satisfy the definition of a valid instrument given above.}\end{myprompt}

In every strategy, the candidates of a run are then converted into standardized \emph{IV cards} (i.e., the variable name and one- or two-sentence rationales for \ref{as:REL}, \ref{as:EX} and \ref{as:IND}), which are written by an independent LLM from the output of the discovery run (Prompt \ref{P_E6}, run with GPT5 in batches of five candidates). This way, the format of the prompt cannot reveal which strategy produced the candidate. As in Section \ref{sec:Prompt-Construction}, each bracketed term is filled in from the domain and the run. In the role-playing strategies, the prompt additionally quotes the role-playing prompt that specifies the covariates (Prompt \ref{P2x-1} in the two-step strategy and Prompt \ref{P_E1} in the one-step strategy) and asks to extract the control variables mentioned in it; the judge does not use this field.

\begin{myprompt}{E6}[IV Card]\label{P_E6}\sloppy\texttt{You are an econometrician. For each candidate instrumental variable listed below,}~\\ \texttt{produce a structured "IV card" with exactly the following fields.}~\\ \texttt{Context:}~\\ \texttt{- Treatment (D): {[}treatment{]}}~\\ \texttt{- Outcome (Y): {[}outcome{]}}~\\ \texttt{- Unobserved confounders (U): {[}confounders{]}}~\\ \texttt{Candidate instruments from the discovery procedure:}~\\ \texttt{{[}list of the candidates of the run, each with the rationale given by the LLM{]}}~\\ \texttt{For EACH instrument, output a JSON object with these fields:}~\\ \texttt{"name": short instrument label (string),}~\\ \texttt{"relevance": 1-2 sentences on the mechanism Z -> D (string),}~\\ \texttt{"exclusion": 1-2 sentences on why Z does NOT directly affect Y except through D (string),}~\\ \texttt{"independence": 1-2 sentences on why Z is independent of U conditional on controls X (string),}~\\ \texttt{"controls": candidate control variables X (list of strings)}~\\ \texttt{Output a JSON array of objects, one per instrument.}~\\ \texttt{Output ONLY valid JSON, no markdown fences, no commentary.}\end{myprompt}

\subsection{The LLM Judge Construction\label{app:judge-prompt}}

The judge (GPT5) receives the system message in Prompt \ref{P_E7} and the user message in Prompt \ref{P_E8}, and returns the two scores as a JSON object. The setting lines of Prompt \ref{P_E8} are shown for the returns to college example; for the two peer-effects domains, they include the caution that ``\texttt{The variable should be understood as factors of peers (not factors of individual receiving peer effects).}'' The expert-scored examples are all IVs of the candidate's domain in the survey of Section \ref{sec:Expert-Evaluation-of} (seven to sixteen per domain), listed in increasing order of denoised validity, each with the rationale shown to the experts and its denoised scores rounded to integers. The message asks the judge to return the two integer scores in a fixed format.

In the evaluation stage, each of the 70 IVs in the expert survey is scored in turn as a hold-out IV that is removed from the few-shot examples. When each hold-out IV is scored, the judge is presented with the rationale shown to the experts. When evaluated (Figure \ref{fig:judge-scatter}), we find that the rank correlation between the judge's scores and the denoised expert scores is 0.64 for validity and 0.78 for non-novelty. In leave-one-out cross-validation, the mean absolute error of the calibrated judge scores is 0.27 for validity and 0.30 for non-novelty on the 1--5 scale, compared with 0.33 and 0.44 for a naive predictor that assigns every IV the average score of its domain.

\begin{myprompt}{E7}[LLM Judge: System Message]\label{P_E7}\sloppy\texttt{You are an expert academic economist evaluating candidate instrumental variables (IVs).}~\\ \texttt{Here is the definition of a valid instrumental variable (IV): An IV is used to estimate the causal effect of a treatment on an outcome when the treatment is endogenous in the sense that there are unobserved confounders that simultaneously affect the treatment and the outcome. An IV is a variable that satisfies the following conditions: (i) it is correlated with the treatment; (ii) it affects the outcome only through the treatment; and (iii) it is independent of the unobserved confounders.}~\\ \texttt{You will be shown an IV candidate with its rationale, in the context of a specific econometric setting.}~\\ \texttt{Your task is to provide two ratings:}~\\ \texttt{You will rate one candidate IV at a time, in the econometric setting described in the next message. The candidate was proposed during an instrument-discovery exercise; its rationale is shown for reference when available. For each IV, please provide:}~\\ \texttt{A validity grade (1 = definitely not valid, up to 5 = definitely valid)}~\\ \texttt{A confidence rating on whether it already exists in the literature (1 = low confidence it exists or equivalently 1 = most novel, up to 5 = high confidence it exists or equivalently 5 = least novel)}~\\ \texttt{Please focus primarily on the variable name; use the rationale text to inform your judgment of the instrument's plausibility, but grade the instrument, not the quality of the write-up. If no rationale is shown, judge from the variable name alone.}~\\ \texttt{Output ONLY a JSON object with two integer fields:}~\\ \texttt{\{"validity": <int 1-5>, "nonnovelty": <int 1-5>\}}~\\ \texttt{IMPORTANT: Use the full 1-5 scale. Do not compress your ratings to a narrow range. A score of 1 means definitely not valid / most novel, and 5 means definitely valid / high confidence it exists in the literature.}\end{myprompt}

\begin{myprompt}{E8}[LLM Judge: User Message]\label{P_E8}\sloppy\texttt{Setting: Returns to College}~\\ \texttt{Setup: High-school graduate deciding whether to attend college.}~\\ \texttt{Endogeneity: Unobserved ability \& school quality jointly influence attendance and earnings.}~\\ \texttt{--- EXPERT-RATED EXAMPLES (from this same setting, ordered from lowest to highest validity) ---}~\\ \texttt{IV suggested: {[}survey IV name{]}}~\\ \texttt{Rationale Provided: {[}survey IV rationale{]}}~\\ \texttt{Expected rating: \{"validity": {[}denoised validity{]}, "nonnovelty": {[}denoised non-novelty{]}\}}~\\ \texttt{{[}one such block for each survey IV of the domain{]}}~\\ \texttt{--- IV TO EVALUATE ---}~\\ \texttt{IV suggested: {[}candidate name{]}}~\\ \texttt{Rationale Provided: Relevance: {[}relevance{]} Exclusion: {[}exclusion{]} Independence: {[}independence{]}}~\\ \texttt{Provide your rating as a JSON object:}\end{myprompt}

\subsection{Additional Tables\label{app:judge-tables-list}}

Tables \ref{tab:judge-eval-1-two_step_roleplay}--\ref{tab:judge-eval-1-direct_novel} list, for each strategy, the instrument families with non-novelty score $k=1$ or $2$ whose average calibrated validity is at least the benchmark of their domain (the LLM-judge analog of Table \ref{tab:eval-1}). No candidate from the one-step direct strategy met this criterion, so \emph{no table is reported for it}. Tables \ref{tab:data-1}--\ref{tab:data-6} list, for each domain, the datasets found for the novel and valid candidates in Section \ref{subsec:data-availability}.

\begin{table}
\noindent \begin{centering}
\begin{tabular}{>{\centering}p{4.2cm}>{\centering\arraybackslash}p{9.2cm}}
\toprule 
Domain$^{\dagger}$ (Benchmark)$^{\dagger\dagger}$ & Novel IVs Suggested$^{\dagger\dagger\dagger}$\tabularnewline
\midrule 
1 (2.59) & {\scriptsize{}Campus Safety Statistics (+0.13)}\tabularnewline
\midrule
2 (2.28) & {\scriptsize{}School Start Times (+0.25)}\tabularnewline
 & {\scriptsize{}Access to Online Learning (+0.06)}\tabularnewline
 & {\scriptsize{}Library Book Count (+0.06)}\tabularnewline
 & {\scriptsize{}School Wi-Fi Quality (+0.06)}\tabularnewline
 & {\scriptsize{}Access to School Lunch Programs (+0.06)}\tabularnewline
 & {\scriptsize{}Mental Health Support (+0.06)}\tabularnewline
 & {\scriptsize{}Access to Internships (+0.06)}\tabularnewline
 & {\scriptsize{}Art/Music Program Availability (+0.06)}\tabularnewline
 & {\scriptsize{}School Sports Team Success (+0.06)}\tabularnewline
 & {\scriptsize{}Extracurricular Offerings (+0.06)}\tabularnewline
 & {\scriptsize{}Availability of Electives (+0.06)}\tabularnewline
 & {\scriptsize{}Community Support (+0.06)}\tabularnewline
 & {\scriptsize{}Mentorship Opportunities (+0.06)}\tabularnewline
 & {\scriptsize{}Access to Tutoring Services (+0.06)}\tabularnewline
 & {\scriptsize{}Number of Guidance Counselor Meetings (+0.06)}\tabularnewline
 & {\scriptsize{}Annual Field Trips (+0.06)}\tabularnewline
 & {\scriptsize{}School Anti-Bullying Policies (+0.06)}\tabularnewline
 & {\scriptsize{}Participation in College Fairs (+0.06)}\tabularnewline
 & {\scriptsize{}School Events/Activities (+0.06)}\tabularnewline
 & {\scriptsize{}Exposure to Role Models (+0.06)}\tabularnewline
 & {\scriptsize{}Cultural Values on Education (+0.06)}\tabularnewline
 & {\scriptsize{}School Discipline Policies (+0.06)}\tabularnewline
 & {\scriptsize{}Parent Satisfaction Surveys (+0.06)}\tabularnewline
\midrule
5 (2.53) & {\scriptsize{}Collaborations with Experts (+0.20)}\tabularnewline
 & {\scriptsize{}Media Presence (+0.20)}\tabularnewline
 & {\scriptsize{}Number of Demonstration Plots (+0.20)}\tabularnewline
 & {\scriptsize{}Number of Training Sessions Conducted (+0.20)}\tabularnewline
 & {\scriptsize{}Number of Technology Demonstrations Conducted (+0.20)}\tabularnewline
 & {\scriptsize{}Positive Testimonials Count (+0.20)}\tabularnewline
 & {\scriptsize{}Successful Case Studies Shared (+0.20)}\tabularnewline
 & {\scriptsize{}Workshops Hosted (+0.20)}\tabularnewline
 & {\scriptsize{}Public Speaking Engagements (+0.20)}\tabularnewline
\bottomrule
\end{tabular}
\par\end{centering}
\renewcommand{\thetable}{A4}\caption{(a) LLM Judge: IVs with $k=1$ or $2$ Outperforming Benchmark: Two-step role-play (Domains 1, 2, 5) ($k=1$: most novel; $k=5$: least novel)}
\label{tab:judge-eval-1-two_step_roleplay}

\noindent\begin{minipage}[t]{1\columnwidth}%
\vspace{-0.2cm}

$^{\dagger}${\scriptsize{}Domain ID: 1. Returns to College Attendance; 2. Returns to Years of Education; 3. Demand Estimation; 4. Production Functions; 5. Peer Effects (technology adoption); 6. Peer Effects (teenage smoking).} $^{\dagger}$$^{\dagger}${\scriptsize{}The benchmark is the average calibrated validity of the strategy's candidates with non-novelty $\ge4$ in the domain, in parenthesis. }$^{\dagger}$$^{\dagger}$$^{\dagger}${\scriptsize{}An instrument family is listed if the average calibrated validity of its novel candidates ($k=1$ or $2$) across runs is at least the benchmark; the improvement over the benchmark is in parenthesis. Domains without qualifying candidates are omitted. Candidate names are as generated by the LLM.}{\scriptsize\par}%
\end{minipage}
\end{table}

\medskip{}
\begin{table}
\noindent \begin{centering}
\begin{tabular}{>{\centering}p{4.2cm}>{\centering\arraybackslash}p{9.2cm}}
\toprule 
Domain$^{\dagger}$ (Benchmark)$^{\dagger\dagger}$ & Novel IVs Suggested$^{\dagger\dagger\dagger}$\tabularnewline
\midrule 
6 (2.32) & {\scriptsize{}Content Reach (+0.41)}\tabularnewline
 & {\scriptsize{}Social Media Advertising (+0.41)}\tabularnewline
 & {\scriptsize{}Visual Appeal (+0.41)}\tabularnewline
 & {\scriptsize{}Frequency of Posts (+0.41)}\tabularnewline
 & {\scriptsize{}Platform Diversity (+0.41)}\tabularnewline
 & {\scriptsize{}Story Views (+0.41)}\tabularnewline
 & {\scriptsize{}Influence Score (+0.41)}\tabularnewline
 & {\scriptsize{}Content Quality (+0.41)}\tabularnewline
 & {\scriptsize{}Engagement Rate (+0.41)}\tabularnewline
 & {\scriptsize{}Sentiment of Comments (+0.41)}\tabularnewline
 & {\scriptsize{}Hashtag Usage (+0.36)}\tabularnewline
 & {\scriptsize{}Content Variety (+0.34)}\tabularnewline
 & {\scriptsize{}Number of Followers (+0.24)}\tabularnewline
 & {\scriptsize{}Use of Influencers/Endorsements (+0.11)}\tabularnewline
 & {\scriptsize{}School Popularity (+0.02)}\tabularnewline
 & {\scriptsize{}School Social Group Reach (+0.02)}\tabularnewline
 & {\scriptsize{}Cultural Influence Index (+0.02)}\tabularnewline
 & {\scriptsize{}Peer Networks (+0.02)}\tabularnewline
 & {\scriptsize{}Communication Skills (+0.02)}\tabularnewline
 & {\scriptsize{}Language Use (+0.02)}\tabularnewline
 & {\scriptsize{}Leadership Role Count (+0.02)}\tabularnewline
 & {\scriptsize{}Lifestyle Display (+0.02)}\tabularnewline
 & {\scriptsize{}Regional Trends (+0.02)}\tabularnewline
 & {\scriptsize{}Local Celebrity Status (+0.02)}\tabularnewline
 & {\scriptsize{}Risk-Taking Behavior (+0.02)}\tabularnewline
 & {\scriptsize{}Participation in School Activities (+0.02)}\tabularnewline
 & {\scriptsize{}Resilience to Criticism (+0.02)}\tabularnewline
\bottomrule
\end{tabular}
\par\end{centering}
\renewcommand{\thetable}{A4}\caption{(b) LLM Judge: IVs with $k=1$ or $2$ Outperforming Benchmark: Two-step role-play (Domain 6) ($k=1$: most novel; $k=5$: least novel)}
\label{tab:judge-eval-1-two_step_roleplay-b}

\noindent\begin{minipage}[t]{1\columnwidth}%
\vspace{-0.2cm}

$^{\dagger}${\scriptsize{}Domain ID: 1. Returns to College Attendance; 2. Returns to Years of Education; 3. Demand Estimation; 4. Production Functions; 5. Peer Effects (technology adoption); 6. Peer Effects (teenage smoking).} $^{\dagger}$$^{\dagger}${\scriptsize{}The benchmark is the average calibrated validity of the strategy's candidates with non-novelty $\ge4$ in the domain, in parenthesis. }$^{\dagger}$$^{\dagger}$$^{\dagger}${\scriptsize{}An instrument family is listed if the average calibrated validity of its novel candidates ($k=1$ or $2$) across runs is at least the benchmark; the improvement over the benchmark is in parenthesis. Domains without qualifying candidates are omitted. Candidate names are as generated by the LLM.}{\scriptsize\par}%
\end{minipage}
\end{table}

\medskip{}
\begin{table}
\noindent \begin{centering}
\begin{tabular}{>{\centering}p{4.2cm}>{\centering\arraybackslash}p{9.2cm}}
\toprule 
Domain$^{\dagger}$ (Benchmark)$^{\dagger\dagger}$ & Novel IVs Suggested$^{\dagger\dagger\dagger}$\tabularnewline
\midrule 
1 (2.71) & {\footnotesize{}On-Campus Housing Availability (+0.02)}\tabularnewline
 & {\footnotesize{}Campus Safety (+0.02)}\tabularnewline
 & {\footnotesize{}Religious Services (+0.02)}\tabularnewline
\midrule
5 (2.61) & {\footnotesize{}Willingness to Share Knowledge (+0.12)}\tabularnewline
 & {\footnotesize{}Record Keeping of Farming Practices (+0.12)}\tabularnewline
 & {\footnotesize{}Success Stories Shared (+0.12)}\tabularnewline
 & {\footnotesize{}Training or Workshops Conducted (+0.12)}\tabularnewline
 & {\footnotesize{}Number of Demonstrations Held (+0.12)}\tabularnewline
 & {\footnotesize{}Number of Farmers Mentored (+0.12)}\tabularnewline
 & {\footnotesize{}Health and Nutrition Advocacy (+0.12)}\tabularnewline
 & {\footnotesize{}NumberOfWorkshopsConducted (+0.12)}\tabularnewline
\midrule
6 (2.38) & {\footnotesize{}Type of Content (+0.34)}\tabularnewline
 & {\footnotesize{}Visual Creativity Score (+0.34)}\tabularnewline
 & {\footnotesize{}Content Authenticity (+0.34)}\tabularnewline
 & {\footnotesize{}Engagement Rate (+0.34)}\tabularnewline
 & {\footnotesize{}Use of Story Features (+0.34)}\tabularnewline
 & {\footnotesize{}Engagement Tools (+0.34)}\tabularnewline
 & {\footnotesize{}Number of Social Media Platforms Used (+0.34)}\tabularnewline
 & {\footnotesize{}Hashtag Usage (+0.34)}\tabularnewline
 & {\footnotesize{}Frequency of Posts (+0.32)}\tabularnewline
 & {\footnotesize{}Content Quality (+0.30)}\tabularnewline
 & {\footnotesize{}Follower Count (+0.29)}\tabularnewline
 & {\footnotesize{}Influence Score (+0.25)}\tabularnewline
 & {\footnotesize{}Language Use (+0.09)}\tabularnewline
 & {\footnotesize{}Peer Mentions (+0.03)}\tabularnewline
 & {\footnotesize{}Mutual Friends' Support (+0.01)}\tabularnewline
\bottomrule
\end{tabular}
\par\end{centering}
\renewcommand{\thetable}{A5}\caption{LLM Judge: IVs with $k=1$ or $2$ Outperforming Benchmark: One-step role-play ($k=1$: most novel; $k=5$: least novel)}
\label{tab:judge-eval-1-one_step_roleplay}

\noindent\begin{minipage}[t]{1\columnwidth}%
\vspace{-0.2cm}

$^{\dagger}${\scriptsize{}Domain ID: 1. Returns to College Attendance; 2. Returns to Years of Education; 3. Demand Estimation; 4. Production Functions; 5. Peer Effects (technology adoption); 6. Peer Effects (teenage smoking).} $^{\dagger}$$^{\dagger}${\scriptsize{}The benchmark is the average calibrated validity of the strategy's candidates with non-novelty $\ge4$ in the domain, in parenthesis. }$^{\dagger}$$^{\dagger}$$^{\dagger}${\scriptsize{}An instrument family is listed if the average calibrated validity of its novel candidates ($k=1$ or $2$) across runs is at least the benchmark; the improvement over the benchmark is in parenthesis. Domains without qualifying candidates are omitted. Candidate names are as generated by the LLM.}{\scriptsize\par}%
\end{minipage}
\end{table}

\medskip{}
\begin{table}
\noindent \begin{centering}
\begin{tabular}{>{\centering}p{4.2cm}>{\centering\arraybackslash}p{9.2cm}}
\toprule 
Domain$^{\dagger}$ (Benchmark)$^{\dagger\dagger}$ & Novel IVs Suggested$^{\dagger\dagger\dagger}$\tabularnewline
\midrule 
5 (2.67) & {\footnotesize{}CharitableActivities (+0.06)}\tabularnewline
 & {\footnotesize{}Attendance at Local Government Meetings (+0.06)}\tabularnewline
\bottomrule
\end{tabular}
\par\end{centering}
\renewcommand{\thetable}{A6}\caption{LLM Judge: IVs with $k=1$ or $2$ Outperforming Benchmark: Two-step direct ($k=1$: most novel; $k=5$: least novel)}
\label{tab:judge-eval-1-two_step_direct}

\noindent\begin{minipage}[t]{1\columnwidth}%
\vspace{-0.2cm}

$^{\dagger}${\scriptsize{}Domain ID: 1. Returns to College Attendance; 2. Returns to Years of Education; 3. Demand Estimation; 4. Production Functions; 5. Peer Effects (technology adoption); 6. Peer Effects (teenage smoking).} $^{\dagger}$$^{\dagger}${\scriptsize{}The benchmark is the average calibrated validity of the strategy's candidates with non-novelty $\ge4$ in the domain, in parenthesis. }$^{\dagger}$$^{\dagger}$$^{\dagger}${\scriptsize{}An instrument family is listed if the average calibrated validity of its novel candidates ($k=1$ or $2$) across runs is at least the benchmark; the improvement over the benchmark is in parenthesis. Domains without qualifying candidates are omitted. Candidate names are as generated by the LLM.}{\scriptsize\par}%
\end{minipage}
\end{table}

\medskip{}
\begin{table}
\noindent \begin{centering}
\begin{tabular}{>{\centering}p{4.2cm}>{\centering\arraybackslash}p{9.2cm}}
\toprule 
Domain$^{\dagger}$ (Benchmark)$^{\dagger\dagger}$ & Novel IVs Suggested$^{\dagger\dagger\dagger}$\tabularnewline
\midrule 
6 (2.73) & {\footnotesize{}Local Cybersecurity Incidents (+0.19)}\tabularnewline
 & {\footnotesize{}Shared Viral Content Access (+0.00)}\tabularnewline
 & {\footnotesize{}Time Zone Engagement (+0.00)}\tabularnewline
 & {\footnotesize{}Local Social Media Workshops (+0.00)}\tabularnewline
\bottomrule
\end{tabular}
\par\end{centering}
\renewcommand{\thetable}{A7}\caption{LLM Judge: IVs with $k=1$ or $2$ Outperforming Benchmark: One-step direct (novelty-encouraged) ($k=1$: most novel; $k=5$: least novel)}
\label{tab:judge-eval-1-direct_novel}

\noindent\begin{minipage}[t]{1\columnwidth}%
\vspace{-0.2cm}

$^{\dagger}${\scriptsize{}Domain ID: 1. Returns to College Attendance; 2. Returns to Years of Education; 3. Demand Estimation; 4. Production Functions; 5. Peer Effects (technology adoption); 6. Peer Effects (teenage smoking).} $^{\dagger}$$^{\dagger}${\scriptsize{}The benchmark is the average calibrated validity of the strategy's candidates with non-novelty $\ge4$ in the domain, in parenthesis. }$^{\dagger}$$^{\dagger}$$^{\dagger}${\scriptsize{}An instrument family is listed if the average calibrated validity of its novel candidates ($k=1$ or $2$) across runs is at least the benchmark; the improvement over the benchmark is in parenthesis. Domains without qualifying candidates are omitted. Candidate names are as generated by the LLM.}{\scriptsize\par}%
\end{minipage}
\end{table}


\clearpage  
\begingroup\linespread{1}\selectfont\setlength{\arrayrulewidth}{0.2pt}\renewcommand{\thetable}{A8}
\begin{longtable}{>{\footnotesize}r>{\footnotesize\raggedright}p{0.29\textwidth}>{\footnotesize}c>{\footnotesize}c>{\footnotesize\raggedright}p{0.24\textwidth}>{\footnotesize\raggedright\arraybackslash}p{0.24\textwidth}}
\caption{Datasets for candidate IV families: Returns to College Attendance (Domain 1).}\label{tab:data-1}\\
\multicolumn{6}{@{}p{\textwidth}@{}}{\scriptsize{}Datasets include these variables: treatment (D) = College attendance (whether the person attended college); outcome (Y) = Wages / earnings in adulthood; covariates (X) = family income, race/ethnicity, gender, state of residence, parental education, religion (where available). Unobserved confounders: Students' ability, personality and school quality.}\\[3pt]
\toprule \# & Instrument family (members) & V & N & Data source (link) & How to get it \\ \midrule \endfirsthead
\toprule \# & Instrument family (members) & V & N & Data source (link) & How to get it \\ \midrule \endhead
\midrule \multicolumn{6}{r}{\scriptsize\emph{continued on next page}} \\ \endfoot
\bottomrule
\multicolumn{6}{@{}p{\textwidth}@{}}{\scriptsize{}Notes: 7 candidate IV names from the two-step role-play strategy (pooled across runs; raw LLM-judge non-novelty rating at most 2 and validity rating at least 3 on at least one card), grouped into 4 instrument families by the judge pipeline's embedding clustering; further family members are listed in parentheses. V (N) is the maximum validity (minimum non-novelty) rating over the family's novel-and-valid cards. One row per family and distinct data source; a family whose members found several data sources has several rows. NA = no dataset found for any member.} \\ \endlastfoot
1 & Campus Safety Statistics {\scriptsize(+2: Campus Safety; Safety Index)} & 3 & 2 & \href{https://nces.ed.gov/use-work/dataset/education-longitudinal-study-2002-els2002-third-follow-public-use-data-file}{ELS:2002} & Free public-use download via NCES Online Codebook/DataLab; create extract and download. Restricted-use license only needed for sensitive geography or PETS files. \\
\hline
 & {\scriptsize\textit{same family, other data source}} &  &  & \href{https://nlsinfo.org/content/getting-started/accessing-data}{NLSY97} & NLSY97 public-use via NLS Investigator (free); apply for restricted-use geocode/college-ID file from BLS to enable merging. Campus Safety and Security (Clery) institution-year files are free to download from OPE's Campus Safety site. \\
\hline
2 & Campus Housing Options {\scriptsize(+1: Housing Options)} & 3 & 2 & \href{https://nlsinfo.org/content/cohorts/nlsy97/using-and-understanding-the-data/nlsy97-documentation}{NLSY97} & NLSY97 public-use via NLS Investigator (free). To link colleges, apply for NLSY97 restricted-use transcript/geocode license via ResearchDataGov; IPEDS IC/IC\_AY files are free downloads from the IPEDS Data Center. \\
\hline
 & {\scriptsize\textit{same family, other data source}} &  &  & \href{https://www.bls.gov/nls/request-restricted-data/nlsy79-geocode-data.htm}{NLSY79} & NLSY79 main data are free (public-use); geocode/college identifiers require a restricted-use data application to BLS. IPEDS IC/COST files are free downloads from the NCES IPEDS Data Center. \\
\hline
3 & Meal Plan Costs & 3 & 2 & \href{https://nlsinfo.org/content/getting-started/accessing-data}{NLSY97} & NLSY97 main and PSTRAN: free download via NLS Investigator; IPEDS: free via IPEDS Data Center; NLSY97 IPEDS UNITIDs needed for the merge are on the restricted-use Geocode files (apply for a license). \\
\hline
4 & Weather and Climate & 3 & 2 & \href{https://www.bls.gov/nls/nlsy79.htm}{NLSY79} & NLSY79 public-use: free via NLS Investigator. Restricted geocodes (state/county/ZIP/tract and college identifiers): apply through ResearchDataGov SAP; use at BLS or FSRDCs. NOAA nClimGrid-Daily: free public download from NCEI. \\
\end{longtable}
\endgroup

\clearpage  
\begingroup\linespread{1}\selectfont\setlength{\arrayrulewidth}{0.2pt}\renewcommand{\thetable}{A9}
\begin{longtable}{>{\footnotesize}r>{\footnotesize\raggedright}p{0.29\textwidth}>{\footnotesize}c>{\footnotesize}c>{\footnotesize\raggedright}p{0.24\textwidth}>{\footnotesize\raggedright\arraybackslash}p{0.24\textwidth}}
\caption{Datasets for candidate IV families: Returns to Years of Education (Domain 2).}\label{tab:data-2}\\
\multicolumn{6}{@{}p{\textwidth}@{}}{\scriptsize{}Datasets include these variables: treatment (D) = Years of completed education; outcome (Y) = Wages / earnings in adulthood; covariates (X) = family income, race/ethnicity, gender, state of residence, parental education, religion (where available). Unobserved confounders: Innate ability, personality, school quality and other determinants of earnings.}\\[3pt]
\toprule \# & Instrument family (members) & V & N & Data source (link) & How to get it \\ \midrule \endfirsthead
\toprule \# & Instrument family (members) & V & N & Data source (link) & How to get it \\ \midrule \endhead
\midrule \multicolumn{6}{r}{\scriptsize\emph{continued on next page}} \\ \endfoot
\bottomrule
\multicolumn{6}{@{}p{\textwidth}@{}}{\scriptsize{}Notes: 1 candidate IV name from the two-step role-play strategy (pooled across runs; raw LLM-judge non-novelty rating at most 2 and validity rating at least 3 on at least one card), grouped into 1 instrument family by the judge pipeline's embedding clustering; further family members are listed in parentheses. V (N) is the maximum validity (minimum non-novelty) rating over the family's novel-and-valid cards. One row per family and distinct data source; a family whose members found several data sources has several rows. NA = no dataset found for any member.} \\ \endlastfoot
1 & School Start Times & 3 & 2 & \href{https://www.icpsr.umich.edu/web/DSDR/studies/21600}{Add Health} & Public-use download (registration) for core files; School Administrator files and full linking typically require restricted-use data agreement via Add Health/CPC. \\
\end{longtable}
\endgroup

\clearpage  
\begingroup\linespread{1}\selectfont\setlength{\arrayrulewidth}{0.2pt}\renewcommand{\thetable}{A10}
\begin{longtable}{>{\footnotesize}r>{\footnotesize\raggedright}p{0.29\textwidth}>{\footnotesize}c>{\footnotesize}c>{\footnotesize\raggedright}p{0.24\textwidth}>{\footnotesize\raggedright\arraybackslash}p{0.24\textwidth}}
\caption{Datasets for candidate IV families: Demand Estimation in a Fish Market (Domain 3).}\label{tab:data-3}\\
\multicolumn{6}{@{}p{\textwidth}@{}}{\scriptsize{}Datasets include these variables: treatment (D) = Price of fish (endogenous regressor); outcome (Y) = Quantity of fish demanded / sold; covariates (X) = none specified (where available). Unobserved confounders: Fish market conditions and customer characteristics that shift demand.}\\[3pt]
\toprule \# & Instrument family (members) & V & N & Data source (link) & How to get it \\ \midrule \endfirsthead
\toprule \# & Instrument family (members) & V & N & Data source (link) & How to get it \\ \midrule \endhead
\midrule \multicolumn{6}{r}{\scriptsize\emph{continued on next page}} \\ \endfoot
\bottomrule
\multicolumn{6}{@{}p{\textwidth}@{}}{\scriptsize{}Notes: 20 candidate IV names from the two-step role-play strategy (pooled across runs; raw LLM-judge non-novelty rating at most 2 and validity rating at least 3 on at least one card), grouped into 12 instrument families by the judge pipeline's embedding clustering; further family members are listed in parentheses. V (N) is the maximum validity (minimum non-novelty) rating over the family's novel-and-valid cards. One row per family and distinct data source; a family whose members found several data sources has several rows. NA = no dataset found for any member.} \\ \endlastfoot
1 & Waste Management Costs {\scriptsize(+2: Waste Disposal Costs; Waste Management Fees)} & 3 & 2 & \href{https://catalog.data.metro.tokyo.lg.jp/dataset/t000013d2000000032}{catalog.\allowbreak{}data.\allowbreak{}metro.\allowbreak{}tokyo.\allowbreak{}lg.\allowbreak{}jp/\allowbreak{}dataset/\allowbreak{}t00001.\allowbreak{}.\allowbreak{}.\allowbreak{}} & Free CSV download (Tokyo Metropolitan Government Open Data). Waste-fee data: public pages and CSV via Tokyo 23 Wards Clean Association/wards. \\
\hline
2 & Technology Costs {\scriptsize(+2: Technology Investment; Technological Investment)} & 3 & 2 & \href{https://dataverse.harvard.edu/dataset.xhtml?persistentId=doi:10.7910/DVN/BXSYEL}{doi:10.\allowbreak{}7910/\allowbreak{}dvn/\allowbreak{}bxsyel} & Free download from Harvard Dataverse (open access). \\
\hline
 & {\scriptsize\textit{same family, other data source}} &  &  & \href{https://data.norge.no/en/datasets/e5d67089-38d5-4c20-b467-199251357c29/fangstdata-koblet-med-fartoydata}{data.\allowbreak{}norge.\allowbreak{}no/\allowbreak{}en/\allowbreak{}datasets/\allowbreak{}e5d67089-\allowbreak{}38d5-\allowbreak{}4c20-\allowbreak{}.\allowbreak{}.\allowbreak{}.\allowbreak{}} & Free open data (CSV). Landing page above; documentation with variable list and definitions is on Fiskeridirektoratet; last 12 months of prices are excluded by policy. \\
\hline
3 & Depreciation {\scriptsize(+2: Depreciation of Equipment; Equipment Depreciation)} & 3 & 2 & \href{https://www.fisheries.noaa.gov/west-coast/science-data/economic-data-collection-west-coast-groundfish-trawl-fishery}{fisheries.\allowbreak{}noaa.\allowbreak{}gov/\allowbreak{}west-\allowbreak{}coast/\allowbreak{}science-\allowbreak{}data} & Restricted/confidential microdata; aggregated data public. For research access, contact the NWFSC EDC team listed on the page (data protected under Magnuson--Stevens Act). \\
\hline
 & {\scriptsize\textit{same family, other data source}} &  &  & \href{https://data.norge.no/en/datasets/e5d67089-38d5-4c20-b467-199251357c29/fangstdata-koblet-med-fartoydata}{data.\allowbreak{}norge.\allowbreak{}no/\allowbreak{}en/\allowbreak{}datasets/\allowbreak{}e5d67089-\allowbreak{}38d5-\allowbreak{}4c20-\allowbreak{}.\allowbreak{}.\allowbreak{}.\allowbreak{}} & Open CSV download under NLOD; documentation PDF on variables. Instrument comes from a separate, restricted research dataset (apply to Fiskeridirektoratet). \\
\hline
4 & Equipment Maintenance {\scriptsize(+1: Equipment Maintenance Costs)} & 3 & 2 & \href{https://mcc.icpsr.umich.edu/evaluations/index.php/catalog/17684}{mcc.\allowbreak{}icpsr.\allowbreak{}umich.\allowbreak{}edu/\allowbreak{}evaluations/\allowbreak{}catalog/\allowbreak{}17684} & No public files shown on landing page; request via MCC Evidence Platform/ICPSR contact. \\
\hline
5 & Utilities Expenses {\scriptsize(+1: Utility Costs)} & 3 & 2 & \href{https://www.franceagrimer.fr/chiffre-et-analyses-economiques/observatoires/observatoire-economique-de-la-peche-et-de-0}{franceagrimer.\allowbreak{}fr/\allowbreak{}chiffre-\allowbreak{}et-\allowbreak{}analyses-\allowbreak{}economiq.\allowbreak{}.\allowbreak{}.\allowbreak{}} & FranceAgriMer data: free download (weekly/monthly/annual aggregates; Excel/PDF/ZIP). ENTSO-E prices: free via Transparency Platform (CSV/API; free account/FTP). \\
\hline
 & {\scriptsize\textit{same family, other data source}} &  &  & \href{https://spec.api.metro.tokyo.lg.jp/spec/t000013d0000000003-08e0e9844a13338fdd79f72b580735ec-0?lang=en}{spec.\allowbreak{}api.\allowbreak{}metro.\allowbreak{}tokyo.\allowbreak{}lg.\allowbreak{}jp/\allowbreak{}spec/\allowbreak{}t000013d00000.\allowbreak{}.\allowbreak{}.\allowbreak{}} & Free public data. Fish prices/quantities downloadable via Tokyo Metropolitan Open Data CSV/API; JEPX market data downloadable from JEPX Market Data pages (select date/range, click Data Download). \\
\hline
6 & Insurance Costs & 3 & 2 & \href{https://economics.mit.edu/people/faculty/josh-angrist/angrist-data-archive}{Fulton Fish Market (Angrist-\allowbreak{}Graddy-\allowbreak{}Imbens /\allowbreak{} G.\allowbreak{}.\allowbreak{}.\allowbreak{}} & Free download from Angrist Data Archive; instrument series from NY Department of Insurance 2007 Annual Report PDF (free). \\
\hline
7 & Security Costs & 3 & 2 & \href{https://reports.pfex.org/customreport.asp}{reports.\allowbreak{}pfex.\allowbreak{}org/\allowbreak{}customreport.\allowbreak{}asp} & Free web query; CSV download from the PFEX Custom Report tool. \\
\hline
8 & Packaging Costs & 3 & 2 & \href{https://rdrr.io/cran/fixest/man/fulton.html}{Fulton Fish Market (Angrist-\allowbreak{}Graddy-\allowbreak{}Imbens /\allowbreak{} G.\allowbreak{}.\allowbreak{}.\allowbreak{}} & Free download via R CRAN package fixest; BLS PPI series are free via the PPI Databases page (Series Report/text files). \\
\hline
9 & Quality Control Costs & 3 & 2 & NA & NA \\
\hline
10 & Legal Fees & 3 & 2 & NA & NA \\
\hline
11 & Supplier Reliability & 4 & 2 & NA & NA \\
\hline
12 & Training Programs & 3 & 2 & NA & NA \\
\end{longtable}
\endgroup

\clearpage  
\begingroup\linespread{1}\selectfont\setlength{\arrayrulewidth}{0.2pt}\renewcommand{\thetable}{A11}
\begin{longtable}{>{\footnotesize}r>{\footnotesize\raggedright}p{0.29\textwidth}>{\footnotesize}c>{\footnotesize}c>{\footnotesize\raggedright}p{0.24\textwidth}>{\footnotesize\raggedright\arraybackslash}p{0.24\textwidth}}
\caption{Datasets for candidate IV families: Production Function Estimation (Domain 4).}\label{tab:data-4}\\
\multicolumn{6}{@{}p{\textwidth}@{}}{\scriptsize{}Datasets include these variables: treatment (D) = Labor and capital input choices; outcome (Y) = Firm output (production, revenue or value added); covariates (X) = capital intensity, scale of operations, market share, industry (where available). Unobserved confounders: Unobserved productivity shocks of the firm.}\\[3pt]
\toprule \# & Instrument family (members) & V & N & Data source (link) & How to get it \\ \midrule \endfirsthead
\toprule \# & Instrument family (members) & V & N & Data source (link) & How to get it \\ \midrule \endhead
\midrule \multicolumn{6}{r}{\scriptsize\emph{continued on next page}} \\ \endfoot
\bottomrule
\multicolumn{6}{@{}p{\textwidth}@{}}{\scriptsize{}Notes: 9 candidate IV names from the two-step role-play strategy (pooled across runs; raw LLM-judge non-novelty rating at most 2 and validity rating at least 3 on at least one card), grouped into 7 instrument families by the judge pipeline's embedding clustering; further family members are listed in parentheses. V (N) is the maximum validity (minimum non-novelty) rating over the family's novel-and-valid cards. One row per family and distinct data source; a family whose members found several data sources has several rows. NA = no dataset found for any member.} \\ \endlastfoot
1 & Project Timeline Requirements {\scriptsize(+2: Project Timeline Targets; Project Timelines)} & 3 & 2 & \href{https://www.contractsfinder.service.gov.uk/apidocumentation}{contractsfinder.\allowbreak{}service.\allowbreak{}gov.\allowbreak{}uk/\allowbreak{}apidocumentation} & Free OCDS JSON/CSV via Contracts Finder API or daily CSV on data.gov.uk; merge to Companies House Accounts bulk XBRL files (free) from \url{https://download.companieshouse.gov.uk/en_accountsdata.html}. \\
\hline
2 & Inventory Management & 3 & 2 & \href{https://www.census.gov/programs-surveys/mops/technical-documentation/methodology.html}{US Census of Manufactures /\allowbreak{} ASM} & Restricted-use microdata through the Federal Statistical Research Data Centers (FSRDC); apply for access. MOPS is a supplement to the ASM, enabling linkage to ASM/CMF plant-level inputs and outputs. \\
\hline
3 & Maintenance Costs & 3 & 2 & \href{https://microdata.gov.in/NADA/index.php/catalog/26/study-description}{India Annual Survey of Industries} & Apply for access via MOSPI's National Data Archive (registration + request/fee); approved users download unit-level files. \\
\hline
4 & Operational Risk Assessment & 3 & 2 & \href{https://data.cms.gov/provider-data/dataset/yq43-i98g}{data.\allowbreak{}cms.\allowbreak{}gov/\allowbreak{}provider-\allowbreak{}data/\allowbreak{}dataset/\allowbreak{}yq43-\allowbreak{}i98g} & Free public download (CSV via Provider Data Catalog). Merge with HCRIS cost report data (free public files on CMS). \\
\hline
5 & Sustainability Goals & 3 & 2 & \href{https://sciencebasedtargets.org/target-dashboard}{Compustat} & SBTi: free .xlsx download from the Target Dashboard; Compustat: WRDS subscription/restricted academic access (Compustat Annual Updates -- Fundamentals Annual). \\
\hline
6 & Quality Control Costs & 3 & 2 & NA & NA \\
\hline
7 & Operational Flexibility & 3 & 2 & NA & NA \\
\end{longtable}
\endgroup

\clearpage  
\begingroup\linespread{1}\selectfont\setlength{\arrayrulewidth}{0.2pt}\renewcommand{\thetable}{A12}
\begin{longtable}{>{\footnotesize}r>{\footnotesize\raggedright}p{0.29\textwidth}>{\footnotesize}c>{\footnotesize}c>{\footnotesize\raggedright}p{0.24\textwidth}>{\footnotesize\raggedright\arraybackslash}p{0.24\textwidth}}
\caption{Datasets for candidate IV families: Peer Effects on Technology Adoption (Domain 5).}\label{tab:data-5}\\
\multicolumn{6}{@{}p{\textwidth}@{}}{\scriptsize{}Datasets include these variables: treatment (D) = A farmer's influence on peers (the peer/reference farmer's behaviour or adoption); outcome (Y) = Peers' technology adoption decisions (adoption of a new agricultural technology); covariates (X) = age, gender, crop type (where available). Unobserved confounders: Shared background and preferences that put the farmer and the peers in the same neighbourhood / social network.}\\[3pt]
\toprule \# & Instrument family (members) & V & N & Data source (link) & How to get it \\ \midrule \endfirsthead
\toprule \# & Instrument family (members) & V & N & Data source (link) & How to get it \\ \midrule \endhead
\midrule \multicolumn{6}{r}{\scriptsize\emph{continued on next page}} \\ \endfoot
\bottomrule
\multicolumn{6}{@{}p{\textwidth}@{}}{\scriptsize{}Notes: 13 candidate IV names from the two-step role-play strategy (pooled across runs; raw LLM-judge non-novelty rating at most 2 and validity rating at least 3 on at least one card), grouped into 9 instrument families by the judge pipeline's embedding clustering; further family members are listed in parentheses. V (N) is the maximum validity (minimum non-novelty) rating over the family's novel-and-valid cards. One row per family and distinct data source; a family whose members found several data sources has several rows. NA = no dataset found for any member.} \\ \endlastfoot
1 & Educational Workshops Conducted {\scriptsize(+1: Workshops Hosted)} & 3 & 2 & \href{https://data.cimmyt.org/dataset.xhtml?persistentId=doi:10.71682/10549340}{doi:10.\allowbreak{}71682/\allowbreak{}10549340} & Free download from CIMMYT Dataverse after accepting terms; includes two public data files (household data and lead-farmer data) and questionnaires. \\
\hline
2 & Number of Demonstrations Conducted {\scriptsize(+1: Number of Technology Demonstrations Conducted)} & 3 & 2 & \href{https://www.openicpsr.org/openicpsr/project/130605/version/V1/view?path=/openicpsr/130605/fcr:versions/V1/data}{Beaman et al. Malawi network targeting} & Free download from openICPSR (public-use Stata files: mnw\_public.dta, census.dta, conversation.dta, panel\_seed.dta, etc.). \\
\hline
 & {\scriptsize\textit{same family, other data source}} &  &  & \href{https://microdata.worldbank.org/catalog/2285}{microdata.\allowbreak{}worldbank.\allowbreak{}org/\allowbreak{}catalog/\allowbreak{}2285} & Free public-use microdata via World Bank Microdata Library (registration/login required to download). \\
\hline
3 & Number of Training Sessions Conducted {\scriptsize(+1: Training Sessions Conducted)} & 3 & 2 & \href{https://www.openicpsr.org/openicpsr/project/130605/version/V1/view}{Beaman et al. Malawi network targeting} & Free download (openICPSR replication package). \\
\hline
 & {\scriptsize\textit{same family, other data source}} &  &  & \href{https://data.cimmyt.org/dataset.xhtml?persistentId=doi:10.71682/10549340}{doi:10.\allowbreak{}71682/\allowbreak{}10549340} & Free download after accepting custom terms on CIMMYT Dataverse (files are public). \\
\hline
4 & Collaborations with Experts & 3 & 2 & \href{https://www.openicpsr.org/openicpsr/project/130605/version/V1/view?path=/openicpsr/130605/fcr:versions/V1/data&type=folder}{Beaman et al. Malawi network targeting} & Free public download from openICPSR (AEA Data and Code Repository). \\
\hline
5 & Number of Demonstration Plots & 3 & 2 & \href{https://microdata.worldbank.org/index.php/catalog/2285}{microdata.\allowbreak{}worldbank.\allowbreak{}org/\allowbreak{}catalog/\allowbreak{}2285} & Free public-use microdata via the MCC external repository linked on the World Bank Microdata page (no fee/registration typically required). \\
\hline
6 & Positive Testimonials Count {\scriptsize(+1: Testimonial Count)} & 3 & 2 & NA & NA \\
\hline
7 & Media Presence & 3 & 2 & NA & NA \\
\hline
8 & Public Speaking Engagements & 3 & 2 & NA & NA \\
\hline
9 & Successful Case Studies Shared & 3 & 2 & NA & NA \\
\end{longtable}
\endgroup

\clearpage  
\begingroup\linespread{1}\selectfont\setlength{\arrayrulewidth}{0.2pt}\renewcommand{\thetable}{A13}
\begin{longtable}{>{\footnotesize}r>{\footnotesize\raggedright}p{0.29\textwidth}>{\footnotesize}c>{\footnotesize}c>{\footnotesize\raggedright}p{0.24\textwidth}>{\footnotesize\raggedright\arraybackslash}p{0.24\textwidth}}
\caption{Datasets for candidate IV families: Peer Effects on Teenage Smoking (Domain 6).}\label{tab:data-6}\\
\multicolumn{6}{@{}p{\textwidth}@{}}{\scriptsize{}Datasets include these variables: treatment (D) = A teenager's influence on peers through social media (the peer's behaviour / influence); outcome (Y) = Peers' smoking decisions; covariates (X) = age, gender, urbanicity, education level, family income level (where available). Unobserved confounders: Shared background and preferences that put the teenager and the peers in the same social network.}\\[3pt]
\toprule \# & Instrument family (members) & V & N & Data source (link) & How to get it \\ \midrule \endfirsthead
\toprule \# & Instrument family (members) & V & N & Data source (link) & How to get it \\ \midrule \endhead
\midrule \multicolumn{6}{r}{\scriptsize\emph{continued on next page}} \\ \endfoot
\bottomrule
\multicolumn{6}{@{}p{\textwidth}@{}}{\scriptsize{}Notes: 37 candidate IV names from the two-step role-play strategy (pooled across runs; raw LLM-judge non-novelty rating at most 2 and validity rating at least 3 on at least one card), grouped into 14 instrument families by the judge pipeline's embedding clustering; further family members are listed in parentheses. V (N) is the maximum validity (minimum non-novelty) rating over the family's novel-and-valid cards. One row per family and distinct data source; a family whose members found several data sources has several rows. NA = no dataset found for any member.} \\ \endlastfoot
1 & Frequency of Posts {\scriptsize(+3: Post Timing; Content Frequency; Post Frequency)} & 3 & 2 & \href{https://sites.nd.edu/nethealth/data-2/}{Notre Dame NetHealth} & Free download (CSV/DTA) from project website; optional form/email registration requested. Codebooks linked on the Documentation page. \\
\hline
2 & Platform Diversity {\scriptsize(+3: Platform Choice; Platform Used; Social Media Platform Diversity)} & 3 & 2 & \href{https://sites.nd.edu/nethealth/data-2/}{Notre Dame NetHealth} & Free download (CSV/DTA) from project site; optional registration form; codebooks provided on the Documentation page. \\
\hline
 & {\scriptsize\textit{same family, other data source}} &  &  & \href{https://keck.usc.edu/cana/social-networks-study/}{keck.\allowbreak{}usc.\allowbreak{}edu/\allowbreak{}cana/\allowbreak{}social-\allowbreak{}networks-\allowbreak{}study} & Restricted; contact investigators/USC CANA for access. Related publications note data are available on reasonable request. \\
\hline
3 & Type of Content {\scriptsize(+3: Content Style; Content Type; Content Variety)} & 3 & 2 & \href{https://www.icpsr.umich.edu/web/NAHDAP/studies/36231}{PATH Study} & Restricted-use microdata via ICPSR/NAHDAP; application and data use agreement required (no fee). \\
\hline
4 & Engagement Rate {\scriptsize(+2: Peer Interaction Rate; Video Content Engagement)} & 3 & 2 & \href{https://sites.nd.edu/nethealth/data-2/}{Notre Dame NetHealth} & Free download (CSV/DTA + PDF codebooks) after completing a short notification form; data files are hosted via public Google Drive links. \\
\hline
5 & Content Virality {\scriptsize(+2: Content Virality Score; Content Reach)} & 3 & 2 & \href{https://www.icpsr.umich.edu/sites/dsdr/posts/shared/add-health-data-guide}{Add Health} & Restricted-use contract via UNC CPC Data Portal (network identifiers required); public-use files downloadable via ICPSR. Apply for restricted-use to link friends and construct network measures. \\
\hline
6 & Content Quality {\scriptsize(+1: Content Quality Score)} & 3 & 2 & \href{https://sites.nd.edu/nethealth/data-2/}{Notre Dame NetHealth} & Free download (CSV/DTA) from project site via Google Drive; optional short form to notify usage. \\
\hline
7 & Influence Score & 3 & 2 & \href{https://olizardo.github.io/NetSense-Codebook/download.html}{Notre Dame NetSense} & Free public download of de-identified CSV/R/Stata files from the project's download page. \\
\hline
8 & Hashtag Usage {\scriptsize(+4: Hashtag Strategy; Hashtag Use; Use of Hashtags; Use of Influential Hashtags)} & 3 & 2 & NA & NA \\
\hline
9 & Sentiment Analysis {\scriptsize(+2: Sentiment Analysis Score; Sentiment of Comments)} & 3 & 2 & NA & NA \\
\hline
10 & Ad Spend {\scriptsize(+1: Social Media Advertising)} & 3 & 2 & NA & NA \\
\hline
11 & Influencer Collaborations {\scriptsize(+1: Influencer Status)} & 3 & 2 & NA & NA \\
\hline
12 & Post Aesthetics Quality {\scriptsize(+1: Visual Appeal)} & 3 & 2 & NA & NA \\
\hline
13 & Number of Followers & 3 & 2 & NA & NA \\
\hline
14 & Story Views & 3 & 2 & NA & NA \\
\end{longtable}
\endgroup


{\small
\bibliographystyle{ecta}
\bibliography{mining_causality}

\begin{thebibliography}{66}
\newcommand{\enquote}[1]{``#1''}
\expandafter\ifx\csname natexlab\endcsname\relax\def\natexlab#1{#1}\fi

\bibitem[\protect\citeauthoryear{Ackerberg, Caves, and Frazer}{Ackerberg
  et~al.}{2015}]{ackerberg2015identification}
\textsc{Ackerberg, D.~A., K.~Caves, and G.~Frazer} (2015):
  \enquote{Identification Properties of Recent Production Function Estimators,}
  \emph{Econometrica}, 83, 2411--2451.

\bibitem[\protect\citeauthoryear{Angrist}{Angrist}{2014}]{angrist2014perils}
\textsc{Angrist, J.~D.} (2014): \enquote{The Perils of Peer Effects,}
  \emph{Labour Economics}, 30, 98--108.

\bibitem[\protect\citeauthoryear{Angrist, Graddy, and Imbens}{Angrist
  et~al.}{2000}]{angrist2000interpretation}
\textsc{Angrist, J.~D., K.~Graddy, and G.~W. Imbens} (2000): \enquote{The
  Interpretation of Instrumental Variables Estimators in Simultaneous Equations
  Models with an Application to the Demand for Fish,} \emph{The Review of
  Economic Studies}, 67, 499--527.

\bibitem[\protect\citeauthoryear{Angrist and Krueger}{Angrist and
  Krueger}{1991}]{angrist1991does}
\textsc{Angrist, J.~D. and A.~B. Krueger} (1991): \enquote{Does compulsory
  school attendance affect schooling and earnings?} \emph{The Quarterly Journal
  of Economics}, 106, 979--1014.

\bibitem[\protect\citeauthoryear{Angrist and Pischke}{Angrist and
  Pischke}{2010}]{angrist2010credibility}
\textsc{Angrist, J.~D. and J.-S. Pischke} (2010): \enquote{The Credibility
  Revolution in Empirical Economics: How Better Research Design Is Taking the
  Con Out of Econometrics,} \emph{Journal of Economic Perspectives}, 24, 3--30.

\bibitem[\protect\citeauthoryear{Ashenfelter and Zimmerman}{Ashenfelter and
  Zimmerman}{1997}]{ashenfelter1997estimates}
\textsc{Ashenfelter, O. and D.~J. Zimmerman} (1997): \enquote{Estimates of the
  returns to schooling from sibling data: Fathers, sons, and brothers,}
  \emph{Review of Economics and Statistics}, 79, 1--9.

\bibitem[\protect\citeauthoryear{Athey and Imbens}{Athey and
  Imbens}{2019}]{athey2019machine}
\textsc{Athey, S. and G.~W. Imbens} (2019): \enquote{Machine Learning Methods
  that Economists Should Know About,} \emph{Annual Review of Economics}, 11,
  685--725.

\bibitem[\protect\citeauthoryear{Battistin and Rettore}{Battistin and
  Rettore}{2002}]{battistin2002testing}
\textsc{Battistin, E. and E.~Rettore} (2002): \enquote{Testing for programme
  effects in a regression discontinuity design with imperfect compliance,}
  \emph{Journal of the Royal Statistical Society Series A: Statistics in
  Society}, 165, 39--57.

\bibitem[\protect\citeauthoryear{Berry, Levinsohn, and Pakes}{Berry
  et~al.}{1995}]{berry1995automobile}
\textsc{Berry, S., J.~Levinsohn, and A.~Pakes} (1995): \enquote{Automobile
  Prices in Market Equilibrium,} \emph{Econometrica}, 63, 841--890.

\bibitem[\protect\citeauthoryear{Blundell and Powell}{Blundell and
  Powell}{2003}]{Blundell:2003wi}
\textsc{Blundell, R. and J.~L. Powell} (2003): \enquote{Endogeneity in
  Nonparametric and Semiparametric Regression Models,} \emph{Econometric
  Society Monographs}, 36, 312--357.

\bibitem[\protect\citeauthoryear{Bramoull{\'e}, Djebbari, and
  Fortin}{Bramoull{\'e} et~al.}{2009}]{bramoulle2009identification}
\textsc{Bramoull{\'e}, Y., H.~Djebbari, and B.~Fortin} (2009):
  \enquote{Identification of Peer Effects through Social Networks,}
  \emph{Journal of Econometrics}, 150, 41--55.

\bibitem[\protect\citeauthoryear{Brown et~al.}{Brown
  et~al.}{2020}]{brown2020language}
\textsc{Brown, T.~B. et~al.} (2020): \enquote{Language Models are Few-Shot
  Learners,} \emph{arXiv preprint arXiv:2005.14165}.

\bibitem[\protect\citeauthoryear{Card}{Card}{1995}]{card1995using}
\textsc{Card, D.} (1995): \enquote{Using Geographic Variation in College
  Proximity to Estimate the Return to Schooling,} in \emph{Aspects of Labour
  Market Behaviour: Essays in Honour of John Vanderkamp}, ed. by L.~N.
  Christofides, E.~K. Grant, and R.~Swidinsky, Toronto: University of Toronto
  Press, 201--222.

\bibitem[\protect\citeauthoryear{Card}{Card}{1999}]{card1999causal}
---\hspace{-.1pt}---\hspace{-.1pt}--- (1999): \enquote{The Causal Effect of
  Education on Earnings,} \emph{Handbook of Labor Economics}, 3, 1801--1863.

\bibitem[\protect\citeauthoryear{Card}{Card}{2001}]{card2001estimating}
---\hspace{-.1pt}---\hspace{-.1pt}--- (2001): \enquote{Estimating the return to
  schooling: Progress on some persistent econometric problems,}
  \emph{Econometrica}, 69, 1127--1160.

\bibitem[\protect\citeauthoryear{Card and Krueger}{Card and
  Krueger}{1994}]{card1994minimum}
\textsc{Card, D. and A.~B. Krueger} (1994): \enquote{Minimum Wages and
  Employment: A Case Study of the Fast-Food Industry in New Jersey and
  Pennsylvania,} \emph{American Economic Review}, 84, 772--793.

\bibitem[\protect\citeauthoryear{Card and Shore-Sheppard}{Card and
  Shore-Sheppard}{2004}]{card2004using}
\textsc{Card, D. and L.~D. Shore-Sheppard} (2004): \enquote{Using Discontinuous
  Eligibility Rules to Identify the Effects of the Federal Medicaid Expansions
  on Low-Income Children,} \emph{Review of Economics and Statistics}, 86,
  752--766.

\bibitem[\protect\citeauthoryear{Chen, Tworek, Jun, Yuan, Ponde~de
  Oliveira~Pinto, Kaplan, Edwards, Burda, Joseph, Brockman, Ray, Puri, Krueger,
  Petrov, Khlaaf, Sastry, Mishkin, Chan, Gray, Ryder, Pavlov, Power, Kaiser,
  Bavarian, Winter, Tillet, Such, Cummings, Plappert, Chantzis, Barnes,
  Herbert-Voss, Guss, Nichol, Paino, Tezak, Tang, Babuschkin, Balaji, Jain,
  Saunders, Hesse, Carr, Leike, Achiam, Misra, Morikawa, Radford, Knight,
  Brundage, Murati, Mayer, Welinder, McGrew, Amodei, McCandlish, Sutskever, and
  Zaremba}{Chen et~al.}{2021}]{chen2021evaluating}
\textsc{Chen, M., J.~Tworek, H.~Jun, Q.~Yuan, H.~Ponde~de Oliveira~Pinto,
  J.~Kaplan, H.~Edwards, Y.~Burda, N.~Joseph, G.~Brockman, A.~Ray, R.~Puri,
  G.~Krueger, M.~Petrov, H.~Khlaaf, G.~Sastry, P.~Mishkin, B.~Chan, S.~Gray,
  N.~Ryder, M.~Pavlov, A.~Power, L.~Kaiser, M.~Bavarian, C.~Winter, P.~Tillet,
  F.~P. Such, D.~Cummings, M.~Plappert, F.~Chantzis, E.~Barnes,
  A.~Herbert-Voss, W.~H. Guss, A.~Nichol, A.~Paino, N.~Tezak, J.~Tang,
  I.~Babuschkin, S.~Balaji, S.~Jain, W.~Saunders, C.~Hesse, A.~N. Carr,
  J.~Leike, J.~Achiam, V.~Misra, E.~Morikawa, A.~Radford, M.~Knight,
  M.~Brundage, M.~Murati, K.~Mayer, P.~Welinder, B.~McGrew, D.~Amodei,
  S.~McCandlish, I.~Sutskever, and W.~Zaremba} (2021): \enquote{Evaluating
  Large Language Models Trained on Code,} \emph{arXiv preprint
  arXiv:2107.03374}.

\bibitem[\protect\citeauthoryear{Chernozhukov, Hansen, Kallus, Spindler, and
  Syrgkanis}{Chernozhukov et~al.}{2024}]{chernozhukov2024applied}
\textsc{Chernozhukov, V., C.~Hansen, N.~Kallus, M.~Spindler, and V.~Syrgkanis}
  (2024): \enquote{Applied Causal Inference Powered by ML and AI,} \emph{arXiv
  preprint arXiv:2403.02467}.

\bibitem[\protect\citeauthoryear{Conley, Hansen, and Rossi}{Conley
  et~al.}{2012}]{conley2012plausibly}
\textsc{Conley, T.~G., C.~B. Hansen, and P.~E. Rossi} (2012):
  \enquote{Plausibly exogenous,} \emph{Review of Economics and Statistics}, 94,
  260--272.

\bibitem[\protect\citeauthoryear{Conley and Udry}{Conley and
  Udry}{2010}]{conley2010learning}
\textsc{Conley, T.~G. and C.~R. Udry} (2010): \enquote{Learning about a New
  Technology: Pineapple in Ghana,} \emph{American Economic Review}, 100,
  35--69.

\bibitem[\protect\citeauthoryear{De~Giorgi}{De~Giorgi}{2005}]{degiorgi2005long}
\textsc{De~Giorgi, G.} (2005): \enquote{Long-term effects of a mandatory
  multistage program: The New Deal for young people in the UK,} IFS Working
  Papers W05/08, Institute for Fiscal Studies, London.

\bibitem[\protect\citeauthoryear{Deaner}{Deaner}{2025}]{deaner2025trade}
\textsc{Deaner, B.} (2025): \enquote{The trade-off between flexibility and
  robustness in instrumental variables analysis,} \emph{American Economic
  Review}, 12.

\bibitem[\protect\citeauthoryear{Dell}{Dell}{2025}]{dell2025deep}
\textsc{Dell, M.} (2025): \enquote{Deep learning for economists,} \emph{Journal
  of Economic Literature}, 63, 5--58.

\bibitem[\protect\citeauthoryear{Du, Kanodia, Brunborg, Vafa, and Athey}{Du
  et~al.}{2024}]{du2024labor}
\textsc{Du, T., A.~Kanodia, H.~Brunborg, K.~Vafa, and S.~Athey} (2024):
  \enquote{LABOR-LLM: Language-Based Occupational Representations with Large
  Language Models,} \emph{arXiv preprint arXiv:2406.17972}.

\bibitem[\protect\citeauthoryear{Fithria, Adlim, Jannah, and Tahlil}{Fithria
  et~al.}{2021}]{fithria2021indonesian}
\textsc{Fithria, F., M.~Adlim, S.~R. Jannah, and T.~Tahlil} (2021):
  \enquote{Indonesian Adolescents' Perspectives on Smoking Habits: A
  Qualitative Study,} \emph{BMC Public Health}, 21, 1--8.

\bibitem[\protect\citeauthoryear{Foster and Rosenzweig}{Foster and
  Rosenzweig}{1995}]{foster1995learning}
\textsc{Foster, A.~D. and M.~R. Rosenzweig} (1995): \enquote{Learning by Doing
  and Learning from Others: Human Capital and Technical Change in Agriculture,}
  \emph{Journal of Political Economy}, 103, 1176--1209.

\bibitem[\protect\citeauthoryear{Gaviria and Raphael}{Gaviria and
  Raphael}{2001}]{gaviria2001school}
\textsc{Gaviria, A. and S.~Raphael} (2001): \enquote{School-Based Peer Effects
  and Juvenile Behavior,} \emph{The Review of Economics and Statistics}, 83,
  257--268.

\bibitem[\protect\citeauthoryear{Gentzkow, Kelly, and Taddy}{Gentzkow
  et~al.}{2019}]{gentzkow2019text}
\textsc{Gentzkow, M., B.~Kelly, and M.~Taddy} (2019): \enquote{Text as data,}
  \emph{Journal of Economic Literature}, 57, 535--574.

\bibitem[\protect\citeauthoryear{Ghanem, Sant'Anna, and W{\"u}thrich}{Ghanem
  et~al.}{2022}]{ghanem2022selection}
\textsc{Ghanem, D., P.~H. Sant'Anna, and K.~W{\"u}thrich} (2022):
  \enquote{Selection and Parallel Trends,} \emph{arXiv preprint
  arXiv:2203.09001}.

\bibitem[\protect\citeauthoryear{Griliches}{Griliches}{1977}]{griliches1977estimating}
\textsc{Griliches, Z.} (1977): \enquote{Estimating the returns to schooling:
  Some econometric problems,} \emph{Econometrica: Journal of the Econometric
  Society}, 1--22.

\bibitem[\protect\citeauthoryear{Griliches and Mairesse}{Griliches and
  Mairesse}{1995}]{griliches1998production}
\textsc{Griliches, Z. and J.~Mairesse} (1995): \enquote{Production Functions:
  The Search for Identification,} \emph{NBER Working Paper No. w5067}.

\bibitem[\protect\citeauthoryear{Guryan}{Guryan}{2001}]{guryan2001does}
\textsc{Guryan, J.} (2001): \enquote{Does Money Matter?
  Regression-Discontinuity Estimates from Education Finance Reform in
  Massachusetts,} Working Paper 8269, National Bureau of Economic Research,
  Cambridge, MA.

\bibitem[\protect\citeauthoryear{Heckman}{Heckman}{1979}]{heckman1979sample}
\textsc{Heckman, J.~J.} (1979): \enquote{Sample Selection Bias as a
  Specification Error,} \emph{Econometrica}, 47, 153--161.

\bibitem[\protect\citeauthoryear{Heckman, Ichimura, and Todd}{Heckman
  et~al.}{1998}]{heckman1998matching}
\textsc{Heckman, J.~J., H.~Ichimura, and P.~Todd} (1998): \enquote{Matching as
  an Econometric Evaluation Estimator,} \emph{The Review of Economic Studies},
  65, 261--294.

\bibitem[\protect\citeauthoryear{Heckman and Vytlacil}{Heckman and
  Vytlacil}{2005}]{heckman2005structural}
\textsc{Heckman, J.~J. and E.~Vytlacil} (2005): \enquote{Structural Equations,
  Treatment Effects, and Econometric Policy Evaluation1,} \emph{Econometrica},
  73, 669--738.

\bibitem[\protect\citeauthoryear{Hern{\'a}n and Robins}{Hern{\'a}n and
  Robins}{2006}]{hernan2006instruments}
\textsc{Hern{\'a}n, M.~A. and J.~M. Robins} (2006): \enquote{Instruments for
  Causal Inference: An Epidemiologist's Dream?} \emph{Epidemiology}, 17,
  360--372.

\bibitem[\protect\citeauthoryear{Imbens and Angrist}{Imbens and
  Angrist}{1994}]{imbens1994identification}
\textsc{Imbens, G.~W. and J.~D. Angrist} (1994): \enquote{Identification and
  Estimation of Local Average Treatment Effects,} \emph{Econometrica}, 62,
  467--475.

\bibitem[\protect\citeauthoryear{Jin, Chen, Leeb, Gresele, Kamal, Lyu, Blin,
  Gonzalez~Adauto, Kleiman-Weiner, Sachan et~al.}{Jin
  et~al.}{2023{\natexlab{a}}}]{jin2023cladder}
\textsc{Jin, Z., Y.~Chen, F.~Leeb, L.~Gresele, O.~Kamal, Z.~Lyu, K.~Blin,
  F.~Gonzalez~Adauto, M.~Kleiman-Weiner, M.~Sachan, et~al.}
  (2023{\natexlab{a}}): \enquote{Cladder: Assessing causal reasoning in
  language models,} \emph{Advances in Neural Information Processing Systems},
  36, 31038--31065.

\bibitem[\protect\citeauthoryear{Jin, Liu, Lyu, Poff, Sachan, Mihalcea, Diab,
  and Sch{\"o}lkopf}{Jin et~al.}{2023{\natexlab{b}}}]{jin2023can}
\textsc{Jin, Z., J.~Liu, Z.~Lyu, S.~Poff, M.~Sachan, R.~Mihalcea, M.~Diab, and
  B.~Sch{\"o}lkopf} (2023{\natexlab{b}}): \enquote{Can large language models
  infer causation from correlation?} \emph{arXiv preprint arXiv:2306.05836}.

\bibitem[\protect\citeauthoryear{Kiciman, Ness, Sharma, and Tan}{Kiciman
  et~al.}{2023}]{kiciman2023causal}
\textsc{Kiciman, E., R.~Ness, A.~Sharma, and C.~Tan} (2023): \enquote{Causal
  reasoning and large language models: Opening a new frontier for causality,}
  \emph{Transactions on Machine Learning Research}.

\bibitem[\protect\citeauthoryear{Kim, Garg, Peng, and Garg}{Kim
  et~al.}{2025}]{kim2025correlated}
\textsc{Kim, E.~M., A.~Garg, K.~Peng, and N.~Garg} (2025): \enquote{Correlated
  Errors in Large Language Models,} in \emph{Proceedings of the 42nd
  International Conference on Machine Learning}, PMLR, vol. 267 of
  \emph{Proceedings of Machine Learning Research}, 30038--30066.

\bibitem[\protect\citeauthoryear{Kong, Zhao, Chen, Li, Qin, Sun, Zhou, Wang,
  and Dong}{Kong et~al.}{2023}]{kong2023better}
\textsc{Kong, A., S.~Zhao, H.~Chen, Q.~Li, Y.~Qin, R.~Sun, X.~Zhou, E.~Wang,
  and X.~Dong} (2023): \enquote{Better zero-shot reasoning with role-play
  prompting,} \emph{arXiv preprint arXiv:2308.07702}.

\bibitem[\protect\citeauthoryear{Korinek}{Korinek}{2023}]{korinek2023generative}
\textsc{Korinek, A.} (2023): \enquote{Generative {AI} for Economic Research:
  Use Cases and Implications for Economists,} \emph{Journal of Economic
  Literature}, 61, 1281--1317.

\bibitem[\protect\citeauthoryear{Lee and Lemieux}{Lee and
  Lemieux}{2010}]{lee2010regression}
\textsc{Lee, D.~S. and T.~Lemieux} (2010): \enquote{Regression Discontinuity
  Designs in Economics,} \emph{Journal of Economic Literature}, 48, 281--355.

\bibitem[\protect\citeauthoryear{Levinsohn and Petrin}{Levinsohn and
  Petrin}{2003}]{levinsohn2003estimating}
\textsc{Levinsohn, J. and A.~Petrin} (2003): \enquote{Estimating Production
  Functions Using Inputs to Control for Unobservables,} \emph{The Review of
  Economic Studies}, 70, 317--341.

\bibitem[\protect\citeauthoryear{Li and Tobias}{Li and
  Tobias}{2011}]{li2011bayesian}
\textsc{Li, M. and J.~L. Tobias} (2011): \enquote{Bayesian Methods in
  Microeconometrics,} in \emph{The Oxford Handbook of Bayesian Econometrics},
  ed. by J.~Geweke, G.~Koop, and H.~van Dijk, Oxford: Oxford University Press,
  221--292.

\bibitem[\protect\citeauthoryear{Ludwig and Mullainathan}{Ludwig and
  Mullainathan}{2024}]{ludwig2024machine}
\textsc{Ludwig, J. and S.~Mullainathan} (2024): \enquote{Machine Learning as a
  Tool for Hypothesis Generation,} \emph{The Quarterly Journal of Economics},
  139, 751--827.

\bibitem[\protect\citeauthoryear{Ludwig, Mullainathan, and Rambachan}{Ludwig
  et~al.}{2025}]{ludwig2025large}
\textsc{Ludwig, J., S.~Mullainathan, and A.~Rambachan} (2025): \enquote{Large
  language models: An applied econometric framework,} Tech. rep., National
  Bureau of Economic Research.

\bibitem[\protect\citeauthoryear{Manning, Zhu, and Horton}{Manning
  et~al.}{2024}]{manning2024automated}
\textsc{Manning, B.~S., K.~Zhu, and J.~J. Horton} (2024): \enquote{Automated
  Social Science: Language Models as Scientist and Subjects,} \emph{National
  Bureau of Economic Research}.

\bibitem[\protect\citeauthoryear{Manski}{Manski}{1993}]{manski1993identification}
\textsc{Manski, C.~F.} (1993): \enquote{Identification of Endogenous Social
  Effects: The Reflection Problem,} \emph{The Review of Economic Studies}, 60,
  531--542.

\bibitem[\protect\citeauthoryear{Mogstad and Torgovitsky}{Mogstad and
  Torgovitsky}{2024}]{mogstad2024instrumental}
\textsc{Mogstad, M. and A.~Torgovitsky} (2024): \enquote{Instrumental Variables
  with Unobserved Heterogeneity in Treatment Effects,} \emph{National Bureau of
  Economic Research}.

\bibitem[\protect\citeauthoryear{{Montiel Olea} and Pflueger}{{Montiel Olea}
  and Pflueger}{2013}]{olea2013robust}
\textsc{{Montiel Olea}, J.~L. and C.~Pflueger} (2013): \enquote{A Robust Test
  for Weak Instruments,} \emph{Journal of Business \& Economic Statistics}, 31,
  358--369.

\bibitem[\protect\citeauthoryear{Mullainathan and Rambachan}{Mullainathan and
  Rambachan}{2024}]{mullainathan2024predictive}
\textsc{Mullainathan, S. and A.~Rambachan} (2024): \enquote{From Predictive
  Algorithms to Automatic Generation of Anomalies,} \emph{National Bureau of
  Economic Research}.

\bibitem[\protect\citeauthoryear{Mullainathan and Spiess}{Mullainathan and
  Spiess}{2017}]{mullainathan2017machine}
\textsc{Mullainathan, S. and J.~Spiess} (2017): \enquote{Machine Learning: An
  Applied Econometric Approach,} \emph{Journal of Economic Perspectives}, 31,
  87--106.

\bibitem[\protect\citeauthoryear{Olley and Pakes}{Olley and
  Pakes}{1996}]{olley1996dynamics}
\textsc{Olley, G.~S. and A.~Pakes} (1996): \enquote{The Dynamics of
  Productivity in the Telecommunications Equipment Industry,}
  \emph{Econometrica}, 64, 1263--1297.

\bibitem[\protect\citeauthoryear{Pearl}{Pearl}{2000}]{pearl2000causality}
\textsc{Pearl, J.} (2000): \emph{Causality: Models, Reasoning, and Inference},
  Cambridge, UK: Cambridge University Press.

\bibitem[\protect\citeauthoryear{Rambachan and Roth}{Rambachan and
  Roth}{2023}]{rambachan2023more}
\textsc{Rambachan, A. and J.~Roth} (2023): \enquote{A More Credible Approach to
  Parallel Trends,} \emph{Review of Economic Studies}, 90, 2555--2591.

\bibitem[\protect\citeauthoryear{Si, Yang, and Hashimoto}{Si
  et~al.}{2025}]{si2025novel}
\textsc{Si, C., D.~Yang, and T.~Hashimoto} (2025): \enquote{Can {LLMs} Generate
  Novel Research Ideas? A Large-Scale Human Study with 100+ {NLP} Researchers,}
  in \emph{The Thirteenth International Conference on Learning
  Representations}.

\bibitem[\protect\citeauthoryear{Staiger and Stock}{Staiger and
  Stock}{1997}]{staiger1997instrumental}
\textsc{Staiger, D. and J.~H. Stock} (1997): \enquote{Instrumental Variables
  Regression with Weak Instruments,} \emph{Econometrica}, 65, 557--586.

\bibitem[\protect\citeauthoryear{Stock and Yogo}{Stock and
  Yogo}{2005}]{stock_yogo_2005}
\textsc{Stock, J.~H. and M.~Yogo} (2005): \enquote{Testing for Weak Instruments
  in Linear IV Regression,} in \emph{Identification and Inference for
  Econometric Models: Essays in Honor of Thomas Rothenberg}, ed. by D.~W.
  Andrews and J.~H. Stock, Cambridge: Cambridge University Press, 80--108.

\bibitem[\protect\citeauthoryear{Van~der Klaauw}{Van~der
  Klaauw}{2002}]{vanderklaauw2002estimating}
\textsc{Van~der Klaauw, W.} (2002): \enquote{Estimating the Effect of Financial
  Aid Offers on College Enrollment: A Regression-Discontinuity Approach,}
  \emph{International Economic Review}, 43, 1249--1287.

\bibitem[\protect\citeauthoryear{Wan, Wu, Hu, Chu, and Li}{Wan
  et~al.}{2024}]{wan2024bridging}
\textsc{Wan, G., Y.~Wu, M.~Hu, Z.~Chu, and S.~Li} (2024): \enquote{Large
  Language Models for Causal Discovery: Current Landscape and Future
  Directions,} \emph{arXiv preprint arXiv:2402.11068}.

\bibitem[\protect\citeauthoryear{Wang, Fu, Du, Gao, Huang, Liu, Chandak, Liu,
  Van~Katwyk, Deac et~al.}{Wang et~al.}{2023}]{wang2023scientific}
\textsc{Wang, H., T.~Fu, Y.~Du, W.~Gao, K.~Huang, Z.~Liu, P.~Chandak, S.~Liu,
  P.~Van~Katwyk, A.~Deac, et~al.} (2023): \enquote{Scientific discovery in the
  age of artificial intelligence,} \emph{Nature}, 620, 47--60.

\bibitem[\protect\citeauthoryear{Wu, Terry, and Cai}{Wu
  et~al.}{2022}]{wu2022ai}
\textsc{Wu, T., M.~Terry, and C.~J. Cai} (2022): \enquote{AI Chains:
  Transparent and Controllable Human-AI Interaction by Chaining Large Language
  Model Prompts,} in \emph{Proceedings of the 2022 CHI Conference on Human
  Factors in Computing Systems}, 1--22.

\bibitem[\protect\citeauthoryear{Zhang, Diddee, Holm, Liu, Liu, Samuel, Wang,
  and Ippolito}{Zhang et~al.}{2025}]{zhang2025noveltybench}
\textsc{Zhang, Y., H.~Diddee, S.~Holm, H.~Liu, X.~Liu, V.~Samuel, B.~Wang, and
  D.~Ippolito} (2025): \enquote{{NoveltyBench}: Evaluating Language Models for
  Humanlike Diversity,} in \emph{Second Conference on Language Modeling}.

\end{thebibliography}
}

\end{document}